\documentclass[aps,prb,reprint,notitlepage,noeprint]{revtex4-2}

\usepackage[english]{babel}
\usepackage{amsmath,amssymb,amsfonts,bm,bbm,xcolor,graphicx,comment,txfonts,mathdots}
\definecolor{darkblue}{rgb}{0.,0.,0.5}
\usepackage[colorlinks,linkcolor=darkblue,citecolor=darkblue,urlcolor=darkblue]{hyperref}
\usepackage{orcidlink}

\newcommand{\id}{\mathbbm{1}}

\newcommand{\e}{\mathrm{e}}
\newcommand{\imag}{\mathrm{i}}
\newcommand{\diff}{\mathrm{d}}
\newcommand{\Diff}{\mathrm{D}}

\newcommand{\ket}[1]{\lvert #1 \rangle}
\newcommand{\bra}[1]{\langle #1 \rvert}
\newcommand{\braket}[1]{\langle #1 \rangle}

\newcommand{\svn}[1]{S_{#1}}

\newcommand{\dQG}{\delta Q_{\mathcal{G}}}
\newcommand{\dQS}{\delta Q_{\Sigma}}

\makeatletter
\newcommand*{\transpose}{%
  {\mathpalette\@transpose{}}%
}
\newcommand*{\@transpose}[2]{%
  \raisebox{\depth}{$\m@th#1\intercal$}%
}
\makeatother

\newcommand{\abs}[1]{\left\lvert #1 \right\rvert}
\newcommand{\norm}[1]{\left\lVert #1 \right\rVert}
\newcommand{\tr}{\mathop{\mathrm{tr}}}
\newcommand{\trK}{\mathop{\mathrm{tr}_K}}

\newcommand{\trCR}{\mathop{\mathrm{tr}_{CR}}}
\newcommand{\Tr}{\mathop{\mathrm{Tr}}}

\newcommand{\detR}{\mathop{\mathrm{det}_R}}

\newcommand{\detCR}{\mathop{\mathrm{det}_{CR}}}
\newcommand{\Det}{\mathop{\mathrm{Det}}}
\newcommand{\pf}{\mathop{\mathrm{pf}}}

\newcommand{\pfC}{\mathop{\mathrm{pf}_C}}
\newcommand{\pfCR}{\mathop{\mathrm{pf}_{CR}}}
\newcommand{\pfKC}{\mathop{\mathrm{pf}_{KC}}}
\newcommand{\Pf}{\mathop{\mathrm{Pf}}}
\newcommand{\sgn}{\mathop{\mathrm{sgn}}}

\renewcommand{\Re}{\mathop{\mathrm{Re}}}

\newcommand{\diag}{\mathop{\mathrm{diag}}}

\makeatletter
\DeclareFontFamily{OMX}{MnSymbolE}{}
\DeclareSymbolFont{MnLargeSymbols}{OMX}{MnSymbolE}{m}{n}
\SetSymbolFont{MnLargeSymbols}{bold}{OMX}{MnSymbolE}{b}{n}
\DeclareFontShape{OMX}{MnSymbolE}{m}{n}{
  <-6>  MnSymbolE5
  <6-7>  MnSymbolE6
  <7-8>  MnSymbolE7
  <8-9>  MnSymbolE8
  <9-10> MnSymbolE9
  <10-12> MnSymbolE10
  <12->   MnSymbolE12
}{}
\DeclareFontShape{OMX}{MnSymbolE}{b}{n}{
  <-6>  MnSymbolE-Bold5
  <6-7>  MnSymbolE-Bold6
  <7-8>  MnSymbolE-Bold7
  <8-9>  MnSymbolE-Bold8
  <9-10> MnSymbolE-Bold9
  <10-12> MnSymbolE-Bold10
  <12->   MnSymbolE-Bold12
}{}

\let\llangle\@undefined
\let\rrangle\@undefined
\DeclareMathDelimiter{\llangle}{\mathopen}%
{MnLargeSymbols}{'164}{MnLargeSymbols}{'164}
\DeclareMathDelimiter{\rrangle}{\mathclose}%
{MnLargeSymbols}{'171}{MnLargeSymbols}{'171}
\makeatother

\begin{document}

\title{Replica Keldysh field theory of quantum-jump processes: General formalism and application to imbalanced and inefficient fermion counting}

\author{Felix Kloiber-Tollinger\orcidlink{0009-0009-4799-6048}}

\author{Lukas M. Sieberer\orcidlink{0000-0002-0163-7850}}

\email{lukas.sieberer@uibk.ac.at}

\affiliation{Institute for Theoretical Physics, University of Innsbruck, 6020
Innsbruck, Austria}

\begin{abstract}
  Measurement-induced phase transitions have largely been explored for
  projective or continuous measurements of Hermitian observables, assuming
  perfect detection without information loss. Yet such transitions also arise in
  more general settings, including quantum-jump processes with non-Hermitian
  jump operators, and under inefficient detection. A theoretical framework for
  treating these broader scenarios has been missing. Here we develop a
  comprehensive replica Keldysh field theory for general quantum-jump processes
  in both bosonic and fermionic systems. Our formalism provides a unified
  description of pure-state quantum trajectories under efficient detection and
  mixed-state dynamics emerging from inefficient monitoring, with deterministic
  Lindbladian evolution appearing as a limiting case. It thus establishes a
  direct connection between phase transitions in nonequilibrium steady states of
  driven open quantum matter and in measurement-induced dynamics. As an
  application, we study imbalanced and inefficient fermion counting in a
  one-dimensional lattice system: monitored gain and loss of fermions occurring
  at different rates, with a fraction of gain and loss jumps undetected. For
  imbalanced but efficient counting, we recover the qualitative picture of the
  balanced case: entanglement obeys an area law for any nonzero jump rate, with
  an extended quantum-critical regime emerging between two parametrically
  separated length scales. Inefficient detection introduces a finite correlation
  length beyond which entanglement, as quantified by the fermionic logarithmic
  negativity, obeys an area law, while the subsystem entropy shows volume-law
  scaling. Numerical simulations support our analytical findings. Our results
  offer a general and versatile theoretical foundation for studying
  measurement-induced phenomena across a wide class of monitored and open
  quantum systems.
\end{abstract}

\maketitle

\section{Introduction}
\label{sec:introduction}

In recent years, measurement-induced phase transitions have emerged as a central
theme in the study of nonequilibrium quantum matter. Originally identified in
spin systems evolving under hybrid quantum circuits that combine random unitary
gates with projective measurements~\cite{Li2018, Skinner2019, Li2019,
  Fisher2023}, it is now clear that such transitions arise in far broader
settings. They occur not only in spin but also bosonic and fermionic systems;
not only in discrete circuit models but also in continuous-time Hamiltonian
evolution; and not only under projective measurements of Hermitian observables
but also under continuous monitoring described by quantum state diffusion or
quantum-jump processes, including jump processes with non-Hermitian jump
operators~\cite{Cao2019, Fuji2020, Goto2020, Tang2020, Lang2020, Rossini2020,
  Alberton2021, Bao2021, Buchhold2021, Botzung2021, Turkeshi2021, Jian2022,
  Muller2022, Kells2023, Poboiko2023, Jian2023, Fava2023, Merritt2023,
  Chahine2024, Poboiko2024, Tsitsishvili2024, Fava2024, Turkeshi2024, Guo2025,
  Poboiko2025, Starchl2025, Muller2025, Soares2025, Fan2025,
  Niederegger2025}. At the same time, the stability of measurement-induced
critical phenomena under imperfect detection has become an active topic of
research~\cite{Ladewig2022, Minoguchi2022, Passarelli2024, Leung2025,
  Leung2025a, Paviglianiti2025}.

While these investigations are often numerical in nature, substantial progress
has also been achieved in the development of analytical methods for
understanding measurement-induced phase transitions. These methods typically
rely on the replica trick, familiar from the theory of spin
glasses~\cite{Mezard1986} and disordered electronic systems~\cite{Evers2008}. In
spin systems evolving under circuit dynamics, the replica trick enables mappings
to classical statistical-mechanics models~\cite{Bao2020, Jian2020}; in fermionic
systems, analytical progress has predominantly come from replica field-theoretic
approaches~\cite{Buchhold2021, Ladewig2022, Muller2022, Jian2023, Poboiko2023,
  Poboiko2024, Fava2023, Chahine2024, Fava2024, Guo2025, Poboiko2025,
  Starchl2025, Muller2025, Niederegger2025}. Replica Keldysh field
theory~\cite{Altland2010a, Sieberer2016a, Sieberer2025, Thompson2023,
  Kamenev2023}, in particular, has enabled major progress in understanding
measurement-induced phenomena in both free and interacting fermions. However,
this framework has so far been limited to quantum state diffusion and to random
projective or generalized measurements~\cite{Nielsen2012}, assuming perfect
detection. A general theoretical framework for quantum-jump processes, including
the ability to incorporate imperfect measurements~\cite{Gardiner2014,
  Gardiner2015, Wiseman2010, Jacobs2014}, has remained unavailable.

We close this gap by developing a replica Keldysh field theory for
\emph{general} quantum-jump processes. Our formalism applies to bosonic and
fermionic many-body systems, including interacting ones, evolving under
continuous monitoring with arbitrary jump operators. A central technical
challenge is performing the average over quantum trajectories within the replica
field-theory framework. This requires averaging over both jump times and jump
types. Prior work resolved this issue only in settings with externally imposed
measurement rates, such as random projective~\cite{Poboiko2023} or generalized
measurements~\cite{Starchl2025}. In contrast, the physically relevant jump rates
in quantum-jump trajectories are state-dependent and thus fluctuate in time,
necessitating a new approach, which we provide.

As a key example of measurement imperfections, we incorporate inefficient
detection, where only a fraction of quantum jumps is registered, into our
formalism~\cite{Gardiner2014, Gardiner2015, Wiseman2010, Jacobs2014}. The
detection efficiency $\eta \in [0,1]$ interpolates between fully efficient
monitoring at $\eta = 1$, which yields pure-state trajectories, and fully
inefficient monitoring at $\eta = 0$, where the dynamics reduce to deterministic
Lindbladian evolution. In this limit, Hermitian jump operators, as realized by
random projective and generalized measurements, typically drive the system to an
infinite-temperature steady state, thereby erasing all signatures of the
measurement-induced phenomena present at $\eta = 1$. By contrast, non-Hermitian
jump operators can stabilize nontrivial stationary states even at $\eta = 0$,
which represents the natural setting of driven open quantum matter. The Keldysh
formalism provides a well-established framework for analyzing the resulting
nonequilibrium steady states in this regime~\cite{Sieberer2016a, Sieberer2025,
  Thompson2023, Kamenev2023}.

Incorporating inefficient detection into our framework therefore serves a dual
purpose. First, it enables analytical progress in understanding the impact of
imperfect monitoring on measurement-induced phenomena. Second, it establishes a
conceptual and technical bridge between measurement-induced dynamics and the
nonequilibrium steady states of driven open quantum systems. Providing this link
is a first step toward a unified perspective on phases and phase transitions in
continuously monitored and driven open quantum matter, as well as on broader
questions such as topology and symmetry classifications~\cite{Lieu2020,
  Altland2021, Gneiting2022, Sa2023, Kawabata2023, Xiao2024, Bhuiyan2025}.

As an illustration of the formalism, we analyze imbalanced and inefficient
fermion counting in a one-dimensional (1D) lattice system. We use the term
``fermion counting'' in analogy with photon counting in quantum
optics~\cite{Gardiner2014, Gardiner2015, Wiseman2010, Jacobs2014} to denote
continuously monitored fermion gain and loss. Technically, fermion counting is
described by a quantum-jump process with quantum jumps generated by fermionic
annihilation and creation operators. Reference~\cite{Starchl2025} studied the
\emph{balanced and efficient} case, where gain and loss occur at equal rates,
$\gamma_+ = \gamma_-$, and each quantum jump is detected. For balanced rates,
fermion counting can be formulated as a random generalized measurement. This
formulation formed the basis for the analytical approach of
Ref.~\cite{Starchl2025}. Building on the formalism developed in the present
paper, we extend this analysis in two important directions:

(i)~We consider \emph{imbalanced} fermion counting with
$\gamma_+ \neq \gamma_-$, a setting more natural for potential experimental
implementations where fermion gain and loss would be induced by coupling to two
distinct reservoirs acting as particle drain and source, respectively. In this
scenario, balanced coupling represents a fine-tuned limit. We show analytically
and numerically that the qualitative behavior identified in
Ref.~\cite{Starchl2025} persists without fine tuning: the model exhibits no
measurement-induced phase transition. For any nonzero gain and loss rates,
$\gamma_\pm > 0$, the entanglement entropy obeys an area law. However, area-law
scaling appears only beyond a length scale exponentially large in the inverse
mean rate $\gamma = (\gamma_+ + \gamma_-)/2$.  On intermediate scales, between
$l_0 \sim \gamma^{-1}$ and $l_c \sim \gamma^{-2}$, the system displays
signatures of quantum criticality and emergent conformal invariance. For scales
larger than $l_c$, the system undergoes a slow crossover to the area-law
regime. This behavior is captured by a long-wavelength nonlinear sigma model
(NLSM) describing Goldstone-mode fluctuations associated with breaking an
$\mathrm{SU}(R)$ replica symmetry. In the presence of particle-hole symmetry, as
in the nearest-neighbor hopping model with real amplitudes studied here, the
NLSM target manifold changes from $\mathrm{SU}(R)$ to
$\mathrm{SU}(2R)/\mathrm{Sp}(R)$, but the qualitative physics remains the
same~\cite{Jian2022, Fava2024, Poboiko2025}.

(ii)~We examine \emph{inefficient} fermion counting, where only a fraction
$\eta$ of gain and loss jumps is detected. Undetected jumps generate additional
time-continuous incoherent gain and loss. Interestingly, these processes
explicitly break the $\mathrm{SU}(R)$ symmetry and introduce a mass term in the
NLSM Lagrangian proportional to the detection inefficiency
$\delta \eta = 1 - \eta$. This mass induces a correlation length
$\xi \sim \delta \eta^{-1/2}$, beyond which correlations decay
exponentially. The von Neumann entropy exhibits volume-law scaling, reflecting
the mixed character of the trajectories, while genuine entanglement as
quantified by the fermionic logarithmic negativity~\cite{Shapourian2017,
  Shapourian2019} retains area-law behavior on scales larger than
$\xi$. Numerical simulations are made feasible by the fact that incoherent gain
and loss preserve Gaussianity of the mixed states, allowing us to access system
sizes up to $L = 1000$. This contrasts with inefficient monitoring of occupation
numbers, where Gaussianity is lost and the simulation cost increases
dramatically~\cite{Ladewig2022}.

The paper is organized as follows. In Sec.~\ref{sec:quantum-jump-processes}, we
introduce the Kraus-operator formulation of quantum-jump
processes. Section~\ref{sec:replica-formalism} briefly reviews the replica
formalism used to compute trajectory averages of observables that are nonlinear
in the quantum state such as the von Neumann entropy. Building on this
foundation, Sec.~\ref{sec:replica-Keldysh} presents the derivation of the
replica Keldysh field theory for quantum-jump processes. The resulting theory
applies to pure-state trajectories under efficient detection;
Sec.~\ref{sec:inefficient-detection} explains how to extend it to the case of
inefficient detection. A minimal physical model that leads to discrete-time
dynamics in terms of Kraus operators is discussed in
Appendix~\ref{sec:minimal-physical-model}. This completes the general formal
developments of the paper. In Sec.~\ref{sec:imbalanced-inefficient-counting}, we
apply the general formalism to imbalanced and inefficient fermion counting. We
benchmark the analytical results obtained from replica Keldysh field theory
against direct numerical simulations of quantum-jump trajectories. Technical
details are given in Appendices~\ref{sec:HS-transformation-PHS}
and~\ref{sec:numerical-methods}. Conclusions are presented in
Sec.~\ref{sec:conclusions}.

\section{Quantum-jump processes}
\label{sec:quantum-jump-processes}

We begin by introducing the description of discrete-time quantum-jump processes
in terms of Kraus operators. In the main text, we focus on formal aspects; we
review the physical realization of quantum-jump processes in
Appendix~\ref{sec:minimal-physical-model}.

\subsection{Kraus operators}
\label{sec:kraus-operators}

We consider the dynamics of a fermionic or bosonic quantum system subjected to
continuous measurements. At time $t_0$ the system is initialized in the pure
state $\ket{\psi_0}$. The time interval from $t_0$ to $t = t_0 + T$ is
partitioned into $N$ time steps of duration $\Delta t = T/N$. Evolution during a
single time step is described by a set of Kraus operators $\hat{K}_{\alpha}$
with $\alpha \in \mathsf{A}_0 = \{0,\dotsc,A\}$. The operator $\hat{K}_0$
corresponds to no quantum jump occurring during the time step and is defined
by~\cite{Jacobs2014}
\begin{equation}
  \label{eq:K-0}
  \hat{K}_0 = 1 - \imag \left( \hat{H} \Delta t
    - \frac{\imag}{2} \sum_{a = 1}^A \hat{K}_a^{\dagger} \hat{K}_a \right)
  = 1 - \imag \hat{H}_{\mathrm{eff}} \Delta t ,
\end{equation}
where $\hat{H}$ is the Hamiltonian. No-jump evolution described by $\hat{K}_0$
is continuous in the sense that $\hat{K}_0 \to 1$ for $\Delta t \to 0$. In
contrast, quantum jumps, which are described by the remaining Kraus operators
$\hat{K}_a$ with $a \in \mathsf{A} = \{1,\dotsc,A\}$, induce discontinuous
evolution with
\begin{equation}
  \label{eq:K-a}
  \hat{K}_a = \sqrt{\gamma_a \Delta t} \, \hat{c}_a.
\end{equation}
We do not impose any restrictions on the jump operators $\hat{c}_a$: they can be
arbitrary fermionic or bosonic operators. In terms of the jump operators, the
effective non-Hermitian Hamiltonian introduced in Eq.~\eqref{eq:K-0} is given by
\begin{equation}
  \label{eq:H-eff}
  \hat{H}_{\mathrm{eff}} = \hat{H} - \frac{\imag}{2} \sum_{a = 1}^A \gamma_a
  \hat{c}_a^{\dagger} \hat{c}_a .
\end{equation}
Throughout, Greek indices label the full set of Kraus operators,
while Latin indices label only those corresponding to quantum jumps:
\begin{equation}
  \alpha \in \mathsf{A}_0 = \{0,\dotsc,A\},
  \qquad
  a \in \mathsf{A} = \{1,\dotsc,A\}.
\end{equation}

The Kraus operators satisfy the completeness relation
\begin{equation}
  \label{eq:Kraus-operators-completeness-relation}
  \sum_{\alpha=0}^A \hat{K}_{\alpha}^{\dagger} \hat{K}_{\alpha} = 1 ,
\end{equation}
up to corrections that vanish in the continuous-time limit $N \to \infty$ such
that $\Delta t \to 0$. Consequently, the set of Kraus operators defines a
quantum channel describing the deterministic evolution of the density matrix
over a single time step,
\begin{equation}
  \label{eq:quantum-channel}
  \hat{\rho} \mapsto
  \sum_{\alpha = 0}^A \hat{K}_{\alpha} \hat{\rho} \hat{K}_{\alpha}^{\dagger}
  = \left(1 + \mathcal{L} \Delta t \right) \hat{\rho},
\end{equation}
starting from the initial pure state $\hat{\rho}_0 = \ket{\psi_0}
\bra{\psi_0}$. The Liouville superoperator or Lindbladian is
\begin{equation}
  \label{eq:Liouvillian}
  \mathcal{L} \hat{\rho} = -\imag \left[ \hat{H}, \hat{\rho} \right] + \sum_{a =
    1}^A \gamma_a \left( \hat{c}_a \hat{\rho} \hat{c}_a^{\dagger} - \frac{1}{2}
    \left\{ \hat{c}_a^{\dagger} \hat{c}_a, \hat{\rho} \right\} \right).
\end{equation}
In the continuous-time limit, Eq.~\eqref{eq:quantum-channel} can be recast as a
Markovian quantum master equation in Lindblad form.

\subsection{Quantum trajectories}
\label{sec:quantum-trajectories}

Formally, a quantum-jump process is a stochastic unraveling of the quantum
channel~\eqref{eq:quantum-channel}. As a computational tool, such unravelings
find widespread application in the numerical simulation of open-system
dynamics~\cite{Plenio1998, Daley2014}. We begin by considering the unraveling
into pure-state quantum trajectories corresponding to perfectly efficient
detection of quantum jumps; modifications arising from finite detection
efficiency will be introduced further below in
Sec.~\ref{sec:inefficient-detection}.

In such a stochastic unraveling, one Kraus operator $\hat{K}_\alpha$ is drawn at
random and applied to the state during each time step. For a system in the state
$\ket{\psi}$, the operator $\hat{K}_\alpha$ is chosen with probability
\begin{equation}
  \label{eq:P-alpha}
  P_{\alpha}
  = \norm{\hat{K}_{\alpha} \ket{\psi}}^2
  = \braket{\psi | \hat{K}_{\alpha}^{\dagger} \hat{K}_{\alpha} | \psi}.
\end{equation}
If $\alpha = 0$ meaning that no jump occurs during the time step, the system
evolves continuously according to
\begin{equation}
  \label{eq:no-jump-update}
  \ket{\psi} \mapsto
  \frac{\hat{K}_0 \ket{\psi}}{\sqrt{P_0}}
  = \left[
      1 - \imag
      \left(
        \hat{H}_{\mathrm{eff}}
        - \imag
        \sum_{a=1}^A
          \gamma_a \braket{\psi | \hat{c}_a^{\dagger}\hat{c}_a | \psi}
      \right)\Delta t
    \right] \ket{\psi},
\end{equation}
Otherwise, for $\alpha = a \in \{ 1, \dotsc, A \}$, the system undergoes a
discontinuous quantum jump described by
\begin{equation}
  \label{eq:jump-update}
  \ket{\psi} \mapsto
  \frac{\hat{K}_a \ket{\psi}}{\sqrt{P_a}}
  = \frac{\hat{c}_a \ket{\psi}}{\lVert \hat{c}_a \ket{\psi} \rVert}.
\end{equation}
While we introduce quantum-jump processes here from a formal perspective, a
microscopic derivation is presented in
Appendix~\ref{sec:minimal-physical-model}. There, we consider a minimal setup in
which the system is coupled to independent bosonic or fermionic baths, one for
each jump channel. Bath occupations are monitored continuously: Detection of no
excitation in any bath occurs with probability $P_0$ and yields no-jump
evolution, whereas detection of an excitation in bath $a$, with probability
$P_a$, induces the corresponding jump~\cite{Gardiner2014, Gardiner2015,
  Wiseman2010, Jacobs2014}. Continuously measuring bath occupations thus
produces precisely the stochastic unraveling of Eq.~\eqref{eq:quantum-channel}
described above. For this reason, we refer to the index $\alpha$ specifying the
evolution during a given time step as a measurement outcome.

A quantum trajectory over $N$ time steps is specified by a sequence of
measurement outcomes $\alpha_n$ with $n \in \mathsf{N} = \{1,\dotsc,N\}$,
collected into the vector
$\boldsymbol{\alpha} = (\alpha_1, \dotsc, \alpha_N) \in \mathsf{A}_0^N$. The
probability for this sequence is
\begin{equation}
  \label{eq:P-alpha-trajectory}
  P_{\boldsymbol{\alpha}}
  = \norm{\hat{K}_{\alpha_N} \dotsb \hat{K}_{\alpha_1} \ket{\psi_0}}^2,
\end{equation}
and the corresponding normalized state is
\begin{equation}
  \label{eq:normalized-pure-trajectory}
  \ket{\psi_{\boldsymbol{\alpha}}} = \left. \hat{K}_{\alpha_N} \dotsb
    \hat{K}_{\alpha_1} \ket{\psi_0} \middle/ \sqrt{P_{\boldsymbol{\alpha}}}
  \right..
\end{equation}
This expression describes the evolution of the quantum state \emph{conditioned}
on a specific sequence of measurement outcomes. By contrast,
Eq.~\eqref{eq:quantum-channel} captures the \emph{unconditional} dynamics, in
which one averages over all possible outcomes.

Observables of interest are functions $f_{\boldsymbol{\alpha}}$ of the
normalized state or, more generally, of the measurement record itself. The
average of $f_{\boldsymbol{\alpha}}$ over the ensemble of trajectories is
\begin{equation}
  \label{eq:f-alpha-P-alpha-trajectory}
  \overline{f_{\boldsymbol{\alpha}}}
  = \sum_{\boldsymbol{\alpha} \in \mathsf{A}_0^N}
      P_{\boldsymbol{\alpha}} f_{\boldsymbol{\alpha}}.
\end{equation}
For observables linear in the state, such as the expectation value
$f_{\boldsymbol{\alpha}} = \braket{\psi_{\boldsymbol{\alpha}} | \hat{O} |
  \psi_{\boldsymbol{\alpha}}}$, this reduces to
\begin{equation}
  \overline{\braket{\psi_{\boldsymbol{\alpha}} | \hat{O} |
      \psi_{\boldsymbol{\alpha}}}}
  = \tr(\hat{O}\hat{\rho}),
  \qquad
  \hat{\rho}
  = \sum_{\boldsymbol{\alpha} \in \mathsf{A}_0^N}
      P_{\boldsymbol{\alpha}}
      \ket{\psi_{\boldsymbol{\alpha}}}\bra{\psi_{\boldsymbol{\alpha}}}.
\end{equation}
That is, the expectation value may be expressed in terms of the
ensemble-averaged density matrix. Nonlinear observables, such as the von Neumann
entropy of a subsystem, do not admit such a simplification.

This framework encompasses random projective~\cite{Poboiko2023} or generalized
measurements~\cite{Starchl2025}, performed at a fixed rate $\gamma$, as a
special case. In this setting, the rates $\gamma_a$ in Eq.~\eqref{eq:K-a} are
identical and equal to the measurement rate, $\gamma_a = \gamma$, and the jump
operators are measurement operators and thus obey the completeness relation
\begin{equation}
  \label{eq:measurement-operators-completeness-relation}
  \sum_{a=1}^A \hat{c}_a^{\dagger}\hat{c}_a = 1,
\end{equation}
in addition to the Kraus-operator completeness
relation~\eqref{eq:Kraus-operators-completeness-relation}. The completeness
relation~\eqref{eq:measurement-operators-completeness-relation} leads to the
total probability of a jump occurring in a single time step being
state-independent,
\begin{equation}
  P_{\mathrm{jump}} = \sum_{a \in \mathsf{A}} P_a = \gamma \Delta t.
\end{equation}
Therefore, jumps may indeed be identified with measurements performed at a fixed
rate $\gamma$. For projective measurements, the measurement operators
additionally satisfy $\hat{c}_a = \hat{c}_a^{\dagger} = \hat{c}_a^2$ and project
onto eigenspaces of the measured observable. Then, the
Liouvillian~\eqref{eq:Liouvillian} takes the form
\begin{equation}
  \mathcal{L}\hat{\rho}
  = - \imag \left[ \hat{H},\hat{\rho} \right]
  + \gamma \left(\sum_{a=1}^A \hat{c}_a \hat{\rho}\hat{c}_a - \hat{\rho}\right).
\end{equation}
The completely mixed state $\hat{\rho} = 1/D$, with $D$ the Hilbert-space
dimension, is stationary under evolution generated by this Liouvillian and, for
spatially extended systems with local measurements, is generically the unique
steady state---independent of all system parameters, including the measurement
rate~$\gamma$. Consequently, random projective measurements yield nontrivial
effects only in nonlinear trajectory-dependent observables. By contrast, for
general non-Hermitian jump operators, the steady state of
Eq.~\eqref{eq:Liouvillian} is typically nontrivial, implying that both linear
and nonlinear observables depend sensitively on system parameters.

\subsection{Rescaled Kraus operators and continuous-time limit}
\label{sec:rescaled-Kraus-operators}

We have introduced quantum-jump processes in discrete time, which provides a
convenient starting point for the construction of a functional-integral
representation of the trajectory average~\eqref{eq:f-alpha-P-alpha-trajectory}
presented in Sec.~\ref{sec:replica-Keldysh}. However, in this construction, we
ultimately want to take the continuous-time limit. Therefore, we now discuss how
Eq.~\eqref{eq:f-alpha-P-alpha-trajectory} can be rewritten in a form that is
suitable for taking this limit.

Taking the continuous-time limit is facilitated by introducing rescaled Kraus
operators,
\begin{equation}
  \label{eq:J-alpha-q-0-q-a}
  \hat{J}_{\alpha}
  = \left. \hat{K}_{\alpha} \middle/ \sqrt{q_{\alpha}} \right.,
  \qquad
  q_0 = 1 - \Gamma_0 \Delta t,
  \qquad
  q_a = \Gamma_a \Delta t,
\end{equation}
which take the explicit form
\begin{equation}
  \label{eq:J-alpha}
  \hat{J}_0
  = 1 - \imag \left( \hat{H}_{\mathrm{eff}}
    + \frac{\imag \Gamma_0}{2} \right) \Delta t,
  \qquad
  \hat{J}_a = \sqrt{\frac{\gamma_a}{\Gamma_a}} \, \hat{c}_a.
\end{equation}
The primary purpose of the rescaling is to remove the explicit dependence of the
quantum jump Kraus operators on the time step $\Delta t$. Whereas the operators
$\hat{K}_a$ in Eq.~\eqref{eq:K-a} depend explicitly on $\Delta t$, this
dependence is factored out in the rescaled operators $\hat{J}_a$. As discussed
in Sec.~\ref{sec:replica-formalism}, this construction enables a nontrivial
continuous-time limit for replica numbers $R > 1$. The rates $\Gamma_0$ and
$\Gamma_a$, by contrast, are introduced to ensure that the rescaled Kraus
operators remain dimensionless; the values of these rates may be chosen
freely. At first sight, the most natural choice is $\Gamma_0 = 0$ and
$\Gamma_a = \gamma_a$. Below, however, we consider two alternative schemes: a
\emph{uniform rescaling}, in which all quantum jump Kraus operators are assigned
the same rate, $\Gamma_a = \Gamma_1$, and a \emph{detailed rescaling}, in which
the rates $\Gamma_a$ are arbitrary. The latter includes the choice
$\Gamma_0 = 0$ and $\Gamma_a = \gamma_a$ as a special case. Although these
schemes lead to slightly different intermediate representations of the replica
Keldysh field theory, the differences disappear in the replica limit. We
nevertheless discuss them separately because they admit distinct intermediate
interpretations.

In terms of the rescaled Kraus operators, the trajectory probability
$P_{\boldsymbol{\alpha}}$ in Eq.~\eqref{eq:P-alpha-trajectory} factorizes as
$P_{\boldsymbol{\alpha}} = Q_{\boldsymbol{\alpha}} O_{\boldsymbol{\alpha}}$,
with
\begin{equation}
  \label{eq:Q-alpha-O-alpha}
  Q_{\boldsymbol{\alpha}}
  = \prod_{n=1}^N q_{\alpha_n},
  \qquad
  O_{\boldsymbol{\alpha}}
  = \left\|
      \hat{J}_{\alpha_N} \cdots \hat{J}_{\alpha_1} \ket{\psi_0}
    \right\|^2.
\end{equation}
The average of $f_{\boldsymbol{\alpha}}$ can then be written as
\begin{equation}
  \label{eq:f-alpha-Q-alpha-O-alpha}
  \overline{f_{\boldsymbol{\alpha}}}
  = \sum_{\boldsymbol{\alpha} \in \mathsf{A}_0^N}
  Q_{\boldsymbol{\alpha}}
  O_{\boldsymbol{\alpha}}
  f_{\boldsymbol{\alpha}}.
\end{equation}
This expression naturally suggests interpreting the average as being taken with
respect to the distribution $Q_{\boldsymbol{\alpha}}$, which is defined in
Eq.~\eqref{eq:Q-alpha-O-alpha} as a product of the probabilities $q_0$ and
$q_a$, Eq.~\eqref{eq:J-alpha-q-0-q-a}, for no jump and a jump of type $a$ to
occur in a given time step. Under this interpretation, the quantity being
averaged is the product $O_{\boldsymbol{\alpha}} f_{\boldsymbol{\alpha}}$. It is
important to stress, however, that $Q_{\boldsymbol{\alpha}}$ is \emph{not} the
physical probability of the trajectory $\boldsymbol{\alpha}$; the physical
probability remains $P_{\boldsymbol{\alpha}}$.

We next rewrite Eq.~\eqref{eq:f-alpha-Q-alpha-O-alpha} in a form suitable for
taking the continuous-time limit, first considering the case of uniform
rescaling and subsequently turning to the more general case of detailed
rescaling.

\subsubsection{Uniform rescaling}
\label{sec:uniform-rescaling}

For uniform rescaling, the probabilities $q_a$ all take the same value
$q_1 = \Gamma_1 \Delta t$. Then, the probability $Q_{\boldsymbol{\alpha}}$ in
Eq.~\eqref{eq:Q-alpha-O-alpha} depends only on the total number of jumps
$M_{\boldsymbol{\alpha}}$, that is, the number of measurement outcomes
$\alpha_n \neq 0$ in the sequence $\boldsymbol{\alpha}$,
\begin{equation}
  \label{eq:Q-M-alpha}
  Q_{\boldsymbol{\alpha}} = Q_{M_{\boldsymbol{\alpha}}} = q_0^{N -
    M_{\boldsymbol{\alpha}}} q_1^{M_{\boldsymbol{\alpha}}}.
\end{equation}
To perform the average over trajectories in
Eq.~\eqref{eq:f-alpha-Q-alpha-O-alpha}, we choose a different parameterization
of trajectories: instead of specifying the values of $\alpha_n$ in each time
step, we fully determine a trajectory by specifying the total number of jumps
$M$, the time steps $n_m$ with $m \in \{ 1, \dotsc, M \}$ or corresponding times
$t_m = t_0 + n_m \Delta t$ at which the jumps occur, and the types of jumps
$a_m \in \mathsf{A}$. We collect the time-step and jump indices in vectors
$\mathbf{n} = \left( n_1, \dotsc, n_M \right)$ with $n_1 < \dotsb < n_M$ and
$\mathbf{a} = \left( a_1, \dotsc, a_M \right)$. These vectors uniquely determine
a vector $\boldsymbol{\alpha}_{\mathbf{n}, \mathbf{a}}$ of length $N$: all
elements with index $n$ not contained in $\mathbf{n}$ vanish
$\left( \boldsymbol{\alpha}_{\mathbf{n}, \mathbf{a}} \right)_n = 0$,
corresponding to no-jump evolution, whereas
$\left( \boldsymbol{\alpha}_{\mathbf{n}, \mathbf{a}} \right)_{n_m} = a_m$. We
can thus write the average~\eqref{eq:f-alpha-Q-alpha-O-alpha} as
\begin{equation}
  \label{eq:f-alpha-Q-M-alpha-O-alpha}
  \begin{split}
    \overline{f_{\boldsymbol{\alpha}}} = \sum_{M = 0}^N Q_M \sum_{\mathbf{a} \in
      \mathsf{A}^M} \sum_{\mathbf{n} \in \mathsf{O}_{M, N}}
    O_{\boldsymbol{\alpha}_{\mathbf{n}, \mathbf{a}}}
    f_{\boldsymbol{\alpha}_{\mathbf{n}, \mathbf{a}}},
  \end{split}
\end{equation}
where the third sum is over all ordered $M$-tuples with each element in
$\mathsf{N} = \{ 1, \dotsc, N \}$,
\begin{equation}
  \mathsf{O}_{M, N} = \left\{ \mathsf{N}^M : n_1 < \dotsb < n_M
  \right\}.
\end{equation}
As stated above, we want to bring this in a form that enables taking the
continuous-time limit. For the quantities
$O_{\boldsymbol{\alpha}_{\mathbf{n}, \mathbf{a}}}$ and
$f_{\boldsymbol{\alpha}_{\mathbf{n}, \mathbf{a}}}$, this is achieved by noting
that they are functions of the jump times,
$\mathbf{t} = \left( t_1, \dotsc, t_M \right)$, with
$t_m = t_0 + n \Delta t \in [t_0, t]$, and the jump indices,
\begin{equation}
  \label{eq:O-f-functions}
  O_{\boldsymbol{\alpha}_{\mathbf{n}, \mathbf{a}}} = O_{\mathbf{a}}(\mathbf{t}),
  \qquad f_{\boldsymbol{\alpha}_{\mathbf{n}, \mathbf{a}}} =
  f_{\mathbf{a}}(\mathbf{t}).
\end{equation}
Furthermore, in the sums in Eq.~\eqref{eq:f-alpha-Q-M-alpha-O-alpha}, we first
remove the restriction of the elements of $\mathbf{n}$ being ordered and extend
the summation over all $\mathbf{n} \in \mathsf{N}^M$. When we do this, we have
to include a factor $d_{\mathbf{n}}$ that ensures that all elements $n_m$ are
different, that is, $d_{\mathbf{n}} = 0$ if $n_m = n_{m'}$ for at least one pair
of elements, and $d_{\mathbf{n}} = 1$ otherwise. Further, for any given ordered
sequence $\mathbf{n}$, the unrestricted sum contains all possible $M!$
permutations of the elements $n_m$. Therefore, we have to include a factor
$1/M!$ to compensate the resultant overcounting,
\begin{equation}
  \label{eq:f-alpha-Q-M-alpha-d-n-O-alpha}
  \overline{f_{\boldsymbol{\alpha}}} = \sum_{M = 0}^N \frac{1}{M!} Q_M \sum_{\mathbf{a} \in
    \mathsf{A}^M} \sum_{\mathbf{n} \in \mathsf{N}^M}
  d_{\mathbf{n}} O_{\mathbf{a}}(\mathbf{t}) f_{\mathbf{a}}(\mathbf{t}).
\end{equation}
In the continuous-time limit, the sums over the components of $\mathbf{n}$
become integrals. The factor $d_{\mathbf{n}}$ has support on a set of measure
zero in the domain of integration $[t_0, t]^M$, and may thus be dropped,
\begin{equation}  
  \sum_{\mathbf{n} \in \mathsf{N}^M} d_{\mathbf{n}} = \frac{1}{\Delta
    t^M} \sum_{n_1 = 1}^N \Delta t \dotsb \sum_{n_M = 1}^N \Delta t \,
  d_{\mathbf{n}} \sim N^M \prod_{m = 1}^M \int_{t_0}^t \frac{\diff
    t_m}{T}.
\end{equation}
Combining the prefactor $N^M$ with $Q_M/M!$ in
Eq.~\eqref{eq:f-alpha-Q-M-alpha-d-n-O-alpha}, we obtain in the limit
$N \to \infty$ a Poisson distribution for the number of jumps during the
evolution time $T = t - t_0$,
\begin{equation}
  \frac{N^M}{M!} Q_M = \frac{\left( \Gamma_1 T \right)^M}{M!} \left( 1 -
    \frac{\Gamma_0 T}{N} \right)^{N - M} \to \left( \frac{\Gamma_1}{\Gamma_0}
  \right)^M P_M(T),
\end{equation}
where
\begin{equation}
  \label{eq:P-M-T}
  P_M(T) = \frac{\left( \Gamma_0 T \right)^M}{M!} \e^{- \Gamma_0 T},
\end{equation}
and we have used
\begin{equation}
  \left( 1 - \frac{\Gamma_0 T}{N} \right)^N \to \e^{- \Gamma_0 T}, \qquad \left( 1 -
    \frac{\Gamma_0 T}{N} \right)^{- M} \to 1.
\end{equation}
Therefore, in the continuous-time limit,
Eq.~\eqref{eq:f-alpha-Q-M-alpha-d-n-O-alpha} becomes
\begin{equation}
  \label{eq:uniform-rescaling-continuous-time-limit}
  \overline{f_{\boldsymbol{\alpha}}} \sim \sum_{M = 0}^{\infty} \left(
    \frac{\Gamma_1}{\Gamma_0} \right)^M P_M(T) \left( \prod_{m = 1}^M \int_{t_0}^t
  \frac{\diff t_m}{T} \sum_{a_m = 1}^A \right)
  O_{\mathbf{a}}(\mathbf{t}) f_{\mathbf{a}}(\mathbf{t}).
\end{equation}
This expression is directly analogous to the average over random projective or
generalized measurements performed at a rate $\gamma$~\cite{Poboiko2023,
  Starchl2025}. To make the analogy explicit, consider for concreteness a 1D
fermionic lattice of size $L$ with local measurements.  Quantum jumps correspond
to applying measurement operators, and the jump-type index $a$ becomes a double
index, where $n$ labels the measurement outcome and $l$ labels the lattice
site. We focus here on projective measurements of local occupation numbers with
outcomes $n \in \{ 0, 1 \}$. For the choice $\Gamma_1 = \gamma$ and
$\Gamma_0 = L \Gamma_1$, Eq.~\eqref{eq:uniform-rescaling-continuous-time-limit}
takes the form
\begin{equation}  
  \overline{f_{\boldsymbol{\alpha}}} \sim \sum_{M = 0}^{\infty} P_M(T) \left(
    \prod_{m = 1}^M \int_{t_0}^t \frac{\diff t_m}{L T} \sum_{l_m = 1}^L
    \sum_{n_m = 0}^1 \right) O_{\mathbf{a}}(\mathbf{t})
  f_{\mathbf{a}}(\mathbf{t}).
\end{equation}
This representation has a clear interpretation: when measurements occur at rate
$\gamma$, the number of measurements in a time interval $T$ is Poisson
distributed~\eqref{eq:P-M-T} with mean $\gamma T$. Conditional on $M$
measurements, the measurement times are uniformly distributed in
$t_m \in [t_0, t]$, and one further averages over the measurement locations and
outcomes. This reproduces precisely the ensemble average employed in
Refs.~\cite{Poboiko2023, Starchl2025}.

More generally, an analogous interpretation applies to
Eq.~\eqref{eq:uniform-rescaling-continuous-time-limit} even for non-Hermitian
jump operators, provided one chooses $\Gamma_0 = A \Gamma_1$. In this view, the
expression resembles an average over the number of jumps, their times, and
their types. It is important to stress, however, that this should be regarded
only as an interpretation suggested by the algebraic structure of
Eq.~\eqref{eq:uniform-rescaling-continuous-time-limit}. Physically, jumps do not
occur at the externally chosen rate $\Gamma_0$. The true probability of a
trajectory with a given jump sequence is determined by
Eq.~\eqref{eq:P-alpha-trajectory}, which is manifestly independent of the
artificial rates $\Gamma_0$ and $\Gamma_1$ introduced through the rescaling in
Eq.~\eqref{eq:J-alpha-q-0-q-a}. In the replica Keldysh functional integral
developed in Sec.~\ref{sec:replica-Keldysh}, we will see explicitly that these
rates indeed drop out.

Finally, we consider the special case that we average over an observable that
factorizes over jumps:
\begin{equation}
  O_{\mathbf{a}}(\mathbf{t}) f_{\mathbf{a}}(\mathbf{t}) = \prod_{m = 1}^M g_{a_m}(t_m).
\end{equation}
Then, using
\begin{equation}
  \label{eq:uniform-factorization}
  \prod_{m = 1}^M \int_{t_0}^t \frac{\diff t_m}{T} \sum_{a_m = 1}^A g_{a_m}(t_m)
  = \left[ \int_{t_0}^t \frac{\diff t'}{T} \sum_{a = 1}^A g_a(t') \right]^M,
\end{equation}
Eq.~\eqref{eq:uniform-rescaling-continuous-time-limit} can be rewritten as
\begin{equation}
  \label{eq:uniform-rescaling-continuous-time-limit-factorized}
  \overline{f_{\boldsymbol{\alpha}}} \sim \exp \! \left\{ \int_{t_0}^t \diff t'
    \left[ \Gamma_1 \sum_{a = 1}^A g_a(t') - \Gamma_0 \right] \right\}.
\end{equation}
Note that the factor $T^{-M}$ in Eq.~\eqref{eq:uniform-factorization} is
cancelled upon multiplication with the Poisson distribution~\eqref{eq:P-M-T} in
Eq.~\eqref{eq:uniform-rescaling-continuous-time-limit}. Furthermore, the factor
$T$ appearing in the exponent in Eq.~\eqref{eq:P-M-T} is expressed in
Eq.~\eqref{eq:uniform-rescaling-continuous-time-limit-factorized} as an integral
over time. Therefore, no factor of $T$ appears in
Eq.~\eqref{eq:uniform-rescaling-continuous-time-limit-factorized}.

\subsubsection{Detailed rescaling}
\label{sec:detailed-rescaling}

For detailed rescaling with $\Gamma_a = \gamma_a$, the probability
$Q_{\boldsymbol{\alpha}}$ in Eq.~\eqref{eq:Q-alpha-O-alpha} depends on the
number of jumps $M_{\boldsymbol{\alpha}, a}$ for each individual type of jump
$a \in \mathsf{A}$,
\begin{equation}
  \label{eq:Q-alpha-M-alpha}
  Q_{\boldsymbol{\alpha}} = Q_{\mathbf{M}_{\boldsymbol{\alpha}}} = q_0^{N -
    M_{\boldsymbol{\alpha}}} \prod_{a = 1}^A q_a^{M_{\boldsymbol{\alpha}, a}},
\end{equation}
where
$\mathbf{M}_{\boldsymbol{\alpha}} = \left( M_{\boldsymbol{\alpha}, 1}, \dotsc,
  M_{\boldsymbol{\alpha}, A} \right)$,
and again $M_{\boldsymbol{\alpha}} = \sum_{a = 1}^A M_{\boldsymbol{\alpha}, a}$
is the total number of jumps. We now choose a different parameterization of
trajectories: for each type of jump $a$, we specify the total number of jumps
$M_a$, as well as the time steps
$\mathbf{n}_a = \left( n_{a, 1}, \dotsc, n_{a, M_a} \right)$ with
$n_{a, 1} < \dotsb < n_{a, M}$ or equivalently the times
$t_{a, m_a} = t_0 + n_{a, m_a} \Delta t$ at which the jumps occur. Each sequence
$\mathbf{n}_a$ is ordered, and we again include a factor
$d_{\mathbf{n}_{\mathbf{a}}}$ that is unity when all $n_{a, m_a}$ with
$a \in \mathsf{A}$ and $m_a \in \{ 1, \dotsc, M_a \}$ are different, and zero
otherwise. By
$\mathbf{n}_{\mathbf{a}} = \left( \mathbf{n}_1, \dotsc, \mathbf{n}_A \right)$ we
denote the collection of all vectors $\mathbf{n}_a$. This collection uniquely
determines a vector $\boldsymbol{\alpha}_{\mathbf{n}_{\mathbf{a}}}$ of length
$N$: we set $\left( \boldsymbol{\alpha}_{\mathbf{n}_{\mathbf{a}}} \right)_n = a$
for all $n$ that are contained in $\mathbf{n}_a$. Using this parameterization of
trajectories, we can express the average~\eqref{eq:f-alpha-Q-alpha-O-alpha} as
\begin{multline}  
  \overline{f_{\boldsymbol{\alpha}}} = \sum_{M_1 = 0}^N \sum_{M_2 = 0}^{N - M_1}
  \dotsb \sum_{M_A = 0}^{N - \sum_{a = 2}^A M_a} Q_{\mathbf{M}} \\ \times
  \sum_{\mathbf{n}_1 \in \mathsf{O}_{M_1, N}} \dotsb \sum_{\mathbf{n}_A \in
    \mathsf{O}_{M_a, N}} d_{\mathbf{n}_{\mathbf{a}}}
  O_{\boldsymbol{\alpha}_{\mathbf{n}_{\mathbf{a}}}}
  f_{\boldsymbol{\alpha}_{\mathbf{n}_{\mathbf{a}}}}.
\end{multline}
In analogy to Eq.~\eqref{eq:O-f-functions}, we regard
$O_{\boldsymbol{\alpha}_{\mathbf{n}_{\mathbf{a}}}}$ and
$f_{\boldsymbol{\alpha}_{\mathbf{n}_{\mathbf{a}}}}$ as functions of
$\mathbf{t}_{\mathbf{a}} = \left( \mathbf{t}_1, \dotsc, \mathbf{t}_A \right)$,
where for each type of jump $a$, the vector
$\mathbf{t}_a = \left( t_{a, 1}, \dotsc, t_{a, M_a} \right)$ collects the times
at which jumps occurs:
\begin{equation}
  O_{\boldsymbol{\alpha}_{\mathbf{n}_{\mathbf{a}}}} =
  O(\mathbf{t}_{\mathbf{a}}), \qquad
  f_{\boldsymbol{\alpha}_{\mathbf{n}_{\mathbf{a}}}} = f(\mathbf{t}_{\mathbf{a}}).
\end{equation}
Furthermore, as we did before for uniform rescaling, we can remove the
restrictions on the sums over $\mathbf{n}_a$ if we compensate the resultant
overcounting by introducing factors of $1/M_a!$,
\begin{multline}
  \label{eq:f-alpha-Q-M-alpha-d-n-O-alpha-detailed}
  \overline{f}_{\boldsymbol{\alpha}} = \sum_{M_1 = 0}^N \sum_{M_2 = 0}^{N - M_1}
  \dotsb \sum_{M_A = 0}^{N - \sum_{a = 2}^A M_a} Q_{\mathbf{M}} \left( \prod_{a
      = 1}^A \frac{1}{M_a!} \right) \\ \times \sum_{\mathbf{n}_1 \in
    \mathsf{N}^{M_1}} \dotsb \sum_{\mathbf{n}_A \in \mathsf{N}^{M_A}}
  d_{\mathbf{n}_{\mathbf{a}}} O(\mathbf{t}_{\mathbf{a}})
  f(\mathbf{t}_{\mathbf{a}}).
\end{multline}
In the continuous-time limit $N \to \infty$, the sums over the components of the
vectors $\mathbf{n}_a$ are converted to integrals,
\begin{equation}
  \sum_{\mathbf{n}_a \in \mathsf{N}^{M_a}} \sim N^{M_a} \prod_{m_a =
    1}^{M_a} \int_{t_0}^t \frac{\diff t_m}{T}.
\end{equation}
The factors $N^{M_a}$ can again be combined with other factors appearing in
Eq.~\eqref{eq:f-alpha-Q-M-alpha-d-n-O-alpha-detailed} to yield a product of
Poisson distributions:
\begin{equation}
  \begin{split}
    \left( \prod_{a = 1}^A \frac{N^{M_a}}{M_a!} \right) Q_{\mathbf{M}} & =
    \prod_{a = 1}^A \frac{\left( \Gamma_a T \right)^{M_a}}{M_a!} \left( 1 -
      \frac{\Gamma_0 T}{N} \right)^{N - M_a} \\ & \to \e^{- \left( \Gamma_0 -
        \sum_{a = 1}^A \Gamma_a \right) T} \prod_{a = 1}^A P_{a, M_a}(T),
  \end{split}
\end{equation}
where
\begin{equation}
  \label{eq:P-a-M-a-T}  
  P_{a, M_a}(T) = \frac{\left( \Gamma_a T \right)^{M_a}}{M_a!} \e^{- \Gamma_a T}.
\end{equation}
For the same reason as above, the factor $d_{\mathbf{n}_{\mathbf{a}}}$ can be
dropped in the continuous-time limit $N \to \infty$. Thus,
Eq.~\eqref{eq:f-alpha-Q-M-alpha-d-n-O-alpha-detailed} becomes
\begin{multline}
  \label{eq:detailed-rescaling-continuous-time-limit}
  \overline{f_{\boldsymbol{\alpha}}} \sim \e^{- \left( \Gamma_0 - \sum_{a = 1}^A
      \Gamma_a \right) T} \\ \times \left[ \prod_{a = 1}^A \sum_{M_a =
      0}^{\infty} P_{a, M_a}(T) \prod_{m_a = 1}^{M_a} \int_{t_0}^t \frac{\diff
      t_{a, m_a}}{T} \right] O(\mathbf{t}_{\mathbf{a}})
  f(\mathbf{t}_{\mathbf{a}}).
\end{multline}
The exponential in the first line vanishes for the choice
$\Gamma_0 = \sum_{a = 1}^A \Gamma_a$. Under this condition,
Eq.~\eqref{eq:detailed-rescaling-continuous-time-limit} admits the following
interpretation: jumps of type $a$ may be viewed as occurring with rate
$\Gamma_a$. For each jump type, one then averages over the number of jumps in
the interval $T$, which obeys the Poisson distribution~\eqref{eq:P-a-M-a-T} with
mean $\Gamma_a T$, and over the associated jump times, uniformly distributed in
$t_{a, m_a} \in [t_0, t]$. As in the case of uniform rescaling, this
interpretation is merely suggested by the structure of
Eq.~\eqref{eq:detailed-rescaling-continuous-time-limit}. The actual distribution
of jump types and times is fixed by Eq.~\eqref{eq:P-alpha-trajectory}, which is
insensitive to the auxiliary rates $\Gamma_{\alpha}$ introduced by the rescaling
prescription.

Analogous to the discussion for uniform rescaling, let us examine the special
case in which the observable factorizes over quantum jumps:
\begin{equation}
  O(\mathbf{t}_{\mathbf{a}}) f(\mathbf{t}_{\mathbf{a}}) = \prod_{a = 1}^A
  \prod_{m_a = 1}^{M_a} g_a(t_{a, m_a}).
\end{equation}
Substituting this expression into
Eq.~\eqref{eq:detailed-rescaling-continuous-time-limit} and using
\begin{equation}
  \prod_{m_a = 1}^{M_a} \int_{t_0}^t \frac{\diff t_{a, m_a}}{T} g_a(t_{a, m_a})
  = \left[ \int_{t_0}^t \frac{\diff t'}{T} g_a(t') \right]^{M_a},
\end{equation}
we arrive at
\begin{equation}
  \label{eq:detailed-rescaling-continuous-time-limit-factorized}
  \overline{f_{\boldsymbol{\alpha}}} \sim \exp \! \left\{ \int_{t_0}^t \diff t'
    \left[ \sum_{a = 1}^A \Gamma_a g_a(t') - \Gamma_0 \right] \right\}.
\end{equation}

\section{Replica formalism}
\label{sec:replica-formalism}

Having shown, for both uniform and detailed rescaling, how the trajectory
average in Eq.~\eqref{eq:f-alpha-Q-alpha-O-alpha} can be rewritten in a form
amenable to the continuous-time limit, we now specify the quantities whose
averages we aim to compute.

In the dynamics of quantum many-body systems under continuous measurements,
particular interest lies in observables that depend nonlinearly on the quantum
state~\eqref{eq:normalized-pure-trajectory}. Prominent examples include
subsystem entropies and connected correlation functions. Such observables are
built from trajectory averages of products of quantum expectation values. The
latter can be expressed as expectation values with respect to the $k$-replica
density matrix defined as~\footnote{For fermionic systems, the tensor product is
  understood within the bosonic Hilbert space obtained via a Jordan–Wigner
  transformation. Fermionic operators can then be implemented in the replicated
  Hilbert space~\cite{Fava2023, Fava2024}.}
\begin{equation}
  \label{eq:rho-k}
  \hat{\mu}_k(t) = \overline{\bigotimes_{r=1}^k \hat{\rho}_r(t)},
\end{equation}
where the overbar again denotes the trajectory average, and the index $r$ labels
identical copies or replicas of the original system. For instance, for two
operators $\hat{A}$ and $\hat{B}$,
\begin{equation}
  \label{eq:A-B-replicas}
  \begin{split}
    \overline{\langle \hat{A}(t) \rangle \langle \hat{B}(t) \rangle} & =
    \overline{\tr \! \left[ \hat{A} \hat{\rho}(t) \right] \tr \! \left[ \hat{B}
        \hat{\rho}(t) \right]} \\ & = \tr \! \left\{ \left( \hat{A}_1 \otimes
        \hat{B}_2 \right) \overline{\left[ \hat{\rho}_1(t) \otimes
          \hat{\rho}_2(t) \right]} \right\} \\ & = \tr \! \left[ \left(
        \hat{A}_1 \otimes \hat{B}_2 \right) \hat{\mu}_2 \right].
  \end{split}
\end{equation}

To evaluate the $k$-replica density matrix $\hat{\mu}_k(t)$, we first express a
single-replica density matrix as
\begin{equation}
  \label{eq:rho-D}
  \hat{\rho}_r(t) = \left. \hat{D}_r(t) \middle/ \tr \! \left[ \hat{D}_r(t) \right], \right.
\end{equation}
with the unnormalized density matrix
\begin{equation}
  \label{eq:unnormalized-density-matrix}
  \hat{D}_r(t) = \hat{J}_{r, \alpha_N} \dotsb \hat{J}_{r, \alpha_1}
  \ket{\psi_{r, 0}} \bra{\psi_{r, 0}} \hat{J}_{r, \alpha_1}^{\dagger} \dotsb
  \hat{J}_{r, \alpha_N}^{\dagger},
\end{equation}
where $\hat{J}_{r, \alpha}$ are the Kraus operators~\eqref{eq:J-alpha} acting on
replica $r$, and $\ket{\psi_{r, 0}}$ is the initial state in the corresponding
Hilbert space. Note that since each replica is an identical copy, we have
\begin{equation}
  \tr \! \left[ \hat{D}_1(t) \right] = \dotsb = \tr \! \left[ \hat{D}_R(t)
  \right] = \tr \! \left[ \hat{D}(t) \right].
\end{equation}
The product of Kraus operators in Eq.~\eqref{eq:unnormalized-density-matrix}
describes the evolution from $t_0$ to $t = t_N = t_0 + N \Delta t$, which
justifies denoting this expression by $\hat{D}_r(t)$. For brevity, the explicit
dependence of $\hat{D}_r(t)$ on $\boldsymbol{\alpha}$ is suppressed.

We then insert Eq.~\eqref{eq:rho-D} into Eq.~\eqref{eq:rho-k} and employ
Eq.~\eqref{eq:f-alpha-Q-alpha-O-alpha} to write the trajectory average. Noting
that the denominator in Eq.~\eqref{eq:rho-D} is simply the quantity
$O_{\boldsymbol{\alpha}}$ defined in Eq.~\eqref{eq:Q-alpha-O-alpha}, namely
$O_{\boldsymbol{\alpha}} = \tr[\hat{D}(t)]$, we thus obtain
\begin{equation}
  \label{eq:mu-k-before-replica}
  \hat{\mu}_k(t) = \sum_{\boldsymbol{\alpha} \in \mathsf{A}_0^N}
  Q_{\boldsymbol{\alpha}} \frac{\otimes_{r = 1}^k \hat{D}_r(t)}{\tr \! \left[
      \hat{D}(t) \right]^{k - 1}}.
\end{equation}
We wish to proceed by rewriting the $k$-replica density matrix as a Keldysh
functional integral and taking the average over trajectories using the results
of Sec.~\ref{sec:rescaled-Kraus-operators}. In analogy with the disorder
average in spin glasses~\cite{Mezard1986} and disordered electronic
systems~\cite{Evers2008}, the trajectory average is facilitated by rewriting the right-hand side
of Eq.~\eqref{eq:mu-k-before-replica} such that $\hat{D}(t)$ appears only in
the numerator. This is achieved by means of the replica trick. We first write the denominator as
\begin{equation}
  \tr \! \left[ \hat{D}(t) \right]^{1 - k} = \lim_{R \to 1} \tr \! \left[
    \hat{D}(t) \right]^{R - k}.
\end{equation}
This representation becomes useful once the right-hand side is expressed as a
trace over additional replicas with indices $r \in \{ k + 1, \dotsc, R \}$,
which, strictly speaking, is valid only for integer $R > k$:
\begin{equation}
  \tr \! \left[ \hat{D}(t) \right]^{R - k} = \mathop{\mathrm{tr}_{r = k + 1,
      \dotsc, R}} \! \left[ \bigotimes_{r = k + 1}^R \hat{D}_r(t) \right].
\end{equation}
Inserting the above relations into Eq.~\eqref{eq:mu-k-before-replica} and
interchanging the order of the summation over $\boldsymbol{\alpha}$ and the
limit $R \to 1$---a step whose subtle consequences we discuss below---we obtain
\begin{equation}
  \label{eq:mu-k-after-replica}
  \hat{\mu}_k(t) = \lim_{R \to 1}
  \sum_{\boldsymbol{\alpha} \in \mathsf{A}_0^N} Q_{\boldsymbol{\alpha}}
  \mathop{\mathrm{tr}_{r = k + 1, \dotsc, R}} \! \left[ \bigotimes_{r=1}^R
    \hat{D}_r(t) \right].
\end{equation}
In practice, we evaluate the right-hand side for integer values $R > 1$.
Expectation values such as Eq.~\eqref{eq:A-B-replicas} can then be obtained by
introducing source fields into the Hamiltonian $\hat{H}$ in
Eq.~\eqref{eq:K-0} and differentiating the corresponding $R$-replica Keldysh
partition function with respect to these sources,
\begin{equation}
  \label{eq:Keldysh-partition}
  Z_R(t) = \sum_{\boldsymbol{\alpha} \in \mathsf{A}_0^N}
  Q_{\boldsymbol{\alpha}} \tr \! \left[ \bigotimes_{r = 1}^R \hat{D}_r(t)
  \right] = \sum_{\boldsymbol{\alpha} \in \mathsf{A}_0^N}
  Q_{\boldsymbol{\alpha}} \prod_{r = 1}^R Z_r(t),
\end{equation}
where $Z_r(t) = \tr \! \left[ \hat{D}_r(t) \right]$. After differentiating to
generate the desired observables, the source fields are set to zero, and the
replica limit $R \to 1$ is taken at the end of the calculation.

Before employing the replica trick, the rescaling of the Kraus operators in
Eq.~\eqref{eq:J-alpha-q-0-q-a} cancels term by term in the sum over
$\boldsymbol{\alpha}$ appearing in Eq.~\eqref{eq:mu-k-before-replica}. This
cancellation no longer occurs once the order of the summation over trajectories
and the limit $R \to 1$ is exchanged and the right-hand side of
Eq.~\eqref{eq:mu-k-after-replica} is considered for finite $R > 1$, as in the
$R$-replica Keldysh partition function~\eqref{eq:Keldysh-partition}. Two
important consequences follow. First, for any $R > 1$, the individual terms in
the sum over $\boldsymbol{\alpha}$ retain an explicit dependence on
$\Gamma_{\alpha}$, so different choices of $\Gamma_{\alpha}$ become equivalent
only in the replica limit $R \to 1$. Second, and more importantly, because the
rescaling~\eqref{eq:J-alpha-q-0-q-a} is performed \emph{before} employing the
replica trick, the time step $\Delta t$ enters the replica partition
function~\eqref{eq:Keldysh-partition} exclusively through the factor
$Q_{\boldsymbol{\alpha}}$. This allows the average over trajectories to be
carried out in the continuous-time limit using
Eqs.~\eqref{eq:uniform-rescaling-continuous-time-limit-factorized}
and~\eqref{eq:detailed-rescaling-continuous-time-limit-factorized} for uniform
and detailed rescaling, respectively, as discussed in
Sec.~\ref{sec:replica-Keldysh} below. Had the rescaling been performed
\emph{after} introducing replicas, the factor $Q_{\boldsymbol{\alpha}}$ in
Eq.~\eqref{eq:Keldysh-partition} would acquire an additional exponent
$R$. Consequently,
Eqs.~\eqref{eq:uniform-rescaling-continuous-time-limit-factorized}
and~\eqref{eq:detailed-rescaling-continuous-time-limit-factorized} would contain
a factor $\Delta t^{R-1}$ in the exponent. The exponent would then vanish if the
continuous-time limit $\Delta t \to 0$ were taken before the replica limit
$R \to 1$. Within the replica Keldysh field theory developed below, this would
imply that the contribution of quantum jumps to the Keldysh action disappears
altogether. This observation demonstrates that the rescaling prescription of
Eq.~\eqref{eq:J-alpha-q-0-q-a} is the only one that yields a nontrivial
continuous-time limit for arbitrary $R > 1$.

\section{Replica Keldysh partition function}
\label{sec:replica-Keldysh}

Building on the groundwork laid in Secs.~\ref{sec:quantum-jump-processes} and
\ref{sec:replica-formalism}, we now derive a functional‐integral representation
of the partition function~\eqref{eq:Keldysh-partition} using the Keldysh
technique~\cite{Altland2010a, Kamenev2023, Sieberer2016a, Sieberer2025,
  Thompson2023}. To begin, we consider a single factor $Z_r(t)$ in the product
over replicas in Eq.~\eqref{eq:Keldysh-partition}. Using
Eq.~\eqref{eq:unnormalized-density-matrix}, $Z_r(t)$ takes the form
\begin{equation}
  \label{eq:single-replica-Keldysh-partition-function}
  Z_r(t) = \tr \! \left( \hat{J}_{r,\alpha_N} \dotsb \hat{J}_{r,\alpha_1}
    \ket{\psi_{r, 0}} \bra{\psi_{r, 0}}
    \hat{J}_{r,\alpha_1}^{\dagger} \dotsb \hat{J}_{r,\alpha_N}^{\dagger} \right).
\end{equation}
Within the Keldysh formalism, the products of Kraus operators acting on the ket
and bra components of $\ket{\psi_{r, 0}} \bra{\psi_{r, 0}}$ are interpreted as
describing evolution along the forward and backward branches of the closed time
contour. A functional‐integral representation follows from inserting
coherent-state resolutions of the identity between successive Kraus
operators. This construction yields fields $\psi_{\sigma,r}(t)$ and
$\psi_{\sigma,r}^{*}(t)$, defined on both Keldysh branches $\sigma=\pm$ and for
each replica $r \in \{1,\dotsc,R\}$. Additional indices, such as those labeling
lattice sites, are left implicit. For bosonic systems the two fields are complex
conjugates, while for fermionic systems they are independent Grassmann
variables. We thus obtain
\begin{equation}
  Z_r(t)
  = \int \Diff[\psi_r^{*}, \psi_r] \,
  \e^{\imag \left( S_{r,L}[\psi_r^{*}, \psi_r] + S_{r,D}[\psi_r^{*}, \psi_r] \right)}
  Z_{r,J}[\psi_r^{*}, \psi_r].
\end{equation}
The exponential factor describes continuous evolution and consists of two
contributions: first, a Liouvillian term $S_{r,L}[\psi_r^{*}, \psi_r]$, which
accounts for unitary evolution generated by the Hamiltonian $\hat{H}$ and, as
discussed in Sec.~\ref{sec:inefficient-detection}, acquires additional
dissipative contributions in the case of inefficient detection; second, a decay
term $S_{r,D}[\psi_r^{*}, \psi_r]$ associated with the anti-Hermitian part of
the effective Hamiltonian~\eqref{eq:H-eff}. The remaining factor,
$Z_{r,J}[\psi_r^{*}, \psi_r]$, encodes the effect of quantum jumps. Collecting
the contributions from all replicas, the full replica partition
function~\eqref{eq:Keldysh-partition} can be written as
\begin{equation}
  \label{eq:Z-R-d-psi}
  Z_R(t)
  = \int \Diff[\psi^{*}, \psi] \,
  \e^{\imag \left( S_L[\psi^{*}, \psi] + S_D[\psi^{*}, \psi] + S_J[\psi^{*}, \psi] \right)},
\end{equation}
where
\begin{equation}
  S_L[\psi^{*}, \psi] = \sum_{r=1}^R S_{r,L}[\psi_r^{*}, \psi_r], \quad
  S_D[\psi^{*}, \psi] = \sum_{r=1}^R S_{r,D}[\psi_r^{*}, \psi_r],
\end{equation}
and the jump contribution is defined by
\begin{equation}
  \label{eq:jump-action}
  \e^{\imag S_J[\psi^{*}, \psi]}
  = \sum_{\boldsymbol{\alpha} \in \mathsf{A}_0^N}
    Q_{\boldsymbol{\alpha}}
    \prod_{r=1}^R Z_{r,J}[\psi_r^{*}, \psi_r].
\end{equation}
In the following, we derive explicit expressions for $S_L[\psi^{*}, \psi]$,
$S_D[\psi^{*}, \psi]$, and $S_J[\psi^{*}, \psi]$.

\subsection{Hamiltonian and dissipative Keldysh action}

For continuous evolution, the Hamiltonian part of $\hat{J}_{r, 0}$ in
Eq.~\eqref{eq:J-alpha} produces, upon exponentiation, the standard Hamiltonian
contribution to the Keldysh action, while the remaining terms in $\hat{J}_{r, 0}$
give rise to a decay contribution. Accordingly, the Liouvillian and decay
components of the Keldysh action are
\begin{equation}
  \label{eq:S-L-S-D}
  \begin{split}
    S_L[\psi^{*}, \psi] & = \sum_{\sigma = \pm} \sigma \int_{t_0}^t \diff t'
    \sum_{r = 1}^R \left[ \psi_{\sigma, r}^{*}(t') \imag \partial_t
      \psi_{\sigma, r}(t') - H_{\sigma, r}(t') \right], \\ \imag S_D[\psi^{*},
    \psi] & = - \frac{1}{2} \sum_{\sigma = \pm} \int_{t_0}^t \diff t' \sum_{r =
      1}^R \left[ \sum_{a = 1}^A \gamma_a D_{\sigma, r, a}(t') - \Gamma_0
    \right].
  \end{split}
\end{equation}
In these expressions, $H_{\sigma, r}(t)$ and $D_{\sigma, r, a}(t)$ denote matrix
elements in coherent states,
\begin{align}
  \label{eq:H}  
  H_{\sigma, r}(t) & = \frac{\braket{\zeta_{\sigma} \psi_{\sigma}(t) |
                     \hat{H}_r | \zeta_{\sigma}\psi_{\sigma}(t)}}{\braket{\zeta_{\sigma}
                     \psi_{\sigma}(t) |
                     \zeta_{\sigma}\psi_{\sigma}(t)}}, \\
  \label{eq:D}  
  D_{\sigma, r, a}(t) & = \frac{\braket{\zeta_{\sigma} \psi_{\sigma}(t) |
                        \hat{c}_{r, a}^{\dagger} \hat{c}_{r, a} | \zeta_{\sigma}
                        \psi_{\sigma}(t)}}{\braket{\zeta_{\sigma} \psi_{\sigma}(t) |
                        \zeta_{\sigma} \psi_{\sigma}(t)}},
\end{align}
where $\zeta_{\sigma} = +1$ for bosons and $\zeta_{\sigma} = \sigma$ for fermions; 
that is, as standard in the Keldysh formalism, the sign of fermionic fields on
the backward branch is flipped~\cite{Sieberer2025}.

According to Eq.~\eqref{eq:K-a}, the probability of a quantum jump in a single
time step is $O(\Delta t)$. As a result, a typical trajectory consists of extended
periods of continuous evolution, described by $\hat{J}_{0,r}$ in
Eq.~\eqref{eq:J-alpha}, interrupted by isolated jumps represented by
$\hat{J}_{a,r}$. In the continuous-time limit, the jump times form a set of
measure zero within the evolution interval $[t_0, t]$, so that the integrals in
Eq.~\eqref{eq:S-L-S-D} extend over the entire evolution.

\subsection{Jump Keldysh action}

Having discussed the Liouvillian and decay contributions to the Keldysh action,
which encode the continuous evolution, we next turn to the contribution due to
quantum jumps~\eqref{eq:jump-action}. Here we distinguish between uniform and
detailed rescaling.

\subsubsection{Jump Keldysh action for uniform rescaling}

As discussed in Sec.~\ref{sec:uniform-rescaling}, for uniform rescaling, each
trajectory is specified by the jump times $t_m$ and the jump types $a_m$. In
this case, $Z_{r, J}$ is given by the product of coherent-state matrix elements
of the rescaled jump operators~\eqref{eq:J-alpha},
\begin{equation}
  \label{eq:Z-r-J-uniform}
  Z_{r, J} = \prod_{m = 1}^M \frac{\gamma_{a_m}}{\Gamma_1} V_{r, a_m}(t_m),
\end{equation}
with the vertex
\begin{equation}
  \label{eq:V-a}
  V_{r, a}(t) = c_{+, r, a}(t) c_{-, r, a}^{\dagger}(t),
\end{equation}
where $c_{+, r, a}(t)$ and $c_{-, r, a}^{\dagger}(t)$ are defined analogously to
Eqs.~\eqref{eq:H} and~\eqref{eq:D}:
\begin{equation}
  \label{eq:c-c-dagger}
  \begin{split}
    c_{\sigma, r, a}(t) & = \frac{\braket{\zeta_{\sigma} \psi_{\sigma}(t) |
        \hat{c}_{r, a} | \zeta_{\sigma} \psi_{\sigma}(t)}}{\braket{\zeta_{\sigma}
        \psi_{\sigma}(t) | \zeta_{\sigma} \psi_{\sigma}(t)}}, \\
    c_{\sigma, r, a}^{\dagger}(t) & = \frac{\braket{\zeta_{\sigma}
        \psi_{\sigma}(t) | \hat{c}_{r, a}^{\dagger} | \zeta_{\sigma}
        \psi_{\sigma}(t)}}{\braket{\zeta_{\sigma} \psi_{\sigma}(t) |
        \zeta_{\sigma} \psi_{\sigma}(t)}}.
  \end{split}
\end{equation}
Since Eq.~\eqref{eq:Z-r-J-uniform} factorizes over jumps, we can use
Eq.~\eqref{eq:uniform-rescaling-continuous-time-limit-factorized} to perform the
average over trajectories and thus derive the jump
action~\eqref{eq:jump-action}, resulting in
\begin{equation}
  \label{eq:S-J-uniform-rescaling}
  \imag S_J[\psi^{*}, \psi] = \int_{t_0}^t \diff
  t' \left[ \Gamma_1 \sum_{a = 1}^A  \prod_{r = 1}^R \frac{\gamma_a}{\Gamma_1} V_{r,
      a}(t') - \Gamma_0 \right].
\end{equation}
It is straightforward to verify that terms containing $\Gamma_0$ and $\Gamma_1$
cancel in the replica limit $R \to 1$ when summing the decay and jump actions in
Eq.~\eqref{eq:Z-R-d-psi}.

\subsubsection{Jump Keldysh action for detailed rescaling}
\label{sec:jump-Keldysh-action-detailed}

The parameterization of trajectories for detailed rescaling is given at the
beginning of Sec.~\ref{sec:detailed-rescaling}: for each type of jump $a$, we
specify the number of jumps $M_a$ and the times $t_{a, m_a}$ at which jumps
occur. We thus obtain
\begin{equation}
  Z_{r, J} = \prod_{a = 1}^A \prod_{m_a = 1}^{M_a} \frac{\gamma_a}{\Gamma_a}
  V_{r, a}(t_{a, m_a}).
\end{equation}
Inserting this in Eq.~\eqref{eq:jump-action} and using
Eq.~\eqref{eq:detailed-rescaling-continuous-time-limit-factorized} to take the
average over trajectories, we find
\begin{equation}
  \label{eq:S-J-detailed-rescaling}
  \imag S_J[\psi^{*}, \psi] = \int_{t_0}^t \diff t' \left[ \sum_{a = 1}^A \Gamma_a
    \prod_{r = 1}^R \frac{\gamma_a}{\Gamma_a} V_{r, a}(t') - \Gamma_0 \right].
\end{equation}
This expression reduces to Eq.~\eqref{eq:S-J-uniform-rescaling} for
$\Gamma_1 = \dotsb = \Gamma_A$. Therefore, in the following, we continue to work
with Eq.~\eqref{eq:S-J-detailed-rescaling}.

\section{Inefficient detection}
\label{sec:inefficient-detection}

In studies of measurement-induced phase transitions, measurements are usually
assumed to be perfect, meaning that they extract complete information about the
measured observable. Under such idealized conditions, the dynamics preserve the
purity of the quantum state. In any experimental realization, however,
imperfections are unavoidable~\cite{Ladewig2022, Minoguchi2022, Passarelli2024,
  Leung2025, Leung2025a, Paviglianiti2025}: states decohere due to imperfect
measurements, classical noise, and unwanted coupling to the environment, leading
to mixed-state dynamics. A particularly relevant example is inefficient
detection, where only a fraction of the quantum jumps is
registered~\cite{Gardiner2014, Gardiner2015, Wiseman2010, Jacobs2014}. The
fraction of detected jumps is quantified by the detection efficiency
$\eta \in [0,1]$, interpolating between the limit of fully efficient detection
with $\eta = 1$, which yields pure-state trajectories, and that of fully
inefficient detection with $\eta = 0$, where the dynamics become deterministic
and are governed by the Liouvillian~\eqref{eq:Liouvillian}.

The fate of measurement-induced phase transitions at finite $\eta$ depends
sensitively on the structure of the jump operators. As discussed in
Sec.~\ref{sec:quantum-trajectories}, when jump operators represent random
projective measurements, the stationary state of the deterministic Lindbladian
dynamics is typically a featureless infinite-temperature state, regardless of
the measurement rate. Consequently, signatures of measurement-induced phase
transitions vanish entirely in the limit $\eta \to 0$. By contrast, for
non-Hermitian jump operators, the steady state is generically nontrivial even at
$\eta = 0$. The parameter $\eta$ therefore provides a natural interpolation
between measurement-induced phase transitions of continuously monitored systems
and phase transitions in the nonequilibrium steady states of driven open quantum
matter.

We model a detector of finite efficiency $\eta \in [0,1]$ by splitting each time
step $\Delta t$ into two substeps of duration $\eta \Delta t$ and
$\left( 1 - \eta \right) \Delta t$, respectively~\cite{Jacobs2014}. The Kraus
operators in Eqs.~\eqref{eq:K-0} and~\eqref{eq:K-a} are then modified as
follows: For the first time substep,
\begin{equation}
  \label{eq:K-1-0-a}
  \hat{K}_{1, 0} = 1 - \imag \eta \hat{H}_{\mathrm{eff}} \Delta t, \qquad \hat{K}_{1, a} =
  \sqrt{\eta \gamma_a \Delta t} \, \hat{c}_a,
\end{equation}
and for the second substep,
\begin{equation}
  \label{eq:K-2-0-a}
  \hat{K}_{2, 0} = 1 - \imag \left( 1 - \eta \right) \hat{H}_{\mathrm{eff}} 
  \Delta t, \quad \hat{K}_{2, a} = \sqrt{\left( 1 - \eta \right) \gamma_a \Delta
    t} \, \hat{c}_a.
\end{equation}
Inefficient detection is implemented by stochastically unraveling the evolution
during the first substep, while in the second substep we apply deterministic
averaged evolution. The discussion of trajectories in
Sec.~\ref{sec:quantum-trajectories} generalizes accordingly: For a system in a
mixed state $\hat{\rho}$, the probability for an evolution of type $\alpha$ to
occur in a single time step is
\begin{equation}
  \label{eq:P-alpha-inefficient}
  P_{\alpha} = \tr \! \left( \mathcal{K}_2 \mathcal{K}_{1, \alpha} \hat{\rho}
  \right) = \tr \! \left( \mathcal{K}_{1, \alpha} \hat{\rho} \right),
\end{equation}
where
\begin{equation}
  \label{eq:K-1-superoperator}
  \mathcal{K}_{1, \alpha} \hat{\rho} = \hat{K}_{1, \alpha} \hat{\rho}
  \hat{K}_{1, \alpha}^{\dagger},
\end{equation}
and
\begin{equation}
  \label{eq:K-2}
  \mathcal{K}_2 \hat{\rho} = \sum_{a = 1}^A \hat{K}_{2, a} \hat{\rho}
  \hat{K}_{2, a}^{\dagger} = \left[ 1 + \left( 1 - \eta \right) \mathcal{L}
    \Delta t \right] \hat{\rho},
\end{equation}
with the Liouvillian superoperator~\eqref{eq:Liouvillian}. Note that evolution
with $\mathcal{K}_2$ conserves the trace; this is why $\mathcal{K}_2$ may be
omitted after the second equality in Eq.~\eqref{eq:P-alpha-inefficient}.

A trajectory over $N$ time steps is specified by a sequence of outcomes
$\boldsymbol{\alpha}$ of the measurements performed in each first time substep;
the probability for this sequence is
\begin{equation}
  \label{eq:P-alpha-trajectory-inefficient}
  P_{\boldsymbol{\alpha}} = \tr \! \left( \mathcal{K}_2 \mathcal{K}_{1, \alpha_N}
    \dotsb \mathcal{K}_2 \mathcal{K}_{1, \alpha_1} \hat{\rho}_0 \right),
\end{equation}
and the normalized state after $N$ time steps is
\begin{equation}
  \label{eq:normalized-mixed-trajectory}
  \hat{\rho}_{\boldsymbol{\alpha}} = \left. \mathcal{K}_2 \mathcal{K}_{1, \alpha_N}
    \dotsb \mathcal{K}_2 \mathcal{K}_{1, \alpha_1} \hat{\rho}_0 \middle/
    \sqrt{P_{\boldsymbol{\alpha}}} \right..
\end{equation}

Having specified the discrete-time quantum jump process for inefficient
detection, we now discuss how the associated Keldysh field theory is modified as
compared to the one for efficient detection. As a preliminary step, it will
prove convenient to modify the factors $q_{\alpha}$ appearing in the
definition~\eqref{eq:J-alpha-q-0-q-a} of rescaled Kraus operators for the first
substep by replacing $\Gamma_{\alpha} \to \eta \Gamma_{\alpha}$. In the second
substep, where the evolution is averaged, no rescaling is required.

We begin by examining how inefficient detection modifies the Liouvillian
contribution to the Keldysh action~\eqref{eq:S-L-S-D}. Most importantly, the
factors $\mathcal{K}_2$ appearing in Eq.~\eqref{eq:normalized-mixed-trajectory}
generate additional terms in the action: evolution implemented by
$\mathcal{K}_2$ in Eq.~\eqref{eq:K-2} is governed by the Liouvillian
superoperator~\eqref{eq:Liouvillian}, which contains both coherent and
dissipative terms. The latter yield a contribution that can be obtained as
described in Refs.~\cite{Kamenev2023, Sieberer2016a, Sieberer2025,
  Thompson2023}. We find
\begin{multline}
  \label{eq:S-L-eta}
  \!\!\! S_L[\psi^{*}, \psi] = \int_{t_0}^t \diff t' \sum_{r = 1}^R \left\{
    \sum_{\sigma = \pm} \sigma \left[ \psi_{\sigma, r}^{*}(t') \imag \partial_t
      \psi_{\sigma, r}(t') - H_{\sigma, r}(t') \right] \right. \\ \left.  -
    \imag \left( 1 - \eta \right) \sum_{a = 1}^A \gamma_a \left[ V_{r, a}(t') -
      \frac{1}{2} \sum_{\sigma = \pm} D_{\sigma, r, a}(t') \right] \right\}.
\end{multline}
Note that the Hamiltonian $\hat{H}$ appears in the Kraus operators in both time
substeps in Eqs.~\eqref{eq:K-1-0-a} and~\eqref{eq:K-2-0-a} with prefactors
$\eta$ and $1 - \eta$, respectively. Thus, $\eta$ cancels in the sum of the
respective contributions to the Keldysh action. In contrast, the dissipative
part stemming from the Liouvillian~\eqref{eq:Liouvillian} appears only in the
second time substep~\eqref{eq:K-2}. Therefore, in Eq.~\eqref{eq:S-L-eta} this
contribution appears with a prefactor $1 - \eta$.

A second element of the Keldysh action associated with continuous evolution is
the decay term in Eq.~\eqref{eq:S-L-S-D}. Under inefficient detection, this term
originates solely from the non-Hermitian part of the effective Hamiltonian in
the first time substep, Eq.~\eqref{eq:K-1-0-a}. Although an analogous term also
appears in the second substep~\eqref{eq:K-2-0-a}, its effect is already encoded
in Eq.~\eqref{eq:S-L-eta}. Consequently, the decay part of the Keldysh action is
simply multiplied by a factor of $\eta$:
\begin{equation}
  \label{eq:S-D-eta}
  \imag S_D[\psi^{*}, \psi] = - \frac{\eta}{2} \sum_{\sigma =
    \pm} \int_{t_0}^t \diff t' \sum_{r = 1}^R \left[ \sum_{a = 1}^A
    \gamma_a D_{\sigma, r, a}(t') - \Gamma_0 \right].
\end{equation}

We now turn to the jump action. The discussion in
Sec.~\ref{sec:rescaled-Kraus-operators} of how to perform the trajectory average
in the continuous-time limit for uniform and detailed rescaling relies on
different ways of parameterizing quantum trajectories, which remain valid for
inefficient detection. The only modification is that due to the rescaling
$\Gamma_{\alpha} \to \eta \Gamma_{\alpha}$, the exponents in
Eqs.~\eqref{eq:uniform-rescaling-continuous-time-limit-factorized}
and~\eqref{eq:detailed-rescaling-continuous-time-limit-factorized} acquire a
prefactor $\eta$. However, $\eta$ drops out of the rescaled Kraus operators
corresponding to quantum jumps. The action that describes jumps is thus
\begin{equation}
  \label{eq:S-J-eta}
  \imag S_J[\psi^{*}, \psi] = \eta \int_{t_0}^t \diff t' \left[ \sum_{a = 1}^A \Gamma_a \prod_{r
      = 1}^R \frac{\gamma_a}{\Gamma_a} V_{r, a}(t') - \Gamma_0 \right].
\end{equation}

The Keldysh field theory for quantum-jump processes with inefficient detection
is therefore defined by the functional integral in Eq.~\eqref{eq:Z-R-d-psi},
together with the three contributions to the Keldysh action given in
Eqs.~\eqref{eq:S-L-eta}, \eqref{eq:S-D-eta}, and~\eqref{eq:S-J-eta}. This forms
one of the central results of our work. For $R > 1$, the framework allows one to
evaluate trajectory averages of observables that are nonlinear in the quantum
state, such as the product of expectation values in
Eq.~\eqref{eq:A-B-replicas}. Such quantities generally exhibit a nontrivial
dependence on the detection efficiency $\eta$; we will present concrete examples
in Sec.~\ref{sec:imbalanced-inefficient-counting}. In contrast, in the limit
$R \to 1$, the construction reduces by design to the standard Lindblad-Keldysh
field theory of Refs.~\cite{Sieberer2016a, Sieberer2025, Thompson2023,
  Kamenev2023}, which captures the dynamics averaged over all measurement
outcomes. In particular, in the sum of Eqs.~\eqref{eq:S-L-eta},
\eqref{eq:S-D-eta}, and~\eqref{eq:S-J-eta}, both $\eta$ and the auxiliary rates
$\Gamma_{\alpha}$ cancel out.

\section{Imbalanced and inefficient fermion counting}
\label{sec:imbalanced-inefficient-counting}

We now apply the above formalism to the problem of imbalanced and inefficient
fermion counting. Specifically, we consider a 1D fermionic lattice system
subject to monitored fermion gain and loss, where the two occur at different
rates and only a fraction of the corresponding jumps is detected. Our analysis
extends the results of Ref.~\cite{Starchl2025}, which focused on the balanced
and perfectly efficient case. When gain and loss rates coincide, fermion
counting can be cast as a sequence of generalized measurements performed at
random times with a fixed rate, and the associated field theory was derived in
Ref.~\cite{Starchl2025} by adapting the framework introduced in
Ref.~\cite{Poboiko2023} for random projective measurements. In contrast,
handling imbalanced rates lies beyond that construction and requires the more
general approach developed in this work.

\subsection{Model}
\label{sec:model}

We consider free fermions on a 1D lattice of size $L$, with annihilation and
creation operators $\hat{\psi}_l$ and $\hat{\psi}_l^{\dagger}$,
respectively. The Hamiltonian is
\begin{equation}
  \label{eq:Hamiltonian}
  \hat{H} = \sum_{l, l' = 1}^L \hat{\psi}_l^{\dagger} H_{l, l'} \hat{\psi}_{l'}.
\end{equation}
Our final results focus on nearest-neighbor hopping between sites $l$ and
$l \pm 1$, described by
\begin{equation}
  \label{eq:Hamiltonian-matrix}
  H_{l, l'} = - J \left( \delta_{l + 1, l'} + \delta_{l, l' + 1} \right),
\end{equation}
where Kronecker deltas are understood modulo $L$ to impose periodic boundary
conditions. In intermediate steps, however, we keep the discussion general and
leave the Hermitian matrix $H$ unspecified.

Fermion counting involves two types of jump operators on each lattice site
describing fermion gain and loss,
\begin{equation}
  \label{eq:gain-loss-jump-operators}
  \hat{c}_{+, l} = \hat{\psi}_l^{\dagger}, \qquad \hat{c}_{-, l} = \hat{\psi}_l,
\end{equation}
with spatially uniform jump rates $\gamma_{\pm}$. Thus, in the notation of the
previous sections, the jump-type index $a$ is replaced by the pair of indices
$\alpha = \pm$ and $l$.

For balanced rates, $\gamma_+ = \gamma_-$, the stationary state of the
unconditional dynamics is a featureless infinite-temperature state, independent
of the jump rate~\cite{Starchl2025}. Sites are uncorrelated, and each site is
empty or occupied with equal probability, corresponding to a mean fermion
density of $n = 1/2$. When $\gamma_+ \neq \gamma_-$, the steady state remains
factorized over sites,
\begin{equation}
  \label{eq:rho-ss}
  \hat{\rho}_{\mathrm{ss}} = \prod_{l = 1}^L \frac{1}{\gamma_+ + \gamma_-}
  \left[ \gamma_+ \hat{n}_l + \gamma_- \left( 1 - \hat{n}_l \right) \right],
\end{equation}
where $\hat{n}_l = \hat{\psi}_l^{\dagger} \hat{\psi}_l$. However, the fermion
density in the steady state is now generally different from $n = 1/2$ and given
by
\begin{equation}
  \label{eq:n}
  n = \frac{\gamma_+}{\gamma_+ + \gamma_-}.
\end{equation}
Our model therefore provides a minimal example in which the stationary state of
the unconditional dynamics depends nontrivially on the jump rates.

In what follows, instead of specifying the rates $\gamma_{\pm}$, it is often
convenient to use the steady-state density $n$ as a free parameter of the model,
together with the average jump rate 
\begin{equation}
  \label{eq:gamma}
  \gamma = (\gamma_+ + \gamma_-)/2.
\end{equation}

\subsection{Replica Keldysh field theory}

Having defined the model, we now apply the results of
Sec.~\ref{sec:inefficient-detection} to derive the replica Keldysh field theory
for imbalanced and inefficient fermion counting.

\subsubsection{Fermionic replica Keldysh action}
\label{sec:fermionic-Keldysh-action}

The field-theoretic description of imbalanced and inefficient fermion counting
is encoded in the functional-integral representation of the partition
function~\eqref{eq:Z-R-d-psi}, with the action consisting of the
Liouvillian~\eqref{eq:S-L-eta}, decay~\eqref{eq:S-D-eta}, and jump
terms~\eqref{eq:S-J-eta}, where we set $\Gamma_{\pm} = \gamma_{\pm}$ and
$\Gamma_0 = L \gamma$. To obtain the explicit form of the action, we evaluate
the coherent-state matrix elements in Eqs.~\eqref{eq:H}, \eqref{eq:D},
and~\eqref{eq:c-c-dagger}, with the last of these determining the vertex in
Eq.~\eqref{eq:V-a}. The calculation relies on fermionic anticommutation
relations to normal-order operators, and on the fact that coherent bra and ket
states are left and right eigenstates of creation and annihilation operators,
respectively. This yields
\begin{equation}
  \label{eq:H-sigma}
  H_{\sigma, r}(t) = \sum_{l, l' = 1}^L \psi_{\sigma, r, l}^{*}(t) H_{l, l'}
  \psi_{\sigma, r, l'}(t).
\end{equation}
The matrix elements~\eqref{eq:D} entering the Liouvillian and decay parts of the
action take the form
\begin{equation}
  \label{eq:D-counting-before-regularization}
  \begin{split}
    D_{\sigma, r, +, l}(t) & = 1 - \psi^*_{\sigma, r, l}(t) \psi_{\sigma, r, l}(t), \\
    D_{\sigma, r, -, l}(t) & = \psi^*_{\sigma,r,l}(t) \psi_{\sigma,r,l}(t),
  \end{split}
\end{equation}
and the vertices~\eqref{eq:V-a} associated with fermion gain and loss are
\begin{equation}
  \label{eq:V-counting}
  \begin{split}
    V_{r, +, l}(t) & = -\psi^*_{+,r,l}(t) \psi_{-,r,l}(t), \\
    V_{r, -, l}(t) & = -\psi_{+,r,l}(t) \psi^*_{-,r,l}(t).
  \end{split}
\end{equation}

The fermionic replica Keldysh action is the starting point for constructing a
long-wavelength effective field theory, formulated in terms of slow modes built
from bilinears of fermionic fields~\cite{Poboiko2023, Starchl2025}. These
bilinears are introduced as auxiliary bosonic degrees of freedom via a
generalized Hubbard-Stratonovich transformation. Following
Refs.~\cite{Poboiko2023, Starchl2025}, below we define the bilinears in the
symmetrized limit in which both fermionic fields are evaluated at the same time
$t$. To prepare for the Hubbard-Stratonovich decoupling, we therefore rewrite
the action in a strictly time-local form using the ``principal-value''
regularization scheme~\cite{Poboiko2023, Starchl2025}.

We begin with the matrix elements in Eqs.~\eqref{eq:H}
and~\eqref{eq:D}. Although their form appears time-local in the continuous-time
limit, this is not strictly the case: before taking the limit, these expressions
involve matrix elements of $\hat{\psi}^{\dagger}_{r, l} \hat{\psi}_{r, l'}$
evaluated between coherent states separated by a single discrete time step. As a
first step, we follow the procedure of Ref.~\cite{Yang2023} to render these
matrix elements strictly time-local, which introduces an additive shift; as a
second step, we apply the principal-value regularization, which induces an
additional shift. Any such shift in Eq.~\eqref{eq:H-sigma} cancels when summing
over the Keldysh branch index in Eq.~\eqref{eq:S-L-eta}. The situation is
different for the matrix elements in
Eq.~\eqref{eq:D-counting-before-regularization}, which take the form
\begin{equation}
  \label{eq:D-counting}
  D_{\sigma, r, \pm, l}(t)
  = \frac{1}{2} \mp \psi^*_{\sigma,r,l}(t) \psi_{\sigma,r,l}(t).
\end{equation}

Since the jump operators~\eqref{eq:gain-loss-jump-operators} are linear in the
fermionic fields, the procedure of Ref.~\cite{Yang2023} introduces no additive
shift in the matrix elements~\eqref{eq:c-c-dagger}. Moreover, because the
vertices~\eqref{eq:V-counting} involve products of fields residing on different
Keldysh branches, they are unaffected by the principal-value regularization; see
Ref.~\cite{Starchl2025} for details.

In what follows, it is convenient to work in the basis~\cite{Poboiko2025}
\begin{equation}
  \label{eq:modified-branch-basis}
  \psi =
  \begin{pmatrix}
    \psi_+ \\ \psi_-
  \end{pmatrix}, 
  \qquad 
  \psi^{\dagger} = \left( \psi_+^{\dagger}, -\psi_-^{\dagger} \right),
\end{equation}
where $\psi_{\pm}$ and $\psi_{\pm}^{\dagger}$ are column- and row-vectors in
replica space. This choice of basis is motivated by the following observations:
The usual Larkin-Ovchinnikov transformation of fermionic fields combines the
bosonic Keldysh rotation with a sign change of $\psi_-^{*}$~\cite{Altland2010a,
  Kamenev2023}. This sign change, allowed only for fermionic fields because
$\psi_-$ and $\psi_-^{*}$ are independent Grassmann variables, greatly
simplifies the quadratic Hamiltonian part of the Keldysh action by reducing it
to the identity in Keldysh space. Here, we implement only the sign change, which
already provides a useful simplification, while omitting the bosonic
rotation. This choice proves convenient for constructing the NLSM target
manifold, whose structure is more transparent in this representation.

In the basis~\eqref{eq:modified-branch-basis}, the
vertices~\eqref{eq:V-counting} take the compact form
\begin{equation}  
  V_{r, \pm, l}(t) = - \psi_{r, l}^{\dagger}(t) \sigma_{\pm} \psi_{r, l}(t),
\end{equation}
where $\sigma_{\pm} = (\sigma_x \pm \imag \sigma_y)/2$. Moreover, only the sum
over branch indices of the coherent-state matrix elements in
Eq.~\eqref{eq:D-counting} enters the action in Eqs.~\eqref{eq:S-L-eta}
and~\eqref{eq:S-D-eta}. This sum can be written as
\begin{equation}  
  \sum_{\sigma = \pm} D_{\sigma, r, \pm, l}(t) 
  = 1 \mp \psi_{r, l}^{\dagger}(t) \sigma_z \psi_{r, l}(t).
\end{equation}

We now rewrite the full action in Eq.~\eqref{eq:Z-R-d-psi} using these compact
expressions. First, we decompose the action as
\begin{equation}
  \label{eq:S-H-S-M}
  S = S_L + S_D + S_J = S_H + S_M,
\end{equation}
where $S_H$ is the Hamiltonian part, the first line of Eq.~\eqref{eq:S-L-eta},
while $S_M$ contains the remaining terms due to measurements: the second line of
Eq.~\eqref{eq:S-L-eta} together with Eqs.~\eqref{eq:S-D-eta}
and~\eqref{eq:S-J-eta}.

The Hamiltonian action reads
\begin{equation}
  \label{eq:S-H}
  S_H[\psi^{*}, \psi] 
  = \sum_{l, l' = 1}^L \int_{t_0}^t \diff t'\,
    \psi_l^{\dagger}(t')\, G_{0, l, l'}^{-1} \psi_{l'}(t'),
\end{equation}
with
\begin{equation}
  \label{eq:G-0}
  G_0^{-1} = \imag \partial_t - H,
\end{equation}
where $H$ is the Hamiltonian matrix in Eq.~\eqref{eq:Hamiltonian}.

The measurement action takes the form
\begin{equation}
  S_M[\psi^{*}, \psi] 
  = \sum_{l = 1}^L \int_{t_0}^t \diff t'\,
    \mathcal{L}_M[\psi_l^{*}(t'), \psi_l(t')],
\end{equation}
with measurement Lagrangian
\begin{multline}
  \label{eq:L-M}
  \imag \mathcal{L}_M[\psi^{*}, \psi] 
  = - (1 - \eta)\, \psi^{\dagger} \Lambda \psi
    + \frac{\eta \Delta \gamma}{2}\, \psi^{\dagger} \sigma_z \psi \\
  + \eta \sum_{\alpha = \pm} \gamma_{\alpha}
      \prod_{r = 1}^R \left( -\psi_r^{\dagger} \sigma_{\alpha} \psi_r \right)
    - \gamma \left[ R + \eta \left( 1 - R \right) \right],
\end{multline}
where the net gain rate is
\begin{equation}
  \label{eq:Delta-gamma}
  \Delta \gamma = \gamma_+ - \gamma_-.
\end{equation}

In Eq.~\eqref{eq:L-M}, the matrices $\sigma_{\pm}$, $\sigma_z$, and
\begin{equation}
  \label{eq:Lambda}
  \Lambda =
  \begin{pmatrix}
    1 - 2n & 2n \\
    2(1 - n) & -1 + 2n
  \end{pmatrix},
\end{equation}
with density given in Eq.~\eqref{eq:n}, act nontrivially in Keldysh space; the
identity in replica space is left implicit. To obtain the form of
Eq.~\eqref{eq:L-M}, we used
\begin{equation}
  \label{eq:Lambda-gamma}  
  \sum_{\alpha = \pm} 
    \gamma_{\alpha} \left( \sigma_{\alpha} - \frac{\alpha}{2}\sigma_z \right)
  = \gamma\, \Lambda.
\end{equation}

The matrix $\Lambda$ encodes the symmetrized equal-time limit of the exact
Green's functions for a given mean fermion density~\cite{Starchl2025}. It will
play an important role in our analysis of the saddle-point structure of the
Keldysh action.

\subsubsection{Generalized Hubbard-Stratonovich transformation}

The generalized Hubbard-Stratonovich transformation is implemented by inserting
the following identity into the functional
integral~\eqref{eq:Z-R-d-psi}~\cite{Poboiko2023}:
\begin{equation}
  \label{eq:HS-identity}
  1 = \int \Diff[\mathcal{G}, \Sigma] \, \e^{- \imag \Tr \left[ \left(
        \mathcal{G} + \imag \psi \psi^{\dagger} \right) \Sigma \right]},
\end{equation}
where $\mathcal{G}$ and $\Sigma$ are Hermitian $2R \times 2R$ matrix fields. We
use $\Tr$ to denote a trace that acts in Keldysh, replica, lattice, and time
spaces. The integration over $\Sigma$ yields a delta functional that enforces
the identity $\mathcal{G} = - \imag \psi \psi^{\dagger}$. Together with the
principal-value regularization, this ensures that the expectation value
$\langle \mathcal{G}(t) \rangle$ coincides with the fermionic Green's function
in the symmetrized equal-time limit,
\begin{equation}
  \label{eq:G-symmetric-limit}
  \left\langle \mathcal{G}_l(t) \right\rangle = - \frac{\imag}{2} \left(
    \langle \psi_l(t) \psi^{*}_l(t + 0^+) \rangle + \langle \psi_l(t)
    \psi^{*}_l(t - 0^+) \rangle \right).
\end{equation}

The identity $\mathcal{G} = - \imag \psi \psi^{\dagger}$ further allows us to
recast the measurement Lagrangian~\eqref{eq:L-M} entirely in terms of
$\mathcal{G}$. For the first two terms in~\eqref{eq:L-M}, we use
\begin{equation}
  \psi^{\dagger} \Lambda \psi = - \tr \! \left( \Lambda \psi \psi^{\dagger}
  \right) = - \imag \tr(\Lambda \mathcal{G}), 
  \quad 
  \psi^{\dagger} \sigma_z \psi = - \imag \tr(\sigma_z \mathcal{G}).
\end{equation}
Following~\cite{Poboiko2023}, we decouple the product over replicas in the third
term of Eq.~\eqref{eq:L-M} simultaneously in all slow channels by summing over
Wick contractions, which yields
\begin{multline}
  \label{eq:L-M-G}
  \imag \mathcal{L}_M[\mathcal{G}] = \imag \left( 1 - \eta \right)
  \gamma \tr(\Lambda \mathcal{G}) - \frac{\imag \eta \Delta \gamma}{2} \tr(\sigma_z \mathcal{G}) \\
  + \imag^R \eta \sum_{\alpha = \pm} \gamma_{\alpha} \detR \! \left[ \trK \!
    \left( \sigma_{\alpha} \mathcal{G} \right) \right] - \gamma \left[ R + \eta
    \left( 1 - R \right) \right],
\end{multline}
where $\trK$ denotes a partial trace in Keldysh space, and $\detR$ is the
determinant in the remaining replica space. The measurement Lagrangian given in
Ref.~\cite{Starchl2025} for balanced and efficient fermion counting is recovered
by setting $\gamma_+ = \gamma_-$ and $\eta = 1$. In the limit $R \to 1$,
Eq.~\eqref{eq:L-M-G} reduces to
\begin{equation}
  \label{eq:L-M-R=1}
  \imag \mathcal{L}_M[\mathcal{G}] = \gamma \left[ \imag \trK(\Lambda
    \mathcal{G}) - 1 \right].
\end{equation}

With the measurement Lagrangian rewritten in terms of $\mathcal{G}$, the
fermionic integral~\eqref{eq:Z-R-d-psi} becomes Gaussian and can be evaluated
explicitly. This yields the partition function in the form
\begin{equation}
  \label{eq:Z-R-d-G-d-Sigma}
  Z_R(t) = \int \Diff[\mathcal{G}, \Sigma] \, \e^{\imag S[\mathcal{G}, \Sigma]},
\end{equation}
where the action is
\begin{equation}
  \label{eq:S-G-Sigma}
  S[\mathcal{G}, \Sigma] = S_0[\mathcal{G}, \Sigma] 
  + \sum_{l=1}^L \int_{t_0}^t \diff t' \, \mathcal{L}_M[\mathcal{G}_l(t')],
\end{equation}
with the measurement Lagrangian~\eqref{eq:L-M-G} and
\begin{equation}
  \label{eq:S-0-G-Sigma}
  \imag S_0[\mathcal{G}, \Sigma]
  = 
  \Tr \! \left[ 
    \ln\!\left(\imag \partial_t - H + \imag \Sigma \right) 
    - \imag \mathcal{G} \Sigma 
  \right].
\end{equation}

\subsubsection{Continuous symmetries of the replica Keldysh action}
\label{sec:continuous-symmetries-Keldysh}

A prerequisite for understanding the long-wavelength behavior of our model is a
characterization of the global continuous symmetries of the replica Keldysh
action. This analysis will inform both the discussion of the Gaussian theory in
Sec.~\ref{sec:gaussian-theory} and the construction of the NLSM in
Sec.~\ref{sec:NLSM}. Of particular importance are unitary field transformations,
$\psi \mapsto \mathcal{R} \psi$ with $\mathcal{R} \in \mathrm{U}(2 R)$, since
breaking such symmetries leads to the appearance of massless Goldstone modes.

Due to the identity $\mathcal{G} = - \imag \psi \psi^{\dagger}$, the matrix
field $\mathcal{G}$ transforms as
$\mathcal{G} \mapsto \mathcal{R} \mathcal{G} \mathcal{R}^{-1}$. When does such a
transformation constitute a symmetry? The action in Eq.~\eqref{eq:S-0-G-Sigma}
is invariant under arbitrary transformations of $\mathcal{G}$ provided $\Sigma$
is transformed simultaneously as
$\Sigma \mapsto \mathcal{R} \Sigma \mathcal{R}^{-1}$. Nontrivial constraints on
the symmetry group $G \le \mathrm{U}(2 R)$ therefore arise from the measurement
Lagrangian~\eqref{eq:L-M-G} by requiring
$\mathcal{L}_M[\mathcal{G}] = \mathcal{L}_M[\mathcal{R} \mathcal{G}
\mathcal{R}^{-1}]$.
To specify these constraints, we analyze the measurement Lagrangian term by
term.

We first consider efficient detection with $\eta = 1$, in which case the first
term in Eq.~\eqref{eq:L-M-G} vanishes.  The second term, proportional to
$\tr(\sigma_z \mathcal{G})$, is invariant if
\begin{equation}
  \mathcal{R} \sigma_z \mathcal{R}^{-1} = \sigma_z.
\end{equation}
This holds when $\mathcal{R}$ is block-diagonal in Keldysh space,
\begin{equation}
  \label{eq:R-AIII}
  \mathcal{R} =
  \begin{pmatrix}
    \mathcal{V}_+ & 0 \\ 0 & \mathcal{V}_-
  \end{pmatrix},
\end{equation}
with $\mathcal{V}_{\pm} \in \mathrm{U}(R)$, so that
$\mathcal{R} \in \mathrm{U}(R) \times \mathrm{U}(R)$.

The matrices $\mathcal{V}_{\pm}$ are further constrained by the third term in
the measurement Lagrangian~\eqref{eq:L-M-G}.  Consider first the contribution
with $\alpha = -$.  Inserting the transformed $\mathcal{G}$, the determinant in
replica space factorizes as
\begin{equation}
  \detR \! \left[ \trK \! \left( \sigma_- \mathcal{R} \mathcal{G}
    \mathcal{R}^{-1} \right) \right]
    = 
    \frac{\detR(\mathcal{V}_+)}{\detR(\mathcal{V}_-)}
    \detR[\trK(\sigma_- \mathcal{G})].
\end{equation}
The final factor appears also in the original expression in
Eq.~\eqref{eq:L-M-G}. Hence, $\mathcal{R}$ is a symmetry if the prefactor
reduces to unity, which is the case if
$\detR(\mathcal{V}_+) = \detR(\mathcal{V}_-)$.  The same condition follows from
the term with $\alpha = +$.

To determine the symmetry group $G$, note that the unitary group factorizes as
$\mathrm{U}(R) = \mathrm{U}(1) \rtimes \mathrm{SU}(R)$, separating the global
phase from the special-unitary part.  Because $\mathcal{V}_+$ and
$\mathcal{V}_-$ must have equal determinants, only a single global
$\mathrm{U}(1)$ factor remains.  Thus the symmetry group is
\begin{equation}
  \label{eq:G-efficient}
  G \cong \mathrm{U}(1) \rtimes \left[ \mathrm{SU}(R) \times \mathrm{SU}(R) \right].
\end{equation}

The block-diagonal constraint~\eqref{eq:R-AIII} arises from the second term in
Eq.~\eqref{eq:L-M-G}, which vanishes for balanced fermion counting with
$\Delta \gamma = 0$. We did not find a way to exclude the possibility that, for
$\Delta \gamma = 0$, the remaining third term might admit non-block-diagonal
symmetries. However, in particle-number-conserving models, $\mathcal{R}$ can
indeed be shown to be block-diagonal~\cite{Starchl2025, Poboiko2025}. Since
fermion counting breaks particle-number conservation, the symmetry group of the
action is expected to be more restricted in this case. Therefore,
transformations $\mathcal{R}$ should remain block-diagonal.

We now turn to inefficient detection with $\eta < 1$. The first term in
Eq.~\eqref{eq:L-M-G} introduces the additional constraint
\begin{equation}
  \label{eq:Lambda-R}
  \mathcal{R} \Lambda \mathcal{R}^{-1} = \Lambda,
\end{equation}
which forces $\mathcal{V}_+ = \mathcal{V}_-$.  
The symmetry group~\eqref{eq:G-efficient} is therefore reduced to
\begin{equation}
  \label{eq:G-inefficient}
  G \cong \mathrm{U}(R).
\end{equation}

\subsubsection{Continuous symmetries in the operator formalism}
\label{sec:operator-symmetries}

We briefly digress to connect the symmetries of the Keldysh action to weak and
strong symmetries of the time-evolution superoperator~\cite{Buca2012,
  Albert2014}. A constructive approach to identifying operator symmetries is
presented in Ref.~\cite{Starchl2025}; instead, here we derive them from the
field-theoretic formulation.

The identity $\mathcal{G} = - \imag \psi \psi^{\dagger}$, together with the
definition of the field vectors in Eq.~\eqref{eq:modified-branch-basis} and the
block-diagonal form of $\mathcal{R}$ in Eq.~\eqref{eq:R-AIII}, implies that the
two blocks $\mathcal{V}_{\pm}$ act separately on the field components
$\psi_{\pm}$ on the forward and backward branches of the closed time contour.
In general, field transformations that act independently on the two branches
correspond to strong symmetries of the time-evolution superoperator, whereas
transformations acting identically on both branches correspond to weak
symmetries~\cite{Sieberer2025}.

For $\eta = 1$, we found that $\mathcal{V}_{\pm} \in \mathrm{U}(R)$ with the
constraint $\detR(\mathcal{V}_+) = \detR(\mathcal{V}_-)$.  The common
determinant defines a weak $\mathrm{U}(1)$ symmetry.  Aside from sharing this
determinant, the matrices $\mathcal{V}_{\pm}$ remain independent, giving rise to
a strong $\mathrm{SU}(R)$ symmetry.  As we will see in Sec.~\ref{sec:NLSM}, it
is precisely this strong $\mathrm{SU}(R)$ symmetry that underlies the NLSM.

For $\eta < 1$, we obtained the additional constraint
$\mathcal{V}_+ = \mathcal{V}_-$, reducing the symmetry group from
Eq.~\eqref{eq:G-efficient} to Eq.~\eqref{eq:G-inefficient} and thus leaving only
a weak $\mathrm{U}(R)$ symmetry.  That inefficient detection breaks the symmetry
required for the NLSM description was already anticipated for balanced fermion
counting in Ref.~\cite{Starchl2025}. Furthermore, Ref.~\cite{Starchl2025} also
shows that two conditions are necessary for the strong $\mathrm{SU}(R)$ symmetry
underlying the NLSM to emerge in free fermionic systems: (i)~conservation of the
total particle number of the system and reservoirs, and (ii)~conservation of the
purity of the state of the system.  The latter requires keeping a complete
record of measurement outcomes, which is fulfilled only when $\eta = 1$.

\subsection{Gaussian theory}
\label{sec:gaussian-theory}

Following Refs.~\cite{Poboiko2023, Starchl2025}, we first analyze the theory
within the Gaussian approximation. Specifically, we determine the
replica-symmetric saddle point of the Keldysh action and restrict the functional
integral~\eqref{eq:Z-R-d-G-d-Sigma} to quadratic fluctuations around this saddle
point. This approximation is valid for small $\gamma$ and on short to
intermediate length scales.

\subsubsection{Saddle point and fluctuations}
\label{sec:saddle-point-fluctuations}

The saddle-point equations $\delta S/\delta \mathcal{G} = 0$ and
$\delta S/\delta \Sigma = 0$ admit a unique replica-diagonal solution.
Replica-asymmetric or replicon saddle points, which form a finite-dimensional
manifold, will be discussed in Sec.~\ref{sec:NLSM}. Setting $R = 1$ in numerical
prefactors, the replica-symmetric saddle point is
\begin{equation}
  \label{eq:G-Sigma-Q-Lambda}
  \mathcal{G} = - \imag Q/2, \qquad \Sigma = \gamma Q, \qquad Q =
  \Lambda,
\end{equation}
independent of $\eta$, and with $\Lambda$ defined in Eq.~\eqref{eq:Lambda}.
Gaussian fluctuations around this saddle point can be parameterized as
\begin{equation}
  \label{eq:G-Sigma-fluctuations}
  \mathcal{G} = -\imag \left( \Lambda + \dQG{} \right) \! \big/2, \qquad \Sigma
  = \gamma \left( \Lambda + \dQS{} \right) \! \big/2,
\end{equation} 
where $\dQG$ and $\dQS$ are $2 R \times 2 R$ matrices. The Gaussian theory is
obtained by expanding the Keldysh action~\eqref{eq:S-G-Sigma} to second order in
$\dQG$ and $\dQS$ as detailed in Refs.~\cite{Poboiko2023, Starchl2025}.  

A key feature of fluctuations around the saddle point can be anticipated from
the discussion of symmetries in Sec.~\ref{sec:continuous-symmetries-Keldysh}:
since the saddle point~\eqref{eq:G-Sigma-Q-Lambda} partially breaks these
symmetries, certain types of fluctuations become massless Goldstone modes. The
manifold of these modes is the quotient $G/H$, where $G$ is the symmetry group
of the Keldysh action, given by Eqs.~\eqref{eq:G-efficient}
and~\eqref{eq:G-inefficient} for $\eta = 1$ and $\eta < 1$, respectively, and
$H$ is the subgroup of $G$ that leaves the saddle point invariant, as expressed
by Eq.~\eqref{eq:Lambda-R}. In Sec.~\ref{sec:continuous-symmetries-Keldysh}, we
already identified $H \cong \mathrm{U}(R)$ for inefficient detection.

It follows that for $\eta = 1$, the Goldstone manifold $G/H \cong \mathrm{SU}(R)$ is
nontrivial, whereas for $\eta < 1$, $G = H$ and the quotient $G/H$ is
trivial. Hence, massless modes arise for $\eta = 1$ due to the strong
$\mathrm{SU}(R)$ symmetry discussed in Sec.~\ref{sec:operator-symmetries}, and
they acquire a mass when this strong symmetry is broken for $\eta < 1$.

The physics of fluctuations of these Goldstone modes on the nonlinear manifold
is discussed in Sec.~\ref{sec:NLSM}. But first, we focus on linearized
fluctuations. As we show, within this approximation, the massless modes for
$\eta = 1$ generate algebraic correlations and logarithmic growth of
entanglement, whereas the mass term for $\eta < 1$ introduces a finite
correlation length.

\subsubsection{Connected density correlation function}

As discussed in Sec.~\ref{sec:replica-formalism}, the nontrivial effects of
measurements become apparent in nonlinear observables. We focus here on the
equal-time connected density correlation function, defined in the operator
formalism and in terms of field expectation values as~\cite{Poboiko2023}
\begin{equation}
  \label{eq:C-l-l-prime-t}
  \begin{split}
    C_{l, l'}(t) & = \frac{1}{2} \overline{\left\langle \left\{ \hat{n}_l(t),
          \hat{n}_{l'}(t) \right\} \right\rangle} - \overline{\left\langle
        \hat{n}_l(t) \right\rangle \left\langle \hat{n}_{l'}(t) \right\rangle} \\
    & = C_{r, r, l, l'}(t, t) - C_{r, r', l, l'}(t, t),
  \end{split}
\end{equation}
where the correlation function of density fluctuations is
\begin{equation}
  \label{eq:C-r-r-prime}
  C_{r, r', l, l'}(t, t') = \langle \delta \rho_{r, l}(t) \delta \rho_{r',
    l'}(t') \rangle.
\end{equation}
Permutation symmetry among replicas renders the specific choice of $r$ and $r'$
in Eq.~\eqref{eq:C-l-l-prime-t} irrelevant, provided that $r \neq r'$. The
density fluctuations in Eq.~\eqref{eq:C-r-r-prime} are related to the $\dQG$ in
Eq.~\eqref{eq:G-Sigma-fluctuations} via~\cite{Poboiko2023}
\begin{equation}
  \delta \rho_{r, l}(t) = - \frac{1}{4} \trK \! \left[ \sigma_x \delta
    Q_{\mathcal{G}, r, r, l}(t) \right].
\end{equation}

The correlation function~\eqref{eq:C-l-l-prime-t} can be calculated explicitly
in the Gaussian approximation. In the steady state, $C_{l, l'}(t)$ becomes
time-independent and translationally invariant, allowing us to write
$C_{l, l'}(t) = C_{l - l'}$. When calculating $C_l$, boundary effects due to
stopping the measurement process at finite time $t$ must be taken into
account~\cite{Poboiko2023}. This leads to an expression for the correlation
function in momentum space, $C_q$, in terms of the solution of an integral
equation that depends on the two parameters $u = 2 l_0 \sin(q/2)$ and
$\eta$, where
\begin{equation}
  \label{eq:l-0}
  l_0 = \frac{J}{\sqrt{2} \gamma}.
\end{equation}
However, the leading asymptotic behavior of $C_q$ for small momenta and small
detection inefficiency $\delta \eta = 1 - \eta$ is correctly reproduced in the
bulk approximation~\cite{Poboiko2023}, in which the boundary condition in time
is neglected and we rescale $u \to 2 u$ and $\delta \eta \to 4 \delta \eta$,
leading to the following expression for the correlation function:
\begin{equation}
  \label{eq:C-q-Gaussian}
  C_q \approx n \left( 1 - n \right) \tilde{c}(q l_0),
\end{equation}
with
\begin{multline}
  \label{eq:c-tilde}
  \tilde{c}(u) = \frac{2}{\pi} \int_0^{\infty} \diff v \left\{ \Re[b(2 u, v)] -
    \left( 1 - 4 \delta \eta \right) \abs{b(2 u, v)}^2 \right\} \\ \times \left(
    1 - \left( 1 - 4 \delta \eta \right)^2 \abs{b(2 u, v)}^2 - 4 \left( 1 - 4
      \delta \eta \right) n \left( 1 - n \right) \right. \\
  \left. \times \left\{ \Re[b(2 u, v)] - \left( 1 - 4 \delta \eta \right)
      \abs{b(2 u, v)}^2 \right\} \right)^{-1},
\end{multline}
where
\begin{equation}
  b(u, v) = \left[ \left( 1 - \imag v \right)^2 + 2 u^2 \right]^{-1/2}.
\end{equation}
For low momenta $q \to 0$, $C_q$ behaves as
\begin{equation}
  \label{eq:C-q-asymptotic-Gaussian}
  C_q \sim
  \begin{cases}
    g_0 q - 4 n \left( 1 - n \right) \left( q l_0 \right)^2, & \delta \eta = 0, \\
    \frac{g_0}{l_0} \sqrt{\left( q l_0 \right)^2 + \delta \eta}, & \delta \eta
    \to 0,
  \end{cases}
\end{equation}
with
\begin{equation}
  \label{eq:g-0}
  g_0 = \frac{\sqrt{2} n \left( 1 - n \right) l_0}{\sqrt{1 - 2 n \left( 1 - n \right)}}.
\end{equation}
Transforming to real space yields
\begin{equation}
  \label{eq:C-l-Gaussian}
  C_l \sim
  \begin{cases}
    - \frac{g_0}{\pi l^2}, & \delta \eta = 0, \\
    - \frac{g_0}{\sqrt{\pi l_0}} \frac{\delta \eta^{1/4}}{l^{3/2}} \e^{- l/\xi},
    & \delta \eta \to 0,
  \end{cases}
\end{equation}
where
\begin{equation}
  \label{eq:xi}
  \xi = \frac{l_0}{\sqrt{\delta \eta}}.
\end{equation}
For $\delta \eta = 0$, this reproduces the result of Ref.~\cite{Starchl2025}:
Algebraic correlations arise as a manifestation of a massless Goldstone mode,
even though particle-number conservation is broken. For $\delta \eta > 0$,
correlations are cut off at the scale $\xi$~\eqref{eq:xi}.

\subsubsection{Subsystem entropy}
\label{sec:subsystem-entropy-Gaussian}

The second key nonlinear observable we consider is the subsystem entropy. For a
system in the state $\hat{\rho}$, the reduced density matrix of a contiguous
block of $\ell$ sites is obtained by tracing out the remaining $L - \ell$ sites,
$\hat{\rho}_{\ell} = \tr_{L - \ell}(\hat{\rho})$. The corresponding von Neumann
subsystem entropy is then
\begin{equation}
  \label{eq:subsystem-entropy}
  \svn{\ell} = - \overline{\tr \! \left[ \hat{\rho}_{\ell}
      \ln\!\left(\hat{\rho}_{\ell}\right) \right]}.
\end{equation}
Efficient detection with $\eta = 1$ yields pure-state quantum trajectories,
$\hat{\rho} = \ket{\psi}\bra{\psi}$, making $\svn{\ell}$ a genuine measure of
entanglement. However, for inefficient detection with $\eta < 1$, the
conditional state $\hat{\rho}$ becomes mixed. Then, entanglement can be
quantified by the fermionic logarithmic negativity~\cite{Shapourian2017,
  Shapourian2019}, which we discuss further in Sec.~\ref{sec:ferm-logar-negat}.

The subsystem entropy can be computed using another property of quantum
trajectories: evolution under the quadratic Hamiltonian~\eqref{eq:Hamiltonian}
and linear jump operators~\eqref{eq:gain-loss-jump-operators} preserves
Gaussianity of the state. For a Gaussian state, $\svn{\ell}$ is obtained from
fluctuations of the subsystem particle number,
$\hat{N}_{\ell} = \sum_{l = 1}^{\ell} \hat{n}_l$, through a cumulant expansion
with coefficients given by the Riemann zeta function
$\zeta(k)$~\cite{Klich2009}:
\begin{equation}
  \label{eq:entropy-cumulant}
  \svn{\ell} = 2 \sum_{k = 1}^{\infty}  \zeta(2k) C_{\ell}^{(2 k)} = \frac{\pi^2}{3}
  C_{\ell}^{(2)} +  \frac{\pi^4}{45} C_{\ell}^{(4)} + \dotsb.
\end{equation}
We restrict ourselves to the leading term of the expansion, which involves the
second cumulant,
\begin{equation}
  \label{eq:cumulant}
  C_{\ell}^{(2)} = \overline{\langle ( \hat{N}_{\ell} - \langle \hat{N}_{\ell}
    \rangle )^2 \rangle}.
\end{equation}
Inserting here the explicit expression for $\hat{N}_{\ell}$ and using the
definition of the conditional density correlation
function~\eqref{eq:C-l-l-prime-t}, we obtain
\begin{equation}
  \label{eq:subsystem-entropy-correlation-function}
  \svn{\ell} = \frac{\pi^2}{3} \sum_{l, l' =
    1}^{\ell} C_{l - l'} = \frac{2 \pi}{3} \int_0^{\infty} \frac{\diff q}{q^2}
  \, C_q \left[ 1 - \cos(q \ell) \right].
\end{equation}
This relation allows us to infer the entanglement entropy from our results for
the connected density correlation function. We note that an alternative approach
to calculating the entanglement entropy, which is not restricted to Gaussian states,
is provided by introducing appropriate temporal boundary conditions in the
Keldysh functional integral~\cite{Poboiko2025}.

The behavior of the subsystem entropy for large subsystem sizes, $\ell \gg l_0$,
follows by inserting the asymptotic form~\eqref{eq:C-q-asymptotic-Gaussian} of
the density correlation function in
Eq.~\eqref{eq:subsystem-entropy-correlation-function}. For efficient detection,
$\delta \eta = 0$, we find
\begin{equation}
  \label{eq:S-l-asymptotic-Gaussian}
  S_{\ell} \sim \frac{2 \pi g_0}{3} \ln(\ell/l_0), \qquad \delta \eta = 0.
\end{equation}
That is, for efficient detection, the Gaussian theory predicts logarithmic
growth of entanglement on large scales, as in a 1D conformal field theory
(CFT)~\cite{DiFrancesco1997}.

On short scales $\ell \lesssim l_0$, the system is expected to behave like a
Fermi gas at infinite temperature~\cite{Poboiko2023}, leading to volume-law
behavior of the entropy:
\begin{equation}
  \label{eq:S-l-volume-law}
  S_{\ell} \sim - \left[ n \ln(n) + \left( 1 - n \right) \ln \left( 1 - n \right)
  \right] \ell.
\end{equation}
The prefactor is not fully reproduced by
Eq.~\eqref{eq:subsystem-entropy-correlation-function} since higher cumulants in
Eq.~\eqref{eq:entropy-cumulant} become significant in this
case~\cite{Poboiko2023}. We expect volume-law behavior $S_{\ell} \sim \ell$ to
persist for inefficient detection, $\eta < 1$, also at large scales
$\ell \gg l_0$. In particular, in the limit $\eta \to 0$,
Eq.~\eqref{eq:S-l-volume-law} describes the exact subsystem entropy of the fully
separable steady state~\eqref{eq:rho-ss}.

Deviations from the CFT behavior~\eqref{eq:S-l-asymptotic-Gaussian} for
$\eta = 1$ arising at short scales and, as discussed below, also at large scales
due to nonlinear fluctuations, can be characterized by the scale-dependent
effective central charge~\cite{Starchl2025}:
\begin{equation}
  \label{eq:effective-central-charge}
  c_{\ell} = 3 \frac{\partial \svn{\ell}}{\partial \! \ln(\ell)}.
\end{equation}
Using the Gaussian prediction~\eqref{eq:S-l-asymptotic-Gaussian}, one finds
\begin{equation}
  \label{eq:effective-central-charge-Gaussian}
  c_{\ell} \sim 2 \pi g_0, \qquad \ell \to \infty,
\end{equation}
with $g_0$ defined in Eq.~\eqref{eq:g-0}. Nonlinear fluctuations of Goldstone
modes drive a renormalization-group (RG) flow of $g_0$ toward zero on large
scales, causing the effective central charge to vanish and the entanglement
entropy to cross over from logarithmic to area-law scaling. Next, we discuss how
this RG flow can be obtained from an NLSM description.

\subsection{Nonlinear sigma model}
\label{sec:NLSM}

As anticipated in Sec.~\ref{sec:saddle-point-fluctuations}, the
replica-symmetric saddle point~\eqref{eq:G-Sigma-Q-Lambda} breaks global
continuous symmetries of the Keldysh action, thereby giving rise to Goldstone
modes. For $\eta = 1$, these modes span a nonlinear manifold, with fluctuations
within this manifold described by an NLSM, which we derive in the present
section.

In our discussion of the Goldstone-mode manifold, and hence of the target
manifold of the NLSM, we have thus far focused exclusively on continuous
symmetries of the Keldysh action. We have neglected an important discrete
symmetry, to which we now turn: particle-hole symmetry (PHS)~\cite{Jian2022,
  Fava2024, Poboiko2025, Starchl2025, Niederegger2025}.

\subsubsection{Particle-hole symmetry}
\label{sec:PHS}

The quadratic hopping Hamiltonian~\eqref{eq:Hamiltonian} exhibits PHS if there
exists a unitary matrix $U$ with $U U^{*} = \pm 1$ such that the Hamiltonian
matrix $H$ satisfies $U H^{*} U^{\dagger} = - H$~\cite{Ludwig2016}. Owing to the
Hermiticity of $H$, this condition is equivalently written as
$U H^{\transpose} U^{\dagger} = - H$.  Here we focus on the case $U U^{*} = +1$,
which, together with unitarity of $U$, implies that $U$ is symmetric,
$U = U^{\transpose}$. By the Autonne-Takagi factorization, such a matrix can be
expressed as $U = R^{\dagger} R^{*}$, where $R$ is unitary and we have used that
all singular values of $U$ are equal to one. For the transformed Hamiltonian
$H' = R H R^{\dagger}$, we obtain
\begin{equation}
  H^{\prime \transpose}
  = R^{*} H^{\transpose} R^{\transpose}
  = - R^{*} U^{\dagger} H U R^{\transpose}
  = - H'.
\end{equation}
Thus, PHS of $H$ implies that $H'$ is skew-symmetric.

In what follows, we leave the explicit form of $H$ unspecified in intermediate
expressions. However, for any $H$ possessing PHS, we assume that we work in a
basis in which PHS reduces to skew symmetry. Our final results apply to the
Hamiltonian matrix~\eqref{eq:Hamiltonian-matrix} describing nearest-neighbor
hopping with real hopping amplitudes. For this case, the necessary basis
transformation corresponds to a gauge transformation of the field operators,
$\hat{\psi}_l \mapsto \imag^l \hat{\psi}_l$. This renders the hopping
matrix~\eqref{eq:Hamiltonian-matrix} purely imaginary,
\begin{equation}
  H_{l, l'} \mapsto - \imag J \left( \delta_{l + 1, l'} - \delta_{l, l' + 1} \right).
\end{equation}

PHS alters the target manifold of the NLSM~\cite{Jian2022, Fava2024,
  Poboiko2025, Starchl2025, Niederegger2025}. To determine this modification, we
first explicitly symmetrize the action with respect to PHS and then carry out
the generalized Hubbard-Stratonovich transformation on the symmetrized action.

We begin by considering the Hamiltonian contribution to the
action~\eqref{eq:S-H}. Working in the basis of
Eq.~\eqref{eq:modified-branch-basis} and leaving the summation over $\sigma$,
$r$, and $l$ implicit as matrix multiplication, the Hamiltonian term takes the
form
\begin{multline}
  \psi^{\dagger}(t) H \psi(t)
  = \frac{1}{2} \psi^{\dagger}(t) \left( H - H^{\transpose} \right) \psi(t) \\
  = \frac{1}{2} \left( \psi^{\dagger}(t) H \psi(t)
      + \psi^{\transpose}(t) H \psi^{*}(t) \right)
  = \bar{\Psi}(t) H \Psi(t),
\end{multline}
where we have introduced the doubled field vectors
\begin{equation}
  \label{eq:doubled-spinors}
  \Psi = \frac{1}{\sqrt{2}}
  \begin{pmatrix}
    \psi \\ - \imag \sigma_y \psi^{*}
  \end{pmatrix},
  \quad
  \bar{\Psi}
  = \frac{1}{\sqrt{2}}
    \left( \psi^{\dagger}, \psi^{\transpose} \imag \sigma_y \right)
  = \Psi^{\transpose} \mathcal{C}.
\end{equation}
To avoid ambiguity, we clarify the notation:
\begin{equation}
  \psi^{*} = \left( \psi^{\dagger} \right)^{\transpose}
  =
  \begin{pmatrix}
    \psi_+^{*} \\ -\psi_-^{*}
  \end{pmatrix},
  \qquad
  \psi^{\transpose} = \left( \psi_+, \psi_- \right).
\end{equation}
The charge-conjugation matrix $\mathcal{C}$ is defined as
\begin{equation}
  \label{eq:C}
  \mathcal{C}
  = - \sigma_y \otimes \tau_y
  = \mathcal{C}^{\transpose}
  = \mathcal{C}^{*}
  = \mathcal{C}^{-1},
\end{equation}
where $\sigma_y$ and $\tau_y$ are Pauli matrices acting in Keldysh and
charge-conjugation space, respectively; the identity in replica space is
implicit. Note that the doubled field vectors defined in
Eq.~\eqref{eq:doubled-spinors} coincide with those in
Ref.~\cite{Starchl2025}. The difference here is that we work in the basis of
Eq.~\eqref{eq:modified-branch-basis}, which is related to the basis used in
Ref.~\cite{Starchl2025} by a bosonic Keldysh rotation.

In terms of the doubled field vectors, the Hamiltonian contribution to the
action~\eqref{eq:S-H} can thus be written as
\begin{equation}
  \label{eq:S-H-PHS}
  S_H[\Psi]
  = \sum_{l, l' = 1}^{L}
  \int_{t_0}^{t} \diff t'\,
  \bar{\Psi}_l(t')\, G_{0,l,l'}^{-1}\, \Psi_{l'}(t').
\end{equation}

The matrices $\pm \imag \sigma_y$ in Eq.~\eqref{eq:doubled-spinors} are not
strictly required to symmetrize the Hamiltonian part of the action. Their choice
is not unique, and an alternative is used in Ref.~\cite{Poboiko2025} for
projective measurements. However, with the present definition of $\Psi$ and
$\bar{\Psi}$, the the measurement Lagrangian~\eqref{eq:L-M} acquires a form
devoid of structure in charge-conjugation space and is therefore manifestly
particle-hole symmetric: Using
\begin{equation}
  \label{eq:Lambda-sigma-z-sigma-+--PHS}
  \sigma_y \Lambda^{\transpose} \sigma_y = - \Lambda,
  \quad
  \sigma_y \sigma_z \sigma_y = - \sigma_z,
  \quad
  \sigma_y \sigma_{\pm}^{\transpose} \sigma_y = - \sigma_{\pm},
\end{equation}
we obtain
\begin{multline}
  \label{eq:L-M-PHS}
  \imag \mathcal{L}_M[\Psi] = - \left( 1 - \eta \right) \bar{\Psi}\Lambda\Psi
  + \frac{\eta \Delta \gamma}{2} \bar{\Psi}\sigma_z\Psi \\
  + \eta \sum_{\alpha = \pm} \gamma_{\alpha} \prod_{r = 1}^{R} \left( -
    \bar{\Psi}_r \sigma_{\alpha} \Psi_r \right) - \gamma \left[ R + \eta \left(
      1 - R \right) \right].
\end{multline}
This is formally equivalent to Eq.~\eqref{eq:L-M} with the replacements $\psi
\to \Psi$ and $\psi^{\dagger} \to \bar{\Psi}$.

The presence or absence as well as the precise form of PHS depends on the
details of the model. In contrast, a chiral symmetry is always present, as it is
intrinsic to the Keldysh formalism~\cite{Poboiko2025, Guo2025}. Consequently, if
the Hamiltonian matrix $H$ exhibits PHS, the model belongs to the chiral
orthogonal symmetry class BDI; if PHS is broken, it instead falls into the
chiral unitary class AIII.

\subsubsection{Generalized Hubbard-Stratonovich transformation with PHS}

We proceed with a generalized Hubbard-Stratonovich transformation of the
particle-hole symmetrized action. For systems with PHS,
Eq.~\eqref{eq:HS-identity} must be modified to
\begin{equation}
  \label{eq:HS-identity-PHS}
  1 = \int \Diff[\mathcal{G}, \Sigma] \, \e^{- \imag \Tr \left[ \left(
        \mathcal{G} + \imag \Psi \bar{\Psi} \right) \Sigma \right]},
\end{equation}
where $\mathcal{G}$ and $\Sigma$ are now Hermitian $4R \times 4R$ matrix fields.
As before, integrating over $\Sigma$ produces a delta functional enforcing the
identity $\mathcal{G} = - \imag \Psi \bar{\Psi}$. Using the relation
$\bar{\Psi} = \Psi^{\transpose} \mathcal{C}$ from
Eq.~\eqref{eq:doubled-spinors}, this identity implies
\begin{equation}
  \label{eq:G-PHS}
  \bar{\mathcal{G}} = \mathcal{C} \mathcal{G}^{\transpose} \mathcal{C} = - \mathcal{G}.
\end{equation}
By Eq.~\eqref{eq:HS-identity-PHS}, the same condition holds for $\Sigma$.

The identity $\mathcal{G} = - \imag \Psi \bar{\Psi}$ allows us to rewrite the
measurement Lagrangian~\eqref{eq:L-M-PHS} in terms of $\mathcal{G}$. Details of
this derivation are given in Appendix~\ref{sec:HS-transformation-PHS}. We find
\begin{multline}
  \label{eq:L-M-G-PHS}
  \imag \mathcal{L}_M[\mathcal{G}] = \imag \left( 1 - \eta \right) \gamma
  \tr(\Lambda \mathcal{G})
  - \frac{\imag \eta \Delta \gamma}{2} \tr(\sigma_z \mathcal{G}) \\
  + \imag^R \eta \sum_{\alpha = \pm} \gamma_{\alpha} \e^{\beta \trCR \ln \trK
    \left( \frac{1}{\beta} \sigma_{\alpha} \mathcal{G} \right)} - \gamma
  \left[ R + \eta \left( 1 - R \right) \right],
\end{multline}
where $\beta = 1/2$ for class BDI with PHS, and $\trCR$ denotes a partial trace
in charge-conjugation and replica spaces. This result, in a different basis and
for the special case of balanced and efficient fermion counting with
$\gamma_+ = \gamma_-$ and $\eta = 1$, was stated without proof in
Ref.~\cite{Starchl2025}. Formally, it reduces to Eq.~\eqref{eq:L-M-G} when
$\beta = 1$, corresponding to class AIII with broken PHS.

With the measurement Lagrangian expressed in terms of $\mathcal{G}$, performing
the remaining integral over $\Psi$ yields the Keldysh partition function in the
same form as in Eqs.~\eqref{eq:Z-R-d-G-d-Sigma} and~\eqref{eq:S-G-Sigma}, where
now
\begin{equation}
  \label{eq:S-0-G-Sigma-beta}
  \imag S_0[\mathcal{G}, \Sigma]
  = \Tr \! \left[ \beta \ln \!\left( \imag \partial_t - H + \imag \Sigma \right)
    - \imag \mathcal{G} \Sigma \right],
\end{equation}
which again formally reduces to Eq.~\eqref{eq:S-0-G-Sigma} for $\beta = 1$.

\subsubsection{NLSM target manifold for efficient detection, $\eta = 1$}
\label{sec:NLSM-target-manifold-efficient}

In Sec.~\ref{sec:gaussian-theory}, we derived the Gaussian theory by expanding
the action to quadratic order in fluctuations around the saddle
point~\eqref{eq:G-Sigma-Q-Lambda}. For simplicity, we chose the
replica-symmetric saddle point as the reference for this expansion. Applying
symmetry transformations to the replica-symmetric saddle point generates other
saddle points. The NLSM describes long-wavelength fluctuations of the Goldstone
modes within the manifold formed by these saddle points. Therefore, to derive
the NLSM, we first need to determine the structure of its target manifold,
meaning the manifold of saddle points, and obtain an explicit
parameterization. We anticipate that the symmetry underlying the NLSM
description is broken for inefficient detection with $\eta < 1$, and hence we
first focus on the case of efficient detection with $\eta = 1$.

As a preliminary step, we determine the replica-symmetric saddle point of the
particle-hole symmetrized action. Due to the factor $\beta$ in
Eq.~\eqref{eq:S-0-G-Sigma-beta}, the replica-symmetric solution to the
saddle-point equations is now
\begin{equation}
  \label{eq:G-Sigma-Q-Lambda-beta}
  \mathcal{G} = - \imag \beta Q/2, \qquad \Sigma = \gamma Q, \qquad Q = \Lambda,
\end{equation}
where, as before, $\beta = 1/2$ for class BDI. For class AIII, we formally
recover Eq.~\eqref{eq:G-Sigma-Q-Lambda} by setting $\beta = 1$. Note that the
matrix $\Lambda$ defined in Eq.~\eqref{eq:Lambda} satisfies
condition~\eqref{eq:G-PHS}, as follows from
Eq.~\eqref{eq:Lambda-sigma-z-sigma-+--PHS}:
\begin{equation}
  \label{eq:Lambda-C-PHS}
  \bar{\Lambda} = \mathcal{C} \Lambda^{\transpose} \mathcal{C} = \sigma_y
  \Lambda^{\transpose} \sigma_y = - \Lambda.
\end{equation}

We obtain the NLSM target manifold by applying continuous unitary
transformations to the replica-symmetric saddle
point~\eqref{eq:G-Sigma-Q-Lambda-beta}. Specifically, we replace $Q = \Lambda$ by
\begin{equation}
  \label{eq:Q-R}
  Q = \mathcal{R} \Lambda \mathcal{R}^{-1},
\end{equation}
where $\mathcal{R} \in \mathrm{U}(2 R)$ for symmetry class AIII and
$\mathcal{R} \in \mathrm{U}(4 R)$ for class BDI. In the latter case, the
transformations must additionally preserve the
condition~\eqref{eq:Lambda-C-PHS}, so that
$Q = - \mathcal{C} Q^{\transpose} \mathcal{C}$. This requirement is satisfied if
$\mathcal{R}^{-1} = \bar{\mathcal{R}}$, which, together with unitarity
$\mathcal{R}^{-1} = \mathcal{R}^{\dagger}$, implies
\begin{equation}
  \label{eq:R-PHS}
  \mathcal{R} = \mathcal{C} \mathcal{R}^{*} \mathcal{C}.
\end{equation}

Equation~\eqref{eq:G-Sigma-Q-Lambda-beta}, with $Q$ given by
Eq.~\eqref{eq:Q-R}, parameterizes the NLSM manifold if the transformations
$\mathcal{R}$ satisfy two conditions: (i)~They must be symmetries of the Keldysh
action. This ensures that if $Q = \Lambda$ is a saddle point, then
$Q = \mathcal{R} \Lambda \mathcal{R}^{-1}$ is also a saddle point. The set of
such transformations forms a group $G$. (ii)~They must act nontrivially on the
saddle point $\Lambda$. Transformations that leave $\Lambda$ invariant satisfy
Eq.~\eqref{eq:Lambda-R} and form a subgroup $H$. To isolate the transformations
that satisfy condition~(ii), one must factor out the subgroup $H$. The resulting
target manifold of the NLSM is then $G/H$.

We first determine the restrictions on $\mathcal{R}$ imposed by requirement~(i).
Note that symmetries for class AIII have already been discussed in
Sec.~\ref{sec:continuous-symmetries-Keldysh}. Here, we repeat the key points and
indicate the modifications required for class BDI.

The action~\eqref{eq:S-0-G-Sigma-beta} is symmetric under arbitrary 
transformations. Therefore, nontrivial symmetry constraints on the group $G$ 
arise solely from the measurement Lagrangian~\eqref{eq:L-M-G-PHS}, where 
$\beta = 1$ for class AIII and $\beta = 1/2$ for class BDI. The explicit form 
for class AIII is given in Eq.~\eqref{eq:L-M-G}. Symmetry transformations must 
satisfy $\mathcal{L}_M[\mathcal{G}] = \mathcal{L}_M[\mathcal{R} \mathcal{G} 
\mathcal{R}^{-1}]$.

As in Sec.~\ref{sec:continuous-symmetries-Keldysh}, we check this condition term
by term. Considering efficient counting with $\eta = 1$, the first term of the
measurement Lagrangian~\eqref{eq:L-M-G-PHS} vanishes. The second term is
invariant if $\mathcal{R}$ is block-diagonal in Keldysh space, as given in
Eq.~\eqref{eq:R-AIII}. For class AIII, the blocks are arbitrary unitary
matrices, $\mathcal{V}_{\pm} \in \mathrm{U}(R)$; for class BDI,
Eq.~\eqref{eq:R-PHS} imposes the constraint
$\mathcal{V}_{\pm} = \tau_y \mathcal{V}_{\mp}^{*} \tau_y$, so that $\mathcal{R}$
can be parameterized in terms of a single matrix
$\mathcal{V} \in \mathrm{U}(2R)$ as
\begin{equation}
  \label{eq:R-BDI}
  \mathcal{R} =
  \begin{pmatrix}
    \mathcal{V} & 0 \\ 0 & \tau_y \mathcal{V}^{*} \tau_y
  \end{pmatrix}.
\end{equation}

Next, we consider the third term of the measurement Lagrangian, starting with
class AIII. As discussed in Sec.~\ref{sec:continuous-symmetries-Keldysh}, this
term is invariant if $\detR(\mathcal{V}_+) = \detR(\mathcal{V}_-)$. This yields
$\mathcal{R} \in G \cong \mathrm{U}(1) \rtimes \left[ \mathrm{SU}(R) \times
  \mathrm{SU}(R) \right]$.

Turning to class BDI, we examine the contribution to the third term of the
measurement Lagrangian~\eqref{eq:L-M-G-PHS} with $\alpha = -$. Inserting here
the transformed matrix $\mathcal{G}$, the exponential factor becomes
\begin{equation}
  \e^{\beta \trCR \ln \trK \left( \frac{1}{\beta} \sigma_- \mathcal{R}
      \mathcal{G} \mathcal{R}^{-1} \right)} = \e^{\beta \trCR \ln \left( \tau_y
      \mathcal{V}^{\transpose} \tau_y \mathcal{V} \right)} \e^{\beta \trCR \ln
    \trK \left( \frac{1}{\beta} \sigma_- \mathcal{G} \right)}.
\end{equation}
The final factor on the right-hand side is already present in
Eq.~\eqref{eq:L-M-G-PHS}. Hence, invariance of the measurement Lagrangian
requires the first factor to reduce to unity. To rewrite this factor, we use the
following general relation for skew-symmetric $2D \times 2D$ matrices $A$ and
$B$:
\begin{equation}
  \label{eq:pf-pf-exp-tr-ln}
  \pf(A) \pf(B) = \e^{\beta \tr \ln \left( A^{\transpose} B \right)} = \left( -
    1 \right)^D \e^{\beta \tr(A B)},
\end{equation}
where $\pf$ denotes the Pfaffian. Here, we apply this to the $2R \times 2R$
matrices $A = \tau_y$, with $1_R$ left implicit, and
$B = \mathcal{V}^{\transpose} \tau_y \mathcal{V}$. Invariance of the measurement
Lagrangian thus requires
\begin{equation}
  \e^{\beta \trCR \ln \left( \tau_y \mathcal{V}^{\transpose} \tau_y \mathcal{V}
    \right)} = \left( -1 \right)^R \pfCR(\tau_y) \pfCR \! \left(
    \mathcal{V}^{\transpose} \tau_y \mathcal{V} \right) = 1.
\end{equation}
Using
\begin{equation}
  \begin{gathered}
    \pfCR(\tau_y) = \pfC(\tau_y)^R = \left( - \imag \right)^R, \\ \pfCR \!
    \left( \mathcal{V}^{\transpose} \tau_y \mathcal{V} \right) =
    \detCR(\mathcal{V}) \pfCR(\tau_y),
  \end{gathered}
\end{equation}
this reduces to $\detCR(\mathcal{V}) = 1$. The same condition arises for the
contribution with $\alpha = +$. We therefore conclude that
$\mathcal{V} \in \mathrm{SU}(2 R)$ and thus
$\mathcal{R} \in G \cong \mathrm{SU}(2 R)$.

We now turn to condition~(ii) and determine the group $H$ of transformations
$\mathcal{R}$ that leave the saddle point $\Lambda$ invariant. Consider the
explicit parameterization of the NLSM target manifold, obtained using
Eqs.~\eqref{eq:Lambda} and \eqref{eq:Q-R}, with transformations $\mathcal{R}$
given by Eq.~\eqref{eq:R-AIII} for class AIII and by Eq.~\eqref{eq:R-BDI} for
class BDI, respectively:
\begin{equation}
  \label{eq:Q-U}
  Q =
  \begin{pmatrix}
    1 - 2 n & 2 n U \\ 2 \left( 1 - n \right) U^{-1} & - 1 + 2 n
  \end{pmatrix},
\end{equation}
where, for AIII, $U = \mathcal{V}_+ \mathcal{V}_-^{-1}$, and for BDI,
$U = \mathcal{V} \sigma_y \mathcal{V}^{\transpose} \sigma_y$. Evidently,
$Q = \mathcal{R} \Lambda \mathcal{R}^{-1} = \Lambda$ is invariant for class AIII
if $\mathcal{V}_+ = \mathcal{V}_-$.  Transformations~\eqref{eq:R-AIII}
satisfying this condition form the group
$H \cong \mathrm{U}(R) = \mathrm{U}(1) \rtimes \mathrm{SU}(R)$. The NLSM target
manifold is therefore given by $U \in G/H = \mathrm{SU}(R)$.

For class BDI, the saddle point $\Lambda$ is invariant for matrices
$\mathcal{V}$ satisfying
$\mathcal{V}^{\transpose} \sigma_y \mathcal{V} = \sigma_y$, which form the
compact symplectic group, $H \cong \mathrm{Sp}(R)$. The NLSM target manifold is
thus $U \in G/H = \mathrm{SU}(2 R) / \mathrm{Sp}(R)$.

For both symmetry classes AIII and BDI, the NLSM target manifold becomes trivial
in the replica limit $R \to 1$. In other words, the replica-symmetric manifold
is trivial, and only the replicon manifold is nontrivial. For $R = 1$, the
theory describes the unconditional dynamics of free fermions with incoherent
gain and loss, which in the steady state~\eqref{eq:rho-ss} leads to short-range
correlations in space and time. Such behavior corresponds to the theory being
massive even for $\eta = 1$~\cite{Starchl2025}.

\subsubsection{Trivial target manifold for inefficient detection, $\eta < 1$}
\label{sec:trivial-target-manifold-inefficient}

We now turn to inefficient detection with $\eta < 1$. In this case, the first
term in the measurement Lagrangian~\eqref{eq:L-M-G-PHS}, which contains
$\tr(\Lambda \mathcal{G})$ with coefficient $1 - \eta$, is nonzero. Invariance
of this term requires $\mathcal{R}$ to satisfy condition~\eqref{eq:Lambda-R}.
However, this is precisely the condition that determines the group $H$. Thus, as
discussed below Eq.~\eqref{eq:Q-U}, for class AIII we have
$G = H \cong \mathrm{U}(R)$, while for class BDI we have
$G = H \cong \mathrm{Sp}(R)$. In both cases, the coset $G/H$ is the trivial
group. The fact that inefficient detection breaks the symmetries underlying the
NLSM description was anticipated for balanced fermion counting in
Ref.~\cite{Starchl2025}.

\subsubsection{Nonlinear sigma model Lagrangian}
\label{sec:NLSM-Lagrangian}

The general form of the NLSM action for $\eta = 1$ can be obtained by inserting
the parameterization of the NLSM manifold from Eq.~\eqref{eq:Q-U} into
Eq.~\eqref{eq:S-0-G-Sigma-beta}, taking the spatial continuum limit in which the
lattice-site index $l$ is replaced by the continuous coordinate $x$, and
performing a gradient expansion~\cite{Poboiko2025}. This procedure does not,
however, account for the renormalization of the coefficients in the NLSM
Lagrangian that results from integrating out massive modes. To determine this
renormalization, we instead begin with the most general form of the NLSM
obtained from the gradient expansion of Eq.~\eqref{eq:S-0-G-Sigma-beta} but
leave its coefficients unspecified. Linearizing the NLSM action and matching it
to the Gaussian theory, after all massive fluctuations have been integrated out
there, then fixes these coefficients. Implementing this program yields the NLSM
Lagrangian for the replicon manifold:
\begin{equation}
  \label{eq:L-R}
  \imag \mathcal{L}_R[U] = - \frac{\beta g}{2} \trCR \! \left(
    \frac{1}{v} \partial_t U^{-1}  \partial_t U + v \partial_x U^{-1} \partial_x
    U \right),
\end{equation}
where the bare value $g_0$ of the dimensionless NLSM coupling $g$ is given in
Eq.~\eqref{eq:g-0}, and
\begin{equation}  
  v = 2 J \sqrt{1 - 2 n \left( 1 - n \right)}.
\end{equation}
As noted above:
\begin{align}
  & \text{AIII:} & \beta & = 1, & U & \in \mathrm{SU}(R), \\
  & \text{BDI:} & \beta & = 1/2, & U & \in \mathrm{SU}(2 R)/\mathrm{Sp}(R).
\end{align}
Thus, the same NLSM arises for all values of $\gamma_{\pm}$. The physics
discussed in Ref.~\cite{Starchl2025} is therefore not restricted to balanced
gain and loss with $\gamma_+ = \gamma_-$.

For $\eta < 1$, the first term in Eq.~\eqref{eq:L-M-G-PHS} yields a mass
contribution proportional to the detection inefficiency
$\delta \eta = 1 - \eta > 0$, which takes the form
\begin{equation}
  \label{eq:delta-L-R-U}
  \imag \delta \mathcal{L}_R[U] = 2 n \left( 1 - n \right) \delta \eta \gamma
  \left[ \beta \trCR \! \left( U + U^{-1} \right) - 2 R \right].
\end{equation}
This mass term suppresses long-wavelength fluctuations of $U$. We may thus
parameterize $U = \e^{\imag \Phi}$ with a Hermitian matrix $\Phi$ and expand the
NLSM Lagrangian to second order,
\begin{multline}
  \label{eq:L-R-Phi}
  \imag \left( \mathcal{L}_R[\Phi] + \delta \mathcal{L}_R[\Phi] \right) = -
  \beta \trCR \! \left[ \frac{g}{2} \left( \frac{1}{v} \partial_t
      \Phi^{-1} \partial_t \Phi \right. \right. \\
  \left. \left. \vphantom{\frac{1}{v}} + v \partial_x \Phi^{-1} \partial_x \Phi
    \right) + 2 n \left( 1 - n \right) \delta \eta \gamma \Phi^2 \right].
\end{multline}
This form of the NLSM Lagrangian is consistent with the low-momentum
expansion~\eqref{eq:C-q-asymptotic-Gaussian} of the correlation function.

\subsubsection{Renormalization-group flow of the NLSM}

For $\eta = 1$, the large-scale behavior of our model is governed by the RG flow
of the NLSM coupling constant $g$. The NLSM description and thus the RG flow is
identical to that obtained for balanced counting in Ref.~\cite{Starchl2025}. For
completeness, we briefly recall the relevant results. We then analyze how the RG
flow is modified by inefficient detection.

On scales $l \gtrsim l_0$, the NLSM coupling constant acquires a logarithmic
renormalization~\cite{Poboiko2023, Poboiko2025},
\begin{equation}
  \label{eq:g-RG}
  g = g_0 - \frac{1}{4 \pi \beta} \ln(l/l_0),
\end{equation}
where, as above, $\beta = 1$ for symmetry class AIII and $\beta = 1/2$ for class
BDI. This expression is valid in the weak-coupling regime $g \gg 1$, which,
according to Eq.~\eqref{eq:g-0}, corresponds to a low rate of fermion gain and
loss. The flow reaches the strong-coupling regime, where $g$ becomes of order
unity, at the characteristic scale
\begin{equation}
  \label{eq:l-star}
  l_{*} = l_0 \e^{4 \pi \beta g_0}.
\end{equation}

As discussed in Sec.~\ref{sec:subsystem-entropy-Gaussian}, the renormalization
of $g$ controls the long-distance behavior of the effective central charge.
Thus, the flow of $g$ toward small values for $l \gtrsim l_{*}$ implies that the
logarithmic growth of the entanglement
entropy~\eqref{eq:S-l-asymptotic-Gaussian} eventually saturates, leading to
area-law scaling for subsystem sizes $\ell \gtrsim l_{*}$~\cite{Poboiko2023}.
Accordingly, our model exhibits no measurement-induced transition between
critical and area-law phases; all choices of $\gamma_{\pm}$ yield area-law
entanglement beyond the scale $l_{*}$. Nevertheless, because $l_{*}$ grows
exponentially with $g_0 \sim J/\gamma$, the onset of area-law behavior falls far
outside the window accessible to our numerical simulations when $J/\gamma$ is
large---even though such scales can be reached in principle~\cite{Fan2025}; see
Sec.~\ref{sec:numerical-simulations} for details.

For small but finite measurement inefficiency,
$0 < \delta \eta = 1 - \eta \ll 1$, the mass term in Eq.~\eqref{eq:L-R-Phi}
suppresses fluctuations of $\Phi$ at large scales and thereby cuts off the RG
flow. The resulting infrared cutoff is the correlation length $\xi$ of
Eq.~\eqref{eq:xi}, but with the microscopic scale $l_0 \sim g_0$
renormalized. Even though this renormalization modifies prefactors, the
functional dependence $\xi \sim \delta \eta^{-1/2}$ remains correct, as we
confirm numerically in Sec.~\ref{sec:numerical-simulations}. Beyond the scale
$\xi$, the long-wavelength behavior is governed by the Gaussian theory, with $g$
replaced by its renormalized value. We examine this regime in detail below.

\subsection{RG-corrected Gaussian theory}
\label{sec:rg-corrected-gauss}

To assess how the renormalization of $g$ affects physical observables, we
incorporate the RG correction Eq.~\eqref{eq:g-RG} into the Gaussian correlation
function~\eqref{eq:C-q-asymptotic-Gaussian}~\cite{Poboiko2023, Starchl2025}.
The infrared cutoff of the RG flow is set by the larger of the two scales:
$l = 1/q$, where $q$ is the momentum at which the correlation function is
evaluated, or the correlation length~\eqref{eq:xi} associated with the mass
generated by $\delta \eta > 0$.

For $\eta = 1$, inserting Eq.~\eqref{eq:g-RG} into
Eq.~\eqref{eq:C-q-asymptotic-Gaussian} yields
\begin{equation}
  \label{eq:C-q-eta=1-RG-corrected}
  \frac{C_q}{g_0 q} \sim 1 - 4 \sqrt{\frac{1 - 2 n \left( 1 - n \right)}{2}} q l_0
  + \frac{1}{4 \pi \beta g_0} \ln(q l_0).
\end{equation}
In the limit $q \to 0$, the logarithmic renormalization dominates and suppresses
$C_q/(g_0 q)$, whereas the Gaussian approximation predicts
$C_q/(g_0 q) \to 1$ as $q \to 0$. The momentum scale at which the logarithmic
term becomes comparable to the Gaussian contribution marks the onset of the
crossover from the regime in which the Gaussian theory is valid to the
fluctuation-dominated strong-coupling regime. Following
Ref.~\cite{Starchl2025}, we associate this crossover with the maximum of the
rescaled correlation function, which occurs at $q_c \sim \gamma^2$. The
corresponding length scale is $l_c \sim \gamma^{-2}$. Because $l_c$ grows
only algebraically with $\gamma$, it remains accessible in numerical studies.

As shown in Ref.~\cite{Starchl2025}, the length scale $l_c \sim \gamma^{-2}$
serves as the upper boundary of a critical window within which the model
exhibits emergent conformal invariance, as expected for a critical
phase~\cite{Alberton2021}. Its lower boundary is
$l_0 \sim \gamma^{-1}$~\eqref{eq:l-0}. Within this interval, the system displays
signatures of conformal behavior, including algebraic decay of density
correlations, $C_l \sim l^{-2}$, and logarithmic entanglement scaling,
$S_{\ell} \sim \ln(\ell)$. While these features also arise from the Gaussian
approximations in Eqs.~\eqref{eq:C-l-Gaussian}
and~\eqref{eq:S-l-asymptotic-Gaussian}, the Gaussian central charge
Eq.~\eqref{eq:effective-central-charge-Gaussian} is valid only in the combined
limits $\gamma \to 0$ and $\ell \to \infty$. For any finite $\gamma$,
approximately logarithmic entanglement growth appears instead on intermediate
scales where $c_{\ell}$ varies slowly and reaches a maximum at
$\ell = l_m \sim \gamma^{-3/2}$. The scaling of $l_m$ ensures that this length
lies within the critical regime, $l_0 < l_m < l_c$. Logarithmic entanglement
growth and additional hallmarks of conformal symmetry, including the universal
functional form of the mutual information, are captured by the renormalized
value of the effective central charge at its maximum~\cite{Starchl2025}.

For $\delta \eta > 0$, the RG flow is terminated when $l \sim q^{-1} \sim
\xi$. Using Eq.~\eqref{eq:C-q-asymptotic-Gaussian} at this scale gives
\begin{equation}
  \label{eq:C-q-eta<1-RG-corrected}
  C_q \sim \frac{g_0}{\xi} \left[ 1 - \frac{1}{4 \pi \beta g_0} \ln(\xi/l_0)
  \right] \sim \sqrt{\delta \eta}, \quad \delta \eta \to 0.
\end{equation}
This is consistent with Eq.~\eqref{eq:C-q-eta=1-RG-corrected}: substituting
$q \sim \xi^{-1}$ into that expression and retaining only the leading terms for
$\xi \gg l_0$ reproduces the same scaling.

\subsection{Numerical simulations of quantum-jump trajectories}
\label{sec:numerical-simulations}

We validate our analytical results by comparing them with direct numerical
simulations of the model. Before presenting our findings in detail, we briefly
summarize the numerical methods employed, which differ from those used for
balanced and efficient fermion counting~\cite{Starchl2025}.

\subsubsection{Numerical methods}

For balanced and efficient fermion counting, the rates of gain and loss jumps
are state-independent and constant in time. This enables the use of a simulation
scheme that captures the continuous-time limit $\Delta t \to 0$ numerically
exactly~\cite{Starchl2025, Daley2014}. In contrast, for imbalanced and
inefficient fermion counting, the jump probabilities depend on the instantaneous
state of the system and therefore vary in time. In principle, one can address
this by directly implementing the evolution of individual trajectories as
outlined in Secs.~\ref{sec:quantum-jump-processes} and
\ref{sec:inefficient-detection} for efficient and inefficient detection,
respectively. As explained in Appendix~\ref{sec:minimal-physical-model}, this
amounts to integrating the underlying stochastic Schr\"odinger equation to first
order in $\Delta t$. Within this approach, the time step must be chosen such
that the probability for any jump within a single step remains
small~\cite{Daley2014}. However, for local jump operators in an extended lattice
system, the number of such operators scales extensively with system size,
implying $\Delta t \sim L^{-1}$. This scaling severely restricts simulations at
large $L$.

To circumvent this limitation, we treat jumps at different lattice sites as
approximately independent. Concretely, during each time step we sweep through
the lattice and independently determine whether a jump occurs at each site. The
full numerical procedure is detailed in
Appendix~\ref{sec:numerical-methods}.

All simulations begin from the initial state
\begin{equation}
  \label{eq:ket-psi-CDW}
  \ket{\psi_0} = \prod_{l = 1}^{L/2} \hat{\psi}_{2 l - 1}^{\dagger} \ket{0},
\end{equation}
where $\ket{0}$ denotes the vacuum, yielding an initial density $n = N/L = 1/2$.
Purity of the state is preserved only when $\eta = 1$. Nevertheless, the
Gaussian structure of the initial state is maintained under time evolution
generated by the quadratic Hamiltonian~\eqref{eq:Hamiltonian} and the linear
jump operators~\eqref{eq:gain-loss-jump-operators}. Consequently, at all times
$t$ the state is fully characterized by the $L \times L$ single-particle density
matrix
\begin{equation}
  \label{eq:single-particle-density-matrix}
  D_{l, l'}^{}(t) = \tr \! \left[ \hat{\psi}_{l'}^{\dagger} \hat{\psi}_l^{}
    \hat{\rho}(t) \right].
\end{equation}
This representation allows us to simulate systems of size up to $L = 1000$. In
our simulations, we treat the average jump rate $\gamma$ and the fermion density
$n$ as independent parameters. These uniquely determine the gain and loss rates
$\gamma_{\pm}$ via Eqs.~\eqref{eq:n} and~\eqref{eq:gamma}. For each set of
parameters $\gamma$, $n$, and $\eta$, we sampled $N_{\mathrm{traj}} = 160$
trajectories.

\subsubsection{Connected density correlation function}
\label{sec:conn-dens-corr}

\begin{figure}
  \centering  
  \includegraphics{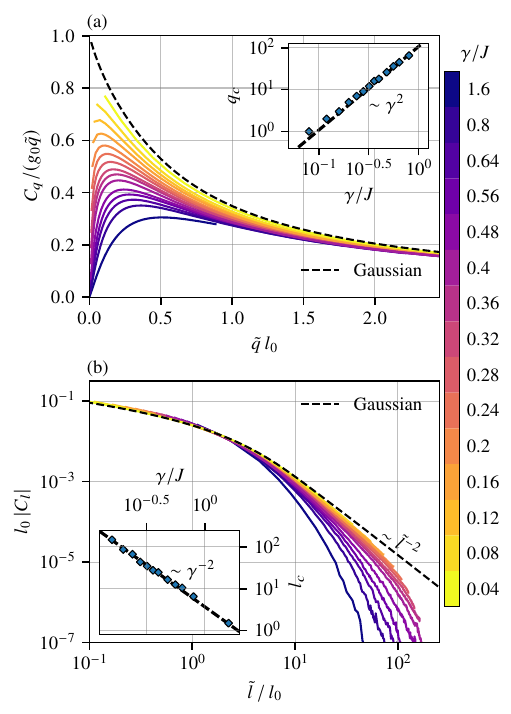}
  \caption{Rescaled density correlation
    function~\eqref{eq:connected-density-correlation-function} in (a)~momentum
    space and (b)~real space for $n = 0.4$ and $\eta = 1$. (a)~The numerical
    data deviate significantly from the Gaussian correlation
    function~\eqref{eq:C-q-Gaussian} (black dashed line) for
    $\tilde{q} l_0 \to 0$: while the Gaussian result approaches a finite value,
    $C_q/(g_0 \tilde{q}) \to 1$, the numerical data exhibit a maximum at $q_c$
    and then decrease toward zero. Inset: The position of the maximum scales as
    $q_c \sim \gamma^2$. (b)~In real space, we observe good agreement with the
    Gaussian prediction on short scales, $\tilde{l} \lesssim l_0$. Significant
    deviations appear for $\tilde{l} \gtrsim l_c$, where the Gaussian result
    decays algebraically, $\abs{C_l} \sim \tilde{l}^{-2}$, while the numerical
    data decay more rapidly. Inset: The crossover scale at which deviations from
    the Gaussian result become pronounced scales as $l_c \sim \gamma^{-2}$.}
  \label{fig:C-l-q-eta=1}
\end{figure}

The first observable we analyze is the connected density correlation
function~\eqref{eq:C-l-l-prime-t}. In the steady state, this quantity becomes
time-independent and translationally invariant, so that
$C_{l, l'}(t) = C_{l - l'}$. Since the system remains in a Gaussian state at all
times, the correlation function can be expressed through the single-particle
density matrix~\eqref{eq:single-particle-density-matrix} using Wick's theorem,
\begin{equation}
  \label{eq:connected-density-correlation-function}
  C_{l - l'} = \overline{\left\langle \hat{n}_l \hat{n}_{l'} \right\rangle} -
  \overline{\left\langle \hat{n}_l \rangle \langle \hat{n}_{l'} \right\rangle}
  = \delta_{l, l'} \overline{D_{l, l}} - \overline{\lvert D_{l, l'} \rvert^2}.
\end{equation}
In our numerical analysis, the overline denotes an average over both quantum
trajectories and over lattice sites $l$ and $l'$ at fixed separation
$l - l'$.

For efficient detection with $\eta = 1$, the results reported in
Ref.~\cite{Starchl2025} for balanced fermion counting remain qualitatively
unchanged for imbalanced counting with $\gamma_+ \neq \gamma_-$ and thus
$n \neq 1/2$.  Figure~\ref{fig:C-l-q-eta=1}(a) displays the rescaled density
correlation function in momentum space, $C_q/(g_0 \tilde{q})$, as a function of
$\tilde{q} l_0$ for fixed density $n = 0.4$ and various values of
$\gamma/J$. Here, $\tilde{q} = 2 \sin(q/2)$ accounts for the finite lattice
spacing~\cite{Poboiko2023}.

At high momenta, the numerical data converge to the Gaussian result, shown as a
black dashed line in the figure. As anticipated in
Sec.~\ref{sec:rg-corrected-gauss}, however, the renormalization of the NLSM
coupling constant $g$ due to strong Goldstone-mode fluctuations leads to
deviations from the Gaussian prediction for $\tilde{q} l_0 \to 0$. The expected
scaling of the crossover momentum, defined as the position of the maximum of the
rescaled correlation function, $q_c \sim \gamma^2$, is well supported by the
numerical results, as illustrated in the inset.

Beyond the corresponding crossover length scale
$l_c \sim q_c^{-1} \sim \gamma^{-2}$, the rescaled correlation function in real
space decays faster than the algebraic form~\eqref{eq:C-l-Gaussian} predicted by
the Gaussian theory. This behavior is evident in Fig.~\ref{fig:C-l-q-eta=1}(b),
which shows the rescaled connected density correlation
function~\eqref{eq:connected-density-correlation-function} as a function of the
chord length
$\tilde{l} = \left( L/\pi \right) \sin(\pi l/L)$~\cite{DiFrancesco1997}.
The corresponding Gaussian prediction, shown again as a black dashed line, is
obtained by transforming Eq.~\eqref{eq:C-q-Gaussian} to real space. Consistent
with Ref.~\cite{Starchl2025}, we identify $l_c$ as the scale beyond which the
deviation between the numerical data and a tangent $\sim \tilde{l}^{-2}$
exceeds $10 \, \%$. As shown in the inset, the numerical results confirm the
expected scaling $l_c \sim q_c^{-1} \sim \gamma^{-2}$.

\begin{figure}
  \centering
  \includegraphics{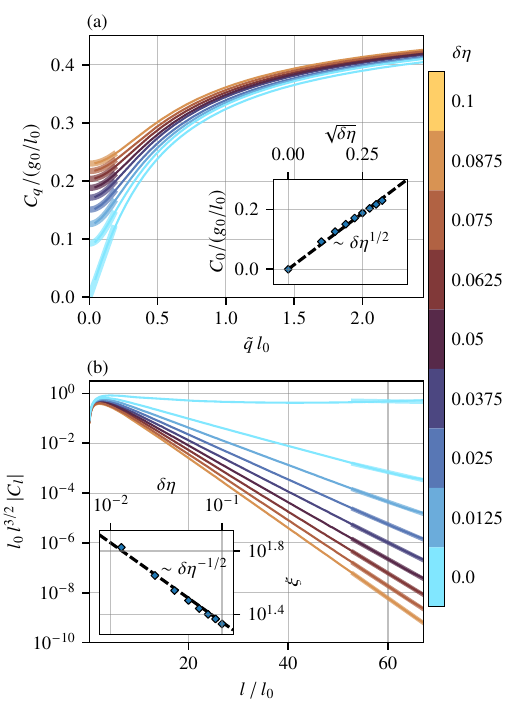}
  \caption{Rescaled density correlation
    function~\eqref{eq:connected-density-correlation-function} in (a)~momentum
    space and (b)~real space for $n = 0.4$ and $\gamma/J = 0.1$. (a)~Finite
    detection inefficiency leads to a nonzero axis intercept for
    $\tilde{q} l_0 \to 0$. Inset: The intercept follows the expected scaling
    $\sim \delta \eta^{1/2}$. (b)~In real space, detection inefficiency induces
    an exponential decay of correlations on large scales. Inset: The fitted
    correlation length $\xi$ agrees well with the predicted behavior
    $\xi \sim \delta \eta^{-1/2}$. Thick lines in (a) and~(b) show fits to
    Eqs.~\eqref{eq:C-q-fit} and~\eqref{eq:C-l-fit}, respectively.}
  \label{fig:C-l-q-eta<1}
\end{figure}

We now discuss how these results are affected by inefficient detection. As
before, we set $n = 0.4$, and we choose $\gamma/J = 0.1$. For these parameters,
when $\eta = 1$, the numerical data shown in Fig.~\ref{fig:C-l-q-eta=1}
qualitatively follow the Gaussian behavior over the entire range of numerically
accessible length scales. Figure~\ref{fig:C-l-q-eta<1}(a) shows the density
correlation function in momentum space for finite $\delta \eta = 1 - \eta$,
without the rescaling by $q^{-1}$ used in Fig.~\ref{fig:C-l-q-eta=1}(a). In this
representation, the RG-corrected Gaussian
theory~\eqref{eq:C-q-eta=1-RG-corrected} predicts that the correlation function
approaches zero for $\tilde{q} l_0 \to 0$ when detection is efficient with
$\delta \eta = 0$. In contrast, for finite detection inefficiency
$\delta \eta > 0$, the axis intercept of the correlation function for
$\tilde{q} l_0 \to 0$ is expected to scale as
$C_0/(g_0/l_0) \sim \delta \eta^{1/2}$, see
Eq.~\eqref{eq:C-q-eta<1-RG-corrected}.

In numerical simulations, the smallest observable momentum is limited by the
inverse system size. To extract the intercept from finite-momentum data, we fit
the functional form given in Eq.~\eqref{eq:C-q-asymptotic-Gaussian} to the data:
\begin{equation}
  \label{eq:C-q-fit}
  C_q/(g_0/l_0) = p_1 \sqrt{\left( \tilde{q} l_0 \right)^2 + p_2},
\end{equation}
with fitting parameters $p_1$ and $p_2$. These fits are shown in
Fig.~\ref{fig:C-l-q-eta<1}(a) as thick lines with reduced opacity. As shown in
the inset, the intercept obtained in this manner is consistent with the expected
scaling $\sim \delta \eta^{1/2}$.

The mass term $\sim \delta \eta$ introduces a finite correlation length
$\xi \sim \delta \eta^{-1/2}$ in real space, as illustrated in
Fig.~\ref{fig:C-l-q-eta<1}(b). For $\delta \eta = 0$, the data exhibit
approximately algebraic decay within the numerically accessible range, with
faster decay expected at even larger scales. However, for $\delta \eta > 0$,
correlations are cut off exponentially. To quantify this behavior, we fit the
functional form of Eq.~\eqref{eq:C-l-Gaussian} to the data:
\begin{equation}
  \label{eq:C-l-fit}
  l_0 l^{3/2} \abs{C_l} = p \e^{- l/\xi},
\end{equation}
with fitting parameters $p$ and $\xi$. The fitted correlation length $\xi$,
shown in the inset, exhibits the expected scaling $\xi \sim \delta \eta^{-1/2}$.

\subsubsection{Subsystem entropy and effective central charge}

\begin{figure}
  \centering
  \includegraphics{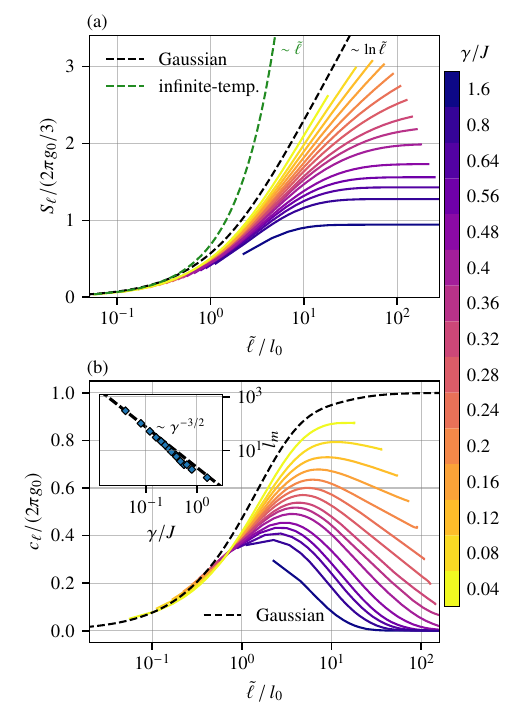}
  \caption{(a)~Rescaled entanglement entropy and (b)~rescaled scale-dependent
    effective central charge for $n = 0.4$ and $\eta = 1$. (a)~For short
    subsystem sizes $\tilde{\ell} \lesssim l_0$, the entanglement entropy
    follows volume-law scaling (green dashed line), consistent with the
    infinite-temperature behavior in Eq.~\eqref{eq:S-l-volume-law}. At larger
    scales, the data show an apparent logarithmic increase for small $\gamma/J$,
    which transitions to area-law behavior as $\gamma/J$ is increased. (b)~The
    effective central charge rises in the volume-law regime, reaches a maximum
    at a characteristic scale $l_m$, and then decreases for
    $\tilde{\ell}/l_0 \gg 1$. This decay demonstrates that the logarithmic
    growth of entanglement observed at small $\gamma/J$ does not persist
    asymptotically. By contrast, the Gaussian approximation (black dashed line)
    for $c_{\ell}$ approaches a constant value, indicating sustained logarithmic
    entanglement growth. Inset: The position of the numerically observed maximum
    of $c_{\ell}$ is consistent with the expected scaling
    $l_m \sim \gamma^{-3/2}$.}
  \label{fig:S-l-c-l-eta=1}
\end{figure}

For the numerical evaluation of the subsystem
entropy~\eqref{eq:subsystem-entropy} in Gaussian states, we employ the formula
\begin{equation}  
  \svn{\ell} = - \sum_{l = 1}^{\ell} \overline{\left[ \lambda_l \ln(\lambda_l) +
      \left( 1 - \lambda_l \right) \ln(1 - \lambda_l) \right]},
\end{equation}
where $\lambda_l$ are the eigenvalues of the reduced single-particle density
matrix $D_{\ell}$ for a subsystem of size $\ell$, obtained by restricting
Eq.~\eqref{eq:single-particle-density-matrix} as
$D_{\ell} = (D_{l, l'})_{l, l' = 1}^{\ell}$~\cite{Peschel2009}. As before, the
overline indicates averaging over both spatial positions and quantum
trajectories.

For efficient detection with $\eta = 1$, each trajectory remains in a pure
state, so the subsystem entropy provides a direct measure of entanglement.
Figure~\ref{fig:S-l-c-l-eta=1}(a) displays the resulting entanglement entropy
for subsystems of size $\ell \in \{1, \dotsc, L/2\}$. For small $\gamma/J$, the
numerical results closely follow the Gaussian prediction obtained by integrating
Eq.~\eqref{eq:subsystem-entropy-correlation-function} with the Gaussian
correlator~\eqref{eq:C-q-Gaussian}. In particular, the data exhibit volume-law
behavior~\eqref{eq:S-l-volume-law} on short scales $\ell \lesssim l_0$ and an
apparent logarithmic increase~\eqref{eq:S-l-asymptotic-Gaussian} on larger
scales. As $\gamma/J$ increases, the entropy crosses over to area-law scaling at
large $\ell$. Although the NLSM analysis indicates that area-law scaling
eventually occurs for any nonzero $\gamma$, this regime sets in only beyond the
exponentially large scale $l_{*}$~\eqref{eq:l-star}. For the system size
$L = 1000$ used in our simulations, this scale is accessible only for
sufficiently large $\gamma/J$. Nonetheless, as discussed below, the onset of the
crossover toward area-law behavior is clearly reflected in the scale-dependent
effective central charge~\eqref{eq:effective-central-charge}.

Figure~\ref{fig:S-l-c-l-eta=1}(b) shows the effective central
charge~\eqref{eq:effective-central-charge} extracted from the numerical
data. For small $\gamma/J$, the numerical results track the Gaussian prediction
on intermediate length scales. However, at large scales the Gaussian result
saturates at the constant value in
Eq.~\eqref{eq:effective-central-charge-Gaussian}, corresponding to persistent
logarithmic growth of the entanglement
entropy~\eqref{eq:S-l-asymptotic-Gaussian}. In contrast, the numerically
obtained central charge remains approximately constant only near its maximum and
decreases for larger $\ell$. This behavior provides clear evidence of the
crossover from logarithmic growth toward area-law scaling. The location $l_m$ of
the maximum of $c_{\ell}$ is shown in the inset of
Fig.~\ref{fig:S-l-c-l-eta=1}(b), and the numerical results are consistent with
the expected scaling $l_m \sim \gamma^{-3/2}$~\cite{Starchl2025}.

\begin{figure}
  \centering
  \includegraphics{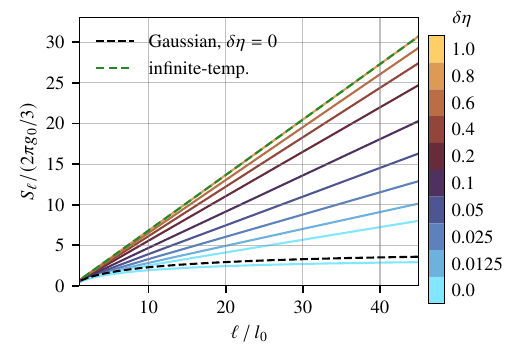}
  \caption{Rescaled subsystem entropy for $\gamma/J = 0.1$ and $n = 0.4$. For
    inefficient detection with $\delta \eta = 1 - \eta > 0$, the subsystem
    entropy exhibits volume-law scaling. In the limit $\delta \eta = 1$, the
    numerical data are consistent with Eq.~\eqref{eq:S-l-volume-law}, as
    expected for a fully separable infinite-temperature state with fixed fermion
    density $n$.}
  \label{fig:S-l-eta<1}
\end{figure}

We now consider the case of inefficient detection, $\eta < 1$, for which
trajectories evolve into mixed states. In this regime, the subsystem
entropy~\eqref{eq:subsystem-entropy} no longer quantifies entanglement; even for
small detection inefficiency $\delta \eta = 1- \eta$, its large-scale behavior
is dominated by the statistical contribution arising from state mixedness. This
trend is evident in Fig.~\ref{fig:S-l-eta<1}. For the comparatively small value
$\gamma/J = 0.1$, where efficient detection yields an apparent logarithmic
entanglement growth in Fig.~\ref{fig:S-l-c-l-eta=1}(a), the subsystem entropy in
Fig.~\ref{fig:S-l-eta<1} instead follows a volume law for $\delta \eta > 0$. The
coefficient of volume-law scaling grows as the detection inefficiency
$\delta \eta$ increases and saturates to the one given in
Eq.~\eqref{eq:S-l-volume-law}, characteristic of the infinite-temperature
state~\eqref{eq:rho-ss} at fixed density $n$, in the limit $\delta \eta \to 1$.

\subsubsection{Fermionic logarithmic negativity}
\label{sec:ferm-logar-negat}

To quantify entanglement in mixed-state dynamics for $\eta < 1$, we employ the
fermionic logarithmic negativity introduced in Refs.~\cite{Shapourian2017,
  Shapourian2019}. For mixed Gaussian states with vanishing anomalous
correlations, $\braket{\hat{\psi}_l \hat{\psi}_{l'}} = 0$, the logarithmic
negativity can be computed as follows~\cite{Alba2023}: We consider a bipartition
into a subsystem $A = \{ 1, \dotsc, \ell \}$ of size $\ell$ and its complement
$B = \{ \ell + 1, \dotsc, L \}$. In terms of the single-particle density
matrix~\eqref{eq:single-particle-density-matrix}, we define the covariance
matrix $G$ as
\begin{equation}
  G = 2 D - 1 =
  \begin{pmatrix}
    G_{AA} & G_{AB} \\
    G_{BA} & G_{BB}
  \end{pmatrix},
\end{equation}
where each block denotes the restriction of the indices of $G$ to the
corresponding subsystems, for example $G_{AA} = ( G_{l,l'} )_{l,l' \in A}$. We
further introduce
\begin{equation}
  G_{\pm} =
  \begin{pmatrix}
    G_{AA} & \pm \imag G_{AB} \\
    \pm \imag G_{BA} & - G_{BB}
  \end{pmatrix},
\end{equation}
and
\begin{equation}
  G^{\mathrm{T}} = \frac{1}{2} \left[ 1 - \left( 1 + G_+ G_- \right)^{-1}
    \left( G_+ + G_- \right) \right].
\end{equation}
Let $\mu_l$ and $\lambda_l$ denote the eigenvalues of $G^{\mathrm{T}}$ and $D$,
respectively, with $l \in \{1, \dotsc, L\}$. The logarithmic negativity,
quantifying entanglement between subsystem $A$ and the remainder of the system,
is then given by
\begin{equation}
  \mathcal{E}_{\ell} = \sum_{l = 1}^L \left\{
    \ln \! \left( \sqrt{\mu_l} + \sqrt{1 - \mu_l} \right)
    + \frac{1}{2} \ln \! \left[ \lambda_l^2 + \left( 1 - \lambda_l \right)^2 \right]
  \right\}.
\end{equation}
Using this expression, the fermionic logarithmic negativity of Gaussian states
can be computed efficiently. An interesting open question is whether this
quantity can also be obtained analytically by introducing appropriate temporal
boundary conditions between replicas in a Keldysh
field-theory~\cite{Shapourian2017, Shapourian2019}.

\begin{figure}
  \centering
  \includegraphics{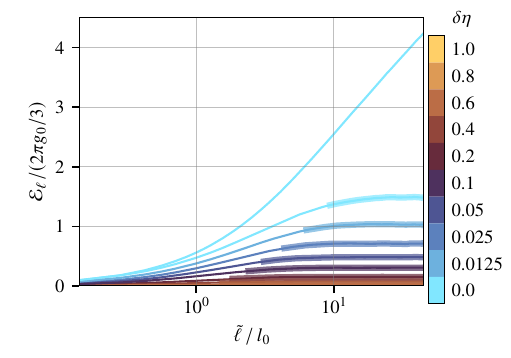}  
  \caption{Rescaled logarithmic negativity for $\gamma/J = 0.1$ and $n = 0.4$ at
    varying measurement inefficiencies $\delta \eta = 1 - \eta$. For
    $\delta \eta = 0$, the data exhibit approximately logarithmic growth with
    subsystem size. Any finite $\delta \eta > 0$ leads to area-law scaling on
    scales $\tilde{\ell} \gtrsim \xi$. This region, with $\xi$ obtained by
    fitting Eq.~\eqref{eq:C-l-fit} to the data in Fig.~\ref{fig:C-l-q-eta<1}(b),
    is highlighted by thicker lines. For $\delta \eta = 1$, the steady
    state~\eqref{eq:rho-ss} is completely disentangled, yielding
    $\mathcal{E}_{\ell} = 0$.}
  \label{fig:E-l-eta<1}
\end{figure}

Figure~\ref{fig:E-l-eta<1} displays the logarithmic negativity for the same
parameters as in Fig.~\ref{fig:S-l-eta<1}. For the pure states obtained for
$\delta \eta = 0$, the logarithmic negativity reduces to the R\'enyi entropy of
order $1/2$ and shows an apparent logarithmic increase with subsystem size
$\ell$, consistent with the von Neumann entanglement entropy in
Fig.~\ref{fig:S-l-c-l-eta=1}(a). In contrast, even a small but finite
measurement inefficiency $\delta \eta$ induces area-law scaling of the
logarithmic negativity once $\ell \gtrsim \xi$. This confirms that the
volume-law behavior of the subsystem entropy in Fig.~\ref{fig:S-l-eta<1} is
entirely of statistical origin and does not indicate genuine quantum
entanglement. In the limit $\eta \to 0$, the steady state~\eqref{eq:rho-ss} of
the deterministic open-system dynamics is fully separable, and the logarithmic
negativity vanishes identically, $\mathcal{E}_{\ell} = 0$.

\section{Conclusions}
\label{sec:conclusions}

In this work, we have developed a comprehensive replica Keldysh field-theory
framework for general quantum-jump processes in bosonic and fermionic many-body
systems. This framework encompasses, in particular, processes involving
non-Hermitian jump operators with state-dependent and thus time-varying
rates---going beyond the paradigmatic scenario of projective measurements
performed at a fixed, externally imposed rate. Moreover, our formalism unifies
the description of pure-state trajectories under efficient detection with the
mixed-state dynamics arising from inefficient monitoring. It thus provides a
versatile framework for exploring a broad range of measurement-induced
phenomena.

Applying this framework to imbalanced and inefficient fermion counting in a 1D
lattice system, we have established several key results. For imbalanced but
efficient detection, we confirmed that the qualitative picture observed
previously in the balanced case persists~\cite{Starchl2025}: there is no
measurement-induced phase transition, and entanglement obeys an area law on
sufficiently large scales, while an extended quantum-critical regime emerges
between parametrically separated scales $l_0 \sim \gamma^{-1}$ and
$l_c \sim \gamma^{-2}$. This regime exhibits signatures of emergent conformal
invariance, including algebraic correlations and approximately logarithmic
entanglement growth.

We have further shown that a finite detection inefficiency $\delta \eta$,
corresponding to a nonzero fraction of undetected fermion gain and loss jumps,
introduces a correlation length $\xi \sim \delta\eta^{-1/2}$. On length scales
larger than $\xi$, quantum entanglement is strongly suppressed, as evidenced by
the fermionic logarithmic negativity, which obeys an area law. In contrast, the
subsystem entropy exhibits volume-law scaling due to the mixedness of individual
quantum trajectories.

More broadly, our results establish a direct conceptual and technical connection
between measurement-induced dynamics and the steady-state physics of driven open
quantum systems~\cite{Sieberer2016a, Sieberer2025, Thompson2023,
  Kamenev2023}. For example, imbalanced and inefficient fermion counting in the
limit of inefficient detection constitutes a representative instance of driven
open quadratic fermionic systems with gain-loss dissipation~\cite{Prosen2008,
  *Prosen2010, Barthel2022}. This class of systems has been studied extensively:
such models are known to host topologically nontrivial phases in their
nonequilibrium steady states~\cite{Bardyn2013} and to exhibit signatures of
non-Hermitian topology~\cite{Bergholtz2021} in their dynamics~\cite{Sayyad2021},
including the non-Hermitian skin effect~\cite{Song2019}. Further paradigmatic
settings include systems with impurities, both in the bulk~\cite{Froml2019} and
at the boundaries~\cite{Turkeshi2024}. The latter give rise to boundary-driven
models that provide a natural platform for studying quantum
transport~\cite{Znidaric2010}. For all of these scenarios, our formalism
furnishes an analytical framework for exploring their unraveling into quantum
trajectories. 

Beyond the class of quadratic fermionic systems, it will be interesting to
study, for example, driven open quantum systems that exhibit a symmetry-breaking
phase transition in the steady state. One might envision a scenario in which the
symmetry is broken in individual pure-state trajectories, while it is restored
in the mixed steady state of the master equation. Using our formalism, it will
be possible to investigate how spontaneous symmetry breaking in trajectories
emerges when the detection efficiency is tuned from $\eta = 0$ to $\eta = 1$.

Through such examples, we hope that our work will shed further light on the
question of whether universal properties exist that transcend the apparent
conceptual divide between measurement-induced pure-state dynamics and the
nonequilibrium steady states of driven open quantum systems.

\begin{acknowledgments}
  L.S.\ acknowledges support from the Austrian Science Fund (FWF) through the
  project 10.55776/COE1, and from the European Union - NextGenerationEU.
\end{acknowledgments}

\subsection*{Note added}

While we were completing this manuscript, a preprint appeared on the arXiv that
derives a replica Keldysh field theory for quantum-jump processes using a
different approach~\cite{Gal2025}.

\subsection*{Data availability}

The data that support the findings of this article are openly
available~\cite{Kloiber-Tollinger2025}.

\appendix

\section{Minimal physical model for quantum-jump trajectories}
\label{sec:minimal-physical-model}

Here we outline a minimal physical model that gives rise to the discrete-time
dynamics governed by the Kraus operators in Eqs.~\eqref{eq:K-0}
and~\eqref{eq:K-a}. The construction follows the standard quantum-optical
treatment of continuous monitoring~\cite{Gardiner2014, Gardiner2015,
  Wiseman2010, Jacobs2014}.

We consider a generic system with Hamiltonian $\hat{H}$ coupled to bosonic or
fermionic baths. The bath annihilation and creation operators are
$\hat{b}_{a,k}$ and $\hat{b}_{a,k}^{\dagger}$, where
$a \in \mathsf{A} = \{ 1,\dotsc, A \}$ indexes the baths and $k$ labels discrete
bath modes of energies $\omega_{a,k}$. The operators satisfy
\begin{equation}
  \left[ \hat{b}_{a,k}, \hat{b}_{a',k'}^{\dagger} \right]_{\zeta}
  = \hat{b}_{a,k} \hat{b}_{a',k'}^{\dagger}
    - \zeta \hat{b}_{a',k'}^{\dagger} \hat{b}_{a,k}
  = \delta_{a,a'} \delta_{k,k'},
\end{equation}
with $\zeta = +1$ for bosons and $\zeta = -1$ for fermions. The Hamiltonian of
bath $a$ is
\begin{equation}
  \label{eq:H-bath}
  \hat{H}_a = \sum_k \omega_{a,k} \hat{b}_{a,k}^{\dagger} \hat{b}_{a,k},
\end{equation}
and the system-bath coupling is described by
\begin{equation}
  \label{eq:H-int}
  \hat{H}_{\mathrm{int}}
  = \imag \sum_{a=1}^A \sum_k 
    \left( g_{a,k}^{*} \hat{c}_a \hat{b}_{a,k}^{\dagger}
         - g_{a,k} \hat{b}_{a,k} \hat{c}_a^{\dagger} \right).
\end{equation}
Bath $a$ inherits its bosonic or fermionic statistics from the system operator
$\hat{c}_a$. For example, in fermionic gain and loss processes governed by jump
operators~\eqref{eq:gain-loss-jump-operators}, Eq.~\eqref{eq:H-int} represents
particle tunneling into and out of fermionic reservoirs~\cite{Starchl2025}.

All baths are assumed to start in the vacuum, $\hat{b}_{a,k} \ket{0_a} = 0$, and
the joint initial state is
$\ket{\Psi(t_0)} = \ket{\psi(t_0), 0_1, \dotsc, 0_A}$.

Following standard derivations of stochastic Schr\"odinger
dynamics~\cite{Gardiner2014, Gardiner2015, Wiseman2010, Jacobs2014}, we move to
an interaction picture in which operators evolve under the bath
Hamiltonians~\eqref{eq:H-bath}, while the state evolves under
\begin{equation}
  \hat{H}_I(t) = \hat{H} + \hat{H}_{\mathrm{int}}(t),
\end{equation}
with
\begin{equation}
  \hat{H}_{\mathrm{int}}(t)
  = \imag \sum_{a=1}^A \left[
    \hat{c}_a \hat{b}_a^{\dagger}(t)
    - \hat{b}_a(t) \hat{c}_a^{\dagger}
  \right],
\end{equation}
and
\begin{equation}
  \hat{b}_a(t)
  = \sum_k g_{a,k} \hat{b}_{a,k} \e^{- \imag \omega_{a,k} t}.
\end{equation}
The interaction-picture bath operators satisfy
\begin{equation}
  \left[ \hat{b}_a(t), \hat{b}_{a'}^{\dagger}(t') \right]_{\zeta}
  = \delta_{a,a'} \Gamma_a(t - t'),
\end{equation}
where
\begin{equation}
  \Gamma_a(t) = \sum_k \abs{g_{a,k}}^2 \e^{- \imag \omega_{a,k} t}.
\end{equation}
We now invoke the Markov approximation, assuming the bath correlation times are
short compared with characteristic system timescales, so that
\begin{equation}
  \Gamma_a(t) = \gamma_a \delta(t).
\end{equation}
In general, $\gamma_a$ is complex. Its imaginary part, the Lamb shift,
contributes to the unitary evolution of the system and is typically absorbed
into $\hat{H}$. In the following, we neglect this Lamb-shift contribution.

The interaction-picture time evolution operator satisfies the integral equation
\begin{equation}
  \label{eq:U-t-t0-integral-equation}
  \hat{U}_I(t, t_0)
  = \id - \imag \int_{t_0}^t \diff t' \, \hat{H}_I(t') \hat{U}_I(t', t_0).
\end{equation}
For simplicity, we set $t_0 = 0$. We integrate the integral equation in small
discrete time steps, such that $t = \Delta t$.

In the weak-coupling or Born approximation,
Eq.~\eqref{eq:U-t-t0-integral-equation} can be solved perturbatively. Iterating
the integral equation to second order yields
\begin{equation}
  \label{eq:U-delta-t-born}
  \ket{\Psi_I(\Delta t)}
  = \hat{U}_I(\Delta t,0) \ket{\Psi_I(0)}
  = \ket{\Psi_I^{(0)}} + \ket{\Psi_I^{(1)}} + \ket{\Psi_I^{(2)}},
\end{equation}
where $\ket{\Psi_I^{(0)}} = \ket{\Psi_I(0)}$. By using
$\hat{b}_a(t) \ket{0_a} = 0$, the first-order contribution can be written as
\begin{equation}
  \begin{split}
    \ket{\Psi_I^{(1)}}
    &= - \imag \int_0^{\Delta t} \diff t \, \hat{H}_I(t) \ket{\Psi_I(0)} \\
    &= \left[ - \imag \hat{H} \Delta t
       + \sum_{a=1}^A \sqrt{\gamma_a} \hat{c}_a \Delta \hat{B}_a^{\dagger}(0)
      \right] \ket{\Psi_I(0)},
  \end{split}
\end{equation}
where the quantum It\^o increments are
\begin{equation}
  \label{eq:quantum-Ito-increment}
  \Delta \hat{B}_a(t)
  = \frac{1}{\sqrt{\gamma_a}}
    \int_t^{t+\Delta t} \diff t' \, \hat{b}_a(t').
\end{equation}
They satisfy the Markov-approximation commutation relation
\begin{equation}
  \label{eq:quantum-Ito-commutation-relation}
  \left[ \Delta \hat{B}_a(t), \Delta \hat{B}_{a'}^{\dagger}(t') \right]_{\zeta}
  = \Delta t \, \delta_{a,a'} \delta_{t,t'}.
\end{equation}
The second-order contribution is
\begin{equation}
  \begin{split}
    \ket{\Psi_I^{(2)}}
    &= - \int_0^{\Delta t} \diff t' \int_0^{t'} \diff t'' \,
       \hat{H}_I(t') \hat{H}_I(t'') \ket{\Psi_I(0)} \\
    &= - \frac{1}{2}
       \sum_{a=1}^A \gamma_a \hat{c}_a^{\dagger} \hat{c}_a \Delta t
       \ket{\Psi_I(0)},
  \end{split}
\end{equation}
where terms generating two reservoir excitations, which occurs with probability
$O(\Delta t^2)$, are omitted~\cite{Gardiner2015}.

So far, we have considered evolution from the initial time $t = 0$ to
$\Delta t$. However, the evolution equation holds for arbitrary time steps from
$t$ to $t + \Delta t$.  This is because the quantum It\^o
increments~\eqref{eq:quantum-Ito-increment} commute for different time steps
according to Eq.~\eqref{eq:quantum-Ito-commutation-relation}. The description in
terms of the commuting operators $\Delta \hat{B}(t)$ in each time interval
suggests a decomposition of each bath Hilbert space into a product of Hilbert
spaces for individual time intervals~~\cite{Gardiner2015}. This leads to an
important simplification of the description of the evolution process: evolution
up to time $t$ does not affect the Hilbert spaces at later times; these remain
in the vacuum state. Thus, from Eq.~\eqref{eq:U-delta-t-born}, we obtain
\begin{equation}
  \label{eq:Psi-I-update}
  \begin{split}
    \ket{\Psi_I(t + \Delta t)}
    &= \left[
        1 - \imag \hat{H}_{\mathrm{eff}} \Delta t
        + \sum_{a=1}^A \sqrt{\gamma_a}
          \hat{c}_a \Delta \hat{B}_a^{\dagger}(t)
      \right] \ket{\Psi_I(t)} \\
    &= \left(
         \hat{K}_0
         + \sum_{a=1}^A
           \hat{K}_a \frac{\Delta \hat{B}_a^{\dagger}(t)}{\sqrt{\Delta t}}
       \right) \ket{\Psi_I(t)},
  \end{split}
\end{equation}
with $\hat{H}_{\mathrm{eff}}$ defined in Eq.~\eqref{eq:H-eff}. The final form
makes explicit the Kraus operators in Eqs.~\eqref{eq:K-0} and~\eqref{eq:K-a}.

We now incorporate continuous monitoring of the bath populations by performing,
in each time step, a projective measurement of
\begin{equation}
  \label{eq:N-a}
  \hat{N}_a(t)
  = \frac{\Delta \hat{B}_a^{\dagger}(t)}{\sqrt{\Delta t}}
    \frac{\Delta \hat{B}_a(t)}{\sqrt{\Delta t}}.
\end{equation}
Since all baths begin in vacuum, the allowed outcomes are either no excitation
or a single excitation in one bath. Multiple excitations occur with probability
$O(\Delta t^2)$ and are neglected.

The probability of detecting no excitation is
\begin{equation}
  P_0 = \norm{\hat{K}_0 \ket{\Psi_I(t)}}^2 = \norm{\hat{K}_0 \ket{\psi(t)}}^2,
\end{equation}
and the system state updates as
$\ket{\psi(t)} \mapsto \hat{K}_0 \ket{\psi(t)}/\sqrt{P_0}$. This corresponds to
continuous, no-jump evolution. The probability of detecting an excitation in
bath $a$ is
\begin{equation}
  P_a = \frac{1}{\Delta t}
        \norm{
          \hat{K}_a \Delta \hat{B}_a^{\dagger}(t)
          \ket{\Psi_I(t)}
        }^2
      = \norm{\hat{K}_a \ket{\psi(t)}}^2,
\end{equation}
and the corresponding update is
$\ket{\psi(t)} \mapsto \hat{K}_a \ket{\psi(t)}/\sqrt{P_a}$, representing a
quantum jump. These probabilities and update rules reproduce precisely the
quantum-trajectory description of Eqs.~\eqref{eq:P-alpha},
\eqref{eq:no-jump-update}, and~\eqref{eq:jump-update}.

For completeness, we note that Eq.~\eqref{eq:Psi-I-update} may also be recast as
a quantum stochastic Schr\"odinger equation for the system alone by introducing
stochastic increments $\Delta N_a(t)$ that count bath excitations in each time
step~\cite{Gardiner2014, Gardiner2015, Wiseman2010, Jacobs2014}.

\section{Generalized Hubbard-Stratonovich transformation with PHS}
\label{sec:HS-transformation-PHS}

In this appendix, we present a detailed and self-contained derivation of the
generalized Hubbard-Stratonovich transformation for systems with PHS,
culminating in Eq.~\eqref{eq:L-M-G-PHS}, which expresses the measurement
Lagrangian in terms of the matrix field $\mathcal{G}$. In
Sec.~\ref{sec:gauss-integ-PHS}, we begin with an elementary evaluation of
Gaussian Grassmann integrals in the presence of PHS.  This serves as the
foundation for the derivation of Wick's theorem in Sec.~\ref{sec:wicks-theorem},
which is in turn required for establishing a useful identity that enables the
simultaneous decoupling of measurement vertices in all slow channels, as
discussed in Sec.~\ref{sec:deco-meas-vert}. The decoupled vertex is obtained by
performing all possible Wick contractions, which can be viewed equivalently as
evaluating a suitable Gaussian Grassmann integral. Our analysis also resolves a
missing prefactor in the corresponding formula reported in
Ref.~\cite{Poboiko2025}. Finally, exploiting the Gaussian integral
representation of the decoupled measurement Lagrangian, we derive
Eq.~\eqref{eq:L-M-G-PHS} in Sec.~\ref{sec:gener-hubb-strat}.

\subsection{Gaussian integrals with PHS}
\label{sec:gauss-integ-PHS}

We aim to show that
\begin{equation}
  \label{eq:Z-J}
  Z = \int \Diff[\psi^{*}, \psi] \, 
  \e^{\imag \left( \bar{\Psi} \mathcal{G}^{-1} \Psi + J^{\transpose} \Psi \right)}
  = \Pf \!\left( - \imag \mathcal{C} \mathcal{G}^{-1} \right)
  \e^{\frac{\imag}{4} J^{\transpose} \mathcal{G} \mathcal{C} J},
\end{equation}
where $\mathcal{G}$ is a matrix in Keldysh, charge-conjugation, replica, and,
when applicable, position and discrete-time spaces. Thus, $\mathcal{G}$ is a
$4 D \times 4 D$ matrix, where $4$ is the dimension of the combined Keldysh and
charge-conjugation space, and $D = R$ is the dimension of replica space or
$D = R L N$ when $N$ discrete time steps are included. In the continuous-time
limit, matrix multiplication in time space becomes integration, and $\Pf$
becomes a functional Pfaffian. When $D = R$, $\Pf$ reduces to the ordinary
Pfaffian $\pf$. The spinors $\Psi$ and $\bar{\Psi}$ are defined in
Eq.~\eqref{eq:doubled-spinors} and satisfy the constraint
$\bar{\Psi} = \Psi^{\transpose} \mathcal{C}$, with $\mathcal{C}$ given in
Eq.~\eqref{eq:C}. The source vector $J$ has $4 D$ Grassmann components. Finally,
the integration measure is
\begin{equation}
  \label{eq:D-psi-star-psi}
  \begin{split}
    \Diff[\psi^{*}, \psi]
    & = \prod_{d = 1}^D
      \diff \psi_{+, d}^{*} \diff \psi_{+, d}
      \diff \psi_{-, d}^{*} \diff \psi_{-, d} \\
    & = \frac{1}{4^D}
      \prod_{d = 1}^D \prod_{i = 1}^4 \diff \Psi_{i, d}
      = \frac{1}{4^D} \Diff[\Psi].
  \end{split}
\end{equation}
This ordering of fields is the natural one arising in the construction of the
Keldysh functional integral~\cite{Altland2010a, Sieberer2016a, Sieberer2025,
  Thompson2023, Kamenev2023}.

To begin, we rewrite the quadratic form in the exponent as
\begin{equation}
  \bar{\Psi} \mathcal{G}^{-1} \Psi = \Psi^{\transpose} \mathcal{C}
  \mathcal{G}^{-1} \Psi = \Psi^{\transpose} \mathcal{P} \Psi, \qquad \mathcal{P}
  = \mathcal{C} \mathcal{G}^{-1}.
\end{equation}
Without loss of generality, $\mathcal{P} = - \mathcal{P}^{\transpose}$ is
skew-symmetric: any symmetric component of $\mathcal{P}$ would cancel in the
quadratic form $\Psi^{\transpose} \mathcal{P} \Psi$ due to the Grassmann nature
of $\Psi$. Note that skew-symmetry of $\mathcal{P}$ is equivalent to the
condition~\eqref{eq:G-PHS} for $\mathcal{G}$. Being skew-symmetric,
even-dimensional, and generally complex, $\mathcal{P}$ can be expressed in the
form
\begin{equation}
  \label{eq:P-U-Sigma}
  \mathcal{P} = U^{\transpose} \Sigma U, \quad U^{-1} = U^{\dagger}, \quad
  \Sigma = \bigoplus_{d = 1}^{2 D} \lambda_d \imag \sigma_y, \quad \lambda_d
  \geq 0.
\end{equation}

We define
\begin{equation}
  \label{eq:Upsilon-Phi}
  \Upsilon = U \Psi = V \Phi, \qquad V = \bigoplus_{d = 1}^{2 D} v, \qquad v =
  \frac{1}{\sqrt{2}}
  \begin{pmatrix}
    1 & 1 \\ \imag & - \imag
  \end{pmatrix}.
\end{equation}
Writing
$\Phi = \left( \phi_1, \phi_1^{*}, \dotsc, \phi_{2 D}, \phi_{2 D}^{*}
\right)^{\transpose}$,
where $\phi_d$ and $\phi_d^{*}$ are independent Grassmann variables, and using
$v^{\transpose} \imag \sigma_y v = \sigma_y$, we may further rewrite the
quadratic form as
\begin{multline}
  \label{eq:Gauss-PHS-qudratic}
  \Psi^{\transpose} \mathcal{P} \Psi = \Upsilon^{\transpose} \Sigma \Upsilon =
  \sum_{d = 1}^{2 D} \lambda_d \left( \Upsilon_{2 d - 1}, \Upsilon_{2 d} \right)
  \imag \sigma_y
  \begin{pmatrix}
    \Upsilon_{2 d - 1} \\ \Upsilon_{2 d}
  \end{pmatrix} \\
  = \sum_{d = 1}^{2 D} \lambda_d \left( \phi_d, \phi_d^{*} \right) \sigma_y
  \begin{pmatrix}
    \phi_d \\ \phi_d^{*}
  \end{pmatrix}
  = 2 \imag \sum_{d = 1}^{2 D} \lambda_d \phi_d^{*} \phi_d = \imag
  \phi^{\dagger} A \phi, 
\end{multline}
where
\begin{equation}
  \label{eq:phi-A}
  \phi = \left( \phi_1, \dotsc, \phi_{2 D} \right)^{\transpose}, \quad A = 2
  \diag(\lambda_1, \dotsc, \lambda_{2 D}).
\end{equation}

Next, we express the source term $J^{\transpose} \Psi$ using $\phi$ and
$\phi^{*}$. Define the $4 D \times 2 D$ matrices $P_{\pm}$ as
\begin{equation}
  \label{eq:P-+-}
  P_+ = \bigoplus_{d = 1}^{2 D}
  \begin{pmatrix}
    1 \\ 0
  \end{pmatrix}, \qquad
  P_- = \bigoplus_{d = 1}^{2 D}
  \begin{pmatrix}
    0 \\ 1
  \end{pmatrix},
\end{equation}
so that $\Phi = P_+ \phi + P_- \phi^{*}$. Then
\begin{equation}
  \label{eq:Gauss-PHS-source}
  J^{\transpose} \Psi = J^{\transpose} U^{\dagger} V \Phi
  = - \imag \left( \mu^{\dagger} \phi - \phi^{\dagger} \nu \right),
\end{equation}
where
\begin{equation}
  \label{eq:mu-nu}
  \mu^{\dagger} = \imag J^{\transpose} U^{\dagger} V P_+, \qquad \nu = - \imag
  P_-^{\transpose} V^{\transpose} U^{*} J.
\end{equation}

We now transform the integration measure~\eqref{eq:D-psi-star-psi} to the
variables $\phi$ and $\phi^{*}$. From Eq.~\eqref{eq:Upsilon-Phi},
\begin{equation}
  \Diff[\psi^{*}, \psi] = \frac{\Det(U)}{4^N \Det(V)} \Diff[\Phi].
\end{equation}
Here $\Det$ denotes a determinant over Keldysh, charge-conjugation, replica, and,
when applicable, position and discrete-time spaces. In the continuous-time
limit, it becomes a functional determinant; otherwise it reduces to an ordinary
determinant $\det$. Using
\begin{equation}
  \Det(V) = \prod_{d = 1}^{2 D} \det(v) = \left( - \imag \right)^{2 D} = \left(
    - 1 \right)^D,
\end{equation}
and
\begin{equation}
  \Diff[\Phi] = \prod_{d = 1}^{2 D} \diff \phi_d \diff \phi_d^{*} = \prod_{d =
    1}^{2 D} \diff \phi_d^{*} \diff \phi_d = \Diff[\phi^{*}, \phi],
\end{equation}
we obtain
\begin{equation}
  \label{eq:Gauss-PHS-measure}
  \Diff[\psi^{*}, \psi] = \left( - \frac{1}{4} \right)^D \Det(U) \Diff[\phi^{*},
  \phi].
\end{equation}

Substituting Eqs.~\eqref{eq:Gauss-PHS-qudratic}, \eqref{eq:Gauss-PHS-source},
and~\eqref{eq:Gauss-PHS-measure} into Eq.~\eqref{eq:Z-J} yields a standard
Gaussian Grassmann integral~\cite{Altland2010a}:
\begin{equation}
  \label{eq:Z-J-Det-U-Det-A}
  \begin{split}
    Z & = \left( - \frac{1}{4} \right)^D \Det(U) \int \Diff[\phi^{*}, \phi] \,
    \e^{- \phi^{\dagger} A \phi + \mu^{\dagger} \phi + \phi^{\dagger} \nu} \\ &
    = \left( - \frac{1}{4} \right)^D \Det(U) \Det(A) \e^{\mu^{\dagger} A^{-1}
      \nu}.
  \end{split}
\end{equation}

We now rewrite the prefactor in terms of the Pfaffian of $\mathcal{P}$. From
Eqs.~\eqref{eq:phi-A} and~\eqref{eq:P-U-Sigma},
\begin{equation}
  \Det(A) = \prod_{d = 1}^{2 D} 2 \lambda_d = 4^D \Pf(\Sigma).
\end{equation}
Thus,
\begin{multline}
  \label{eq:Det-U-Det-A}
  \left( - \frac{1}{4} \right)^D \Det(U) \Det(A) = \left( - 1 \right)^D
  \Det(U) \Pf(\Sigma) \\ = \left( - 1 \right)^D \Pf \! \left( U^{\transpose}
    \Sigma U \right) = \left( - 1 \right)^D \Pf(\mathcal{P}) = \Pf(- \imag
  \mathcal{P}).
\end{multline}

Next, we express the exponent in Eq.~\eqref{eq:Z-J-Det-U-Det-A} in terms of the
source vector $J$. Using Eq.~\eqref{eq:mu-nu},
\begin{equation}
  \mu^{\dagger} A^{-1} \nu = J^{\transpose} U^{\dagger} V P_+ A^{-1} P_-^{\transpose}
  V^{\transpose} U^{*} J.
\end{equation}
With the explicit forms of $A$ and $P_{\pm}$ in Eqs.~\eqref{eq:phi-A}
and~\eqref{eq:P-+-}, respectively,
\begin{equation}
  P_+ A^{-1} P_- = \bigoplus_{d = 1}^{2 D} \frac{1}{2 \lambda_d} \sigma_+.
\end{equation}
Using $\sigma_+ = (1 + \imag \sigma_y)/2$ together with
$v v^{\transpose} = \sigma_z$ and $v \imag \sigma_y v^{\transpose} = \sigma_y$,
we obtain
\begin{equation}
  V P_+ A^{-1} P_- V^{\transpose} = \bigoplus_{d = 1}^{2 D} \frac{1}{4
    \lambda_d} \left( \sigma_z + \sigma_y \right) = \frac{\imag}{4} \left(
    E^{-1} + \Sigma^{-1} \right),
\end{equation}
where $\Sigma$ is defined in Eq.~\eqref{eq:P-U-Sigma}, and
\begin{equation}
  E = \bigoplus_{d = 1}^{2 D} \lambda_d \imag \sigma_z = E^{\transpose}.
\end{equation}
Thus, using $U^{*} = \left( U^{\transpose} \right)^{-1}$, the exponent in
Eq.~\eqref{eq:Z-J-Det-U-Det-A} becomes
\begin{equation}
  \begin{split}
    \mu^{\dagger} A^{-1} \nu & = \frac{\imag}{4} J^{\transpose} U^{-1} \left(
      E^{-1} + \Sigma^{-1} \right) U^{*} J \\ & = \frac{\imag}{4} J^{\transpose}
    \left[ \left( U^{\transpose} E U \right)^{-1} + \mathcal{P}^{-1} \right] J.
  \end{split}
\end{equation}
Since $J$ is a Grassmann vector, the term involving the symmetric matrix
$\left( U^{\transpose} E U \right)^{-1}$ vanishes. Finally, using
$\mathcal{P}^{-1} = \mathcal{G} \mathcal{C}^{-1} = \mathcal{G} \mathcal{C}$ and
inserting the above into Eq.~\eqref{eq:Z-J-Det-U-Det-A}, together with
Eq.~\eqref{eq:Det-U-Det-A}, we recover Eq.~\eqref{eq:Z-J}.

\subsection{Wick's theorem}
\label{sec:wicks-theorem}

We now use Eq.~\eqref{eq:Z-J} to derive Wick's theorem for systems with PHS.  
To begin, we rewrite Eq.~\eqref{eq:Z-J} as
\begin{equation}
  \label{eq:Wick-exponential}
  \langle \e^{\imag J^{\transpose} \Psi} \rangle = \int \frac{\Diff[\psi^{*},
    \psi]}{\Pf \! \left( - \imag \mathcal{C} \mathcal{G}^{-1} \right)} \,
  \e^{\imag \left( \bar{\Psi} \mathcal{G}^{-1} \Psi + J^{\transpose} \Psi
    \right)} = \e^{\frac{\imag}{4}
    J^{\transpose} \mathcal{G} \mathcal{C} J}.
\end{equation}
Expanding both sides in power series and matching terms involving the same total
power of components of $J$, here equal to $2 F$, yields
\begin{equation}
  \label{eq:Wick-J-implicit}
  \frac{1}{(2 F)!} \langle \left( \imag J^{\transpose} \Psi \right)^{2 F} \rangle =
  \frac{1}{F!} \left( \frac{\imag}{4} J^{\transpose} \mathcal{G} \mathcal{C} J
  \right)^F.
\end{equation}

In the following we require Wick's theorem in the setting where all fields share
the same position and time while differing only in their Keldysh,
charge-conjugation, and replica indices. We combine these indices into a single
multiindex $\alpha$. Thus, we can rewrite Eq.~\eqref{eq:Wick-J-implicit} as
\begin{multline}
  \label{eq:Wick-J-explicit}
  \frac{\left( - 1 \right)^F}{(2 F)!} \sum_{\alpha_1, \dotsc, \alpha_{2 F}} \langle \prod_{f
    = 1}^{2 F} J_{\alpha_f} \Psi_{\alpha_f} \rangle \\ = \frac{1}{4^F F!}
  \sum_{\alpha_1, \dotsc, \alpha_{2 f}} \prod_{f = 1}^F J_{\alpha_{2 f - 1}}
  \left( \imag \mathcal{G} C \right)_{\alpha_{2 f - 1}, \alpha_{2 f}}
  J_{\alpha_{2 f}}.
\end{multline}
Consider now a fixed set of indices $\alpha_1, \dotsc, \alpha_{2 F}$.
Equation~\eqref{eq:Wick-J-explicit} must continue to hold when we restrict the
sums to those terms involving only this particular set of indices.  These
indices must be distinct; otherwise, the product over Grassmann variables
vanishes identically.  Under this restriction, the sums run over all
permutations of the fixed indices. Let $\mathrm{S}_{2 F}$ denote the symmetric
group of permutations of $2 F$ elements. Then
\begin{multline}
  \label{eq:Wick-J}
  \frac{\left( - 1 \right)^F}{(2 F)!} \sum_{\sigma \in \mathrm{S}_{2 F}} \langle
  \prod_{f = 1}^{2 F} J_{\alpha_{\sigma_f}} \Psi_{\alpha_{\sigma_f}} \rangle \\
  = \frac{1}{4^F F!} \sum_{\sigma \in \mathrm{S}_{2 F}} \prod_{f = 1}^F
  J_{\alpha_{\sigma_{2 f - 1}}} \left( \imag \mathcal{G} \mathcal{C}
  \right)_{\alpha_{\sigma_{2 f - 1}}, \alpha_{\sigma_{2 f}}}
  J_{\alpha_{\sigma_{2 f}}}.
\end{multline}

We first reorder the product on the left-hand side of Eq.~\eqref{eq:Wick-J}.
Because pairs of Grassmann variables commute, the labels $\alpha_{\sigma_f}$ may
be replaced by $\alpha_f$. The sum over permutations then yields $(2 F)!$ and
cancels the prefactor. Separating the $J$ and $\Psi$ variables gives
\begin{equation}
  \prod_{f = 1}^{2 F} J_{\alpha_f} \Psi_{\alpha_f} = \left( - 1 \right)^{F}
  \left( \prod_{f = 1}^{2 F} J_{\alpha_f} \right) \left( \prod_{f = 1}^{2 F}
    \Psi_{\alpha_f} \right).
\end{equation}

Next, we rearrange the product on the right-hand side of Eq.~\eqref{eq:Wick-J}.
Separating the $J$ factors from the factors of $\imag \mathcal{G} \mathcal{C}$
and using $\sgn(\sigma)$ for the sign of $\sigma$, we obtain
\begin{equation}
  \prod_{f = 1}^F J_{\alpha_{\sigma_{2 f - 1}}} J_{\alpha_{\sigma_{2 f}}} =
  \sgn(\sigma) \prod_{f = 1}^{2 F} J_{\alpha_f}.
\end{equation}

Having aligned both sides of Eq.~\eqref{eq:Wick-J} with respect to the monomial
$\prod_{f = 1}^{2 F} J_{\alpha_f}$, we compare coefficients and arrive at Wick's
theorem:
\begin{equation}
  \label{eq:Wick-theorem}
  \begin{split}
    \langle \prod_{f = 1}^{2 F} \Psi_{\alpha_f} \rangle & = \frac{1}{2^{2 F} F!}
    \sum_{\sigma \in \mathrm{S}_{2 F}} \sgn(\sigma) \prod_{f = 1}^F \left( \imag
      \mathcal{G} \mathcal{C} \right)_{\alpha_{\sigma_{2 f - 1}},
      \alpha_{\sigma_{2 f}}} \\ & = \frac{1}{2^F} \pf \! \left\{ \left[ \left( \imag
        \mathcal{G} \mathcal{C} \right)_{\alpha, \alpha'}
    \right]_{\alpha, \alpha' \in \{\alpha_1, \dotsc, \alpha_{2 F} \}} \right\}.
  \end{split}
\end{equation}

\subsection{Decoupling of measurement vertices with PHS}
\label{sec:deco-meas-vert}

As shown in Ref.~\cite{Poboiko2023} for symmetry class AIII, when $\mathcal{G}$
in the generalized Hubbard-Stratonovich transformation~\eqref{eq:HS-identity} is
used to describe long-wavelength fluctuations, measurement vertices must be
decoupled simultaneously in all possible channels.  This requirement is
satisfied by summing over all Wick contractions of measurement vertices. We
extend the reasoning of Ref.~\cite{Poboiko2023} to symmetry class BDI, which
clarifies the emergence of factors $\beta = 1/2$~\cite{Poboiko2025}.

We denote a point in spacetime by $\mathbf{x} = (l, t)$ and the corresponding
momentum by $\mathbf{q} = (q, \omega)$, with $\mathbf{q} \cdot \mathbf{x} = q l
- \omega t$. For notational convenience, we work in a finite, continuous
spacetime volume so that the momenta $\mathbf{q}$ are discrete. The integration
over $\Sigma$ in Eq.~\eqref{eq:HS-identity-PHS} enforces
\begin{equation}
  \mathcal{G}_{\alpha, \alpha'}(\mathbf{x}) = - \imag \Psi_{\alpha}(\mathbf{x})
  \bar{\Psi}_{\alpha'}(\mathbf{x}).
\end{equation}
To explicitly implement the assumption that $\mathcal{G}$ represents a slow
mode, we rewrite this identity in momentum space and introduce a UV cutoff
$\Lambda$. Moreover, we replace $\bar{\Psi}$ by $\Psi^{\transpose}$ by
multiplying $\mathcal{G}$ by $\mathcal{C}$:
\begin{equation}
  \label{eq:i-C-G-slow}
  \left( \imag \mathcal{G} \mathcal{C} \right)_{\alpha, \alpha'}(\mathbf{q}) = 
  \theta(\Lambda - \abs{\mathbf{q}}) \sum_{\mathbf{k}} \Psi_{\alpha}(\mathbf{k}
  - \mathbf{q}/2) \Psi_{\alpha'}(- \mathbf{k} - \mathbf{q}/2).
\end{equation}

Consider now a local vertex with $2F$ fields, contributing to the action as
\begin{equation}  
  S[\Psi] = \int \diff^2 \mathbf{x} \prod_{f = 1}^{2 F}
  \Psi_{\alpha_f}(\mathbf{x}) = \sum_{\mathbf{k}_1, \dotsc, \mathbf{k}_{2
      F}} \delta_{0, \sum_{f = 1}^{2 F} \mathbf{k}_f} \prod_{f = 1}^{2 F}
  \Psi_{\alpha_f}(\mathbf{k}_f).
\end{equation}
To project this vertex onto slow modes, we retain only those contributions for
which the momenta are pairwise close,
$\abs{\mathbf{k}_f - \mathbf{k}_{f'}} \lesssim \Lambda$. To enumerate all such
pairings, we start from $\abs{\mathbf{k}_{2 f - 1} - \mathbf{k}_{2 f}}
\lesssim \Lambda$ for $f \in \{1, \dotsc, F\}$ and apply all permutations
$\sigma \in \mathrm{S}_{2F}$. This procedure overcounts equivalent pairings:
each $\sigma$ can be followed by $F!$ pair permutations without changing the
pair structure, and reversing the order within each pair also leaves the pairing
unchanged. Correcting for this overcounting requires division by $2^F F!$:
\begin{multline}
  S[\Psi] = \frac{1}{2^F F!} \sum_{\sigma \in \mathrm{S}_{2 F}}
  \sum_{\mathbf{k}_1, \dotsc, \mathbf{k}_{2F}} \delta_{0, \sum_{f = 1}^{2 F}
    \mathbf{k}_f} \\ \times \left[ \prod_{f = 1}^F \theta \! \left( \Lambda -
      \abs{\mathbf{k}_{\sigma_{2 f - 1}} - \mathbf{k}_{\sigma_{2 f}}} \right)
  \right] \left[ \prod_{f = 1}^{2 F} \Psi_{\alpha_f}(\mathbf{k}_f) \right].
\end{multline}

We reorganize the product of fields:
\begin{equation}
  \prod_{f = 1}^{2 F} \Psi_{\alpha_f}(\mathbf{k}_f) = \sgn(\sigma) \prod_{f =
    1}^F \Psi_{\alpha_{\sigma_{2 f - 1}}}(\mathbf{k}_{\sigma_{2 f - 1}})
  \Psi_{\alpha_{\sigma_{2 f}}}(\mathbf{k}_{\sigma_{2 f}}).
\end{equation}
Define the momentum variables
\begin{equation}
  \mathbf{K}_f = \frac{1}{2} \left( \mathbf{k}_{\sigma_{2 f - 1}} -
    \mathbf{k}_{\sigma_{2 f}} \right), \quad \mathbf{q}_f = - \left(
    \mathbf{k}_{\sigma_{2 f - 1}} + \mathbf{k}_{\sigma_{2 f}} \right).
\end{equation}
This yields
\begin{multline}
  S[\Psi] = \frac{1}{2^F F!} \sum_{\sigma \in \mathrm{S}_{2 F}} \sgn(\sigma)
  \sum_{\mathbf{q}_1, \dotsc, \mathbf{q}_{F}} \delta_{0, \sum_{f = 1}^F
    \mathbf{q}_f} \\ \times \prod_{f = 1}^F \theta \! \left( \Lambda -
    \abs{\mathbf{q}_f} \right) \sum_{\mathbf{K}_f}
  \Psi_{\alpha_{\sigma_{2 f - 1}}}(\mathbf{K}_f - \mathbf{q}_f/2)
  \Psi_{\alpha_{\sigma_{2 f}}}(- \mathbf{K}_f - \mathbf{q}_f/2).
\end{multline}
Comparing with Eq.~\eqref{eq:i-C-G-slow}, we identify the structure of the slow
mode on the right-hand side. Transforming back to spacetime, we obtain
\begin{equation}
  \label{eq:S-Psi-slow}
  S[\Psi] = \frac{1}{2^F F!} \sum_{\sigma \in \mathrm{S}_{2 F}} \sgn(\sigma)
  \int \diff^2 \mathbf{x} \prod_{f = 1}^F \left( \imag \mathcal{G} \mathcal{C}
  \right)_{\alpha_{\sigma_{2 f - 1}}, \alpha_{\sigma_{2 f}}}(\mathbf{x}).
\end{equation}

When we now compare Eq.~\eqref{eq:S-Psi-slow} with Wick’s
theorem~\eqref{eq:Wick-theorem}, we observe that the right-hand sides match up
to the factor $1/2^F = \beta^F$ appearing in the latter. This factor may be
absorbed by a redefinition of $\mathcal{G}$. Thus, the projection of the vertex
onto soft modes is equivalent to averaging the original vertex as in
Eq.~\eqref{eq:Wick-exponential}, but with $\mathcal{G}$ replaced by
$\mathcal{G}/\beta$. Writing the vertex as
$S[\Psi] = \int \diff^2 \mathbf{x} \, V[\Psi(\mathbf{x})]$, we obtain
\begin{equation}
  \label{eq:V-G-V-Psi}
  V[\mathcal{G}] = \int \frac{\Diff[\psi^{*}, \psi]}{\pf \! \left(- \imag \beta
      \mathcal{C} \mathcal{G}^{-1} \right)} \, V[\Psi] \e^{\imag \beta
    \bar{\Psi} \mathcal{G}^{-1} \Psi},
\end{equation}
where all fields and matrices are evaluated at the same spacetime point, so the
Pfaffian in the denominator is taken only over Keldysh, charge-conjugation, and
replica spaces. The factor $\beta$ on the right-hand side is missing in
Ref.~\cite{Poboiko2025}.

\subsection{Generalized Hubbard-Stratonovich transformation with PHS}
\label{sec:gener-hubb-strat}

Building on the preceding results, we now apply the generalized
Hubbard-Stratonovich transformation to the particle-hole-symmetrized action
$ S = S_H + S_M $ with the Hamiltonian action~\eqref{eq:S-H-PHS} and the
measurement Lagrangian~\eqref{eq:L-M-PHS}.

We begin by rewriting the measurement Lagrangian~\eqref{eq:L-M-PHS} in terms of
$\mathcal{G} = -\imag \Psi \bar{\Psi}$. For the first two terms, we obtain
\begin{equation}
  \label{eq:Psi-Lambda-sigma-z-G}
  \bar{\Psi} \Lambda \Psi = - \imag \tr(\Lambda \mathcal{G}), \qquad
  \bar{\Psi} \sigma_z \Psi = - \imag \tr(\sigma_z \mathcal{G}).
\end{equation}

To decouple the third term, we use Eq.~\eqref{eq:V-G-V-Psi}. In preparation for
performing the Grassmann integral on the right-hand side, we express the
measurement vertex in exponential form, noting that any exponential containing
$\bar{\Psi}_r \sigma_{\pm} \Psi_r$ truncates after two terms. This yields
\begin{equation}
  V_{\pm}[\Psi] = \prod_{r = 1}^R \left( - \bar{\Psi}_r \sigma_{\pm} \Psi_r \right)
  = \lim_{\epsilon \to 0} (\imag \epsilon)^R
    \e^{\frac{\imag}{\epsilon} \bar{\Psi} \sigma_{\pm} \Psi},
\end{equation}
where the replica index $r$ is implicitly summed in the exponent.  This relation
recasts Eq.~\eqref{eq:V-G-V-Psi} as a Gaussian integral. Using
Eq.~\eqref{eq:Z-J}, we thus obtain
\begin{equation}
  V_{\pm}[\mathcal{G}] = \lim_{\epsilon \to 0} (\imag \epsilon)^R
  \frac{\pf \!\left[ -\imag \beta \mathcal{C} \left( \mathcal{G}^{-1}
        + \frac{1}{\beta \epsilon} \sigma_{\pm} \right) \right]}
       {\pf \!\left( -\imag \beta \mathcal{C} \mathcal{G}^{-1} \right)}.
\end{equation}

To simplify the ratio of Pfaffians, we use that for any skew-symmetric
$4R \times 4R$ matrix $A$, $\pf(A)^{-1} = \pf(A^{-1})$. We then apply
Eq.~\eqref{eq:pf-pf-exp-tr-ln} with $D = 2R$. After that, splitting the trace
over Keldysh, charge-conjugation, and replica spaces as $\tr = \trK \trCR$, and
writing the prefactor as
\begin{equation}
  \epsilon^R = \e^{\beta \trCR[\ln(\epsilon)]},
\end{equation}
we obtain
\begin{equation}
  \label{eq:V-G-lim}
  V_{\pm}[\mathcal{G}] = \imag^R \lim_{\epsilon \to 0} \e^{ \beta
    \trCR \ln \left[ \epsilon \exp \trK \ln \left( 1 +
        \frac{1}{\beta \epsilon} \sigma_{\pm} \mathcal{G} \right) 
    \right]}.
\end{equation}

To take the limit $\epsilon \to 0$, we expand $\exp \trK \ln$ as a power series
in $1/\epsilon$. Note that since $\mathcal{G}$ is a matrix in Keldysh,
charge-conjugation, and replica spaces, $\trK$ is only a partial trace, and
$\exp \trK \ln$ does not reduce to a determinant in Keldysh space. The expansion
of such a determinant in two dimensions would terminate at second order. While
this simplification does not generally occur when $\trK$ is a partial trace, the
appearance of $\sigma_{\pm}$ causes the expansion to terminate already at first
order, as we show in the following.

We obtain a formal series expansion by using the expansions of the exponential
function and the logarithm:
\begin{multline}
  \e^{\trK \ln \left( 1 + \frac{1}{\beta \epsilon} \sigma_{\pm} \mathcal{G} \right)}
  = 1 + \sum_{K = 1}^{\infty} \frac{1}{(\beta \epsilon)^K}
  \sum_{n = 1}^{\infty} \frac{(-1)^{n + K}}{n!} \\
  \times \sum_{k_1, \dotsc, k_n = 1}^{\infty}
  \delta_{K, \sum_{i = 1}^n k_i}
  \prod_{i = 1}^n \frac{1}{k_i}
  \trK \!\left[ (\sigma_{\pm} \mathcal{G})^{k_i} \right].
\end{multline}
To simplify the final product, we note first that for any $2 \times 2$ matrices
$A$ and $B$, one can check by explicit calculation that
\begin{equation}
  \label{eq:trace-factorization}
  \tr(\sigma_{\pm} A \sigma_{\pm} B)
  = \tr(\sigma_{\pm} A)\, \tr(\sigma_{\pm} B).
\end{equation}
This factorization property can be seen to extend to partial traces by using the
Schmidt decomposition of $\mathcal{G}$,
\begin{equation}
  \mathcal{G} = \sum_{\mu = 0}^3 \sigma_{\mu} \otimes \mathcal{G}_{\mu},
\end{equation}
where $\sigma_{\mu}$ are Pauli matrices in Keldysh space $\mathcal{G}_{\mu}$ are
$2 R \times 2 R$ matrices in charge-conjugation and replica spaces. We find
\begin{equation}
  \trK \! \left[ (\sigma_{\pm} \mathcal{G})^k \right]
  = \trK(\sigma_{\pm} \mathcal{G})^k.
\end{equation}
Thus,
\begin{equation}
  \label{eq:exp-tr-ln-c-K}
  \e^{\trK \ln \left( 1 + \frac{1}{\beta \epsilon}
    \sigma_{\pm} \mathcal{G} \right)}
  = 1 + \sum_{K = 1}^{\infty}
    c_K \left[ \frac{1}{\beta \epsilon}
      \trK(\sigma_{\pm} \mathcal{G})^K \right]^K,
\end{equation}
where
\begin{equation}
  c_K = \sum_{n = 1}^{\infty} \frac{(-1)^{n + K}}{n!}
  \sum_{k_1, \dotsc, k_n = 1}^{\infty}
  \delta_{K, \sum_{i = 1}^n k_i}
  \prod_{i = 1}^n \frac{1}{k_i}.
\end{equation}

Now consider a $1 \times 1$ matrix $A$ such that
$\det(1 + A/\epsilon) = 1 + A/\epsilon$. On the other hand, we can write
$\det(1 + A/\epsilon) = \exp \tr \ln(1 + A/\epsilon)$ and use the exact same
expansion as above---obviously, a factorization property analogous to
Eq.~\eqref{eq:trace-factorization} applies also to $1 \times 1$ matrices. Then,
by comparing the coefficients of equal powers of $1/\epsilon$ in
$\det(1 + A/\epsilon) = 1 + A/\epsilon$ and the right-hand side of
Eq.~\eqref{eq:exp-tr-ln-c-K}, we find $c_1 = 1$ and $c_K = 0$ for $K \geq
2$. Therefore,
\begin{equation}
  \e^{\trK \ln \left( 1 + \frac{1}{\beta \epsilon}
      \sigma_{\pm} \mathcal{G} \right)}
  = 1 + \frac{1}{\beta \epsilon} \trK(\sigma_{\pm} \mathcal{G}).
\end{equation}
Inserting this into Eq.~\eqref{eq:V-G-lim} and taking the limit $\epsilon \to 0$
yields
\begin{equation}
  \label{eq:V-G}
  V_{\pm}[\mathcal{G}] = \imag^R \e^{\beta \trCR \ln \trK \left(
      \frac{1}{\beta} \sigma_{\pm} \mathcal{G} \right)}.
\end{equation}
Finally, substituting Eqs.~\eqref{eq:Psi-Lambda-sigma-z-G} and \eqref{eq:V-G}
into Eq.~\eqref{eq:L-M-PHS} yields Eq.~\eqref{eq:L-M-G-PHS}.

With the measurement Lagrangian decoupled, the remaining Grassmann integral in
the Keldysh partition function is quadratic and can be evaluated using
Eq.~\eqref{eq:Z-J}. Leaving the summation over lattice sites and integration
over time implicit, we obtain
\begin{equation}
  \begin{split}
    \int \Diff[\psi^{*}, \psi] \, \e^{\imag \bar{\Psi} G^{-1} \Psi} & = \Pf \!
    \left( - \imag \mathcal{C} G^{-1} \right) \\ & = \left( - 1 \right)^{R L N}
    \e^{\imag \Tr \left[ \beta \ln \left( G^{-1} \right) \right]},
  \end{split}
\end{equation}
where the Pfaffian is taken over Keldysh, charge-conjugation, replica, position,
and time spaces, and where $G^{-1} = G_0^{-1} + \imag \Sigma$ with $G_0^{-1}$
defined in Eq.~\eqref{eq:G-0}. In the second equality we used
Eq.~\eqref{eq:pf-pf-exp-tr-ln} with $A = \sigma_z \mathcal{C}$ and
$B = \mathcal{C} G^{-1}$, and $D = 2 R L N$, where $LN$ is the dimension of
position-time space, that is, lattice size times the number of time steps. The
sign arises from
\begin{equation}
  \Pf(\sigma_z \mathcal{C}) = \pfKC(\sigma_z \mathcal{C})^{R L N} = (-1)^{R L N}.
\end{equation}
We absorb this sign into the integration measure $\Diff[\mathcal{G}, \Sigma]$.
The resulting Keldysh partition function is then given by
Eqs.~\eqref{eq:Z-R-d-G-d-Sigma} and \eqref{eq:S-G-Sigma}, with $S_0$ from
Eq.~\eqref{eq:S-0-G-Sigma-beta}.

\section{Numerical methods}
\label{sec:numerical-methods}

In what follows we provide a description of the numerical procedure we use to
simulate quantum trajectories for imbalanced and inefficient fermion
counting. Our approach builds on the formulation of quantum-jump trajectories
for efficient detection in Sec.~\ref{sec:quantum-jump-processes} together with
its extension to inefficient detection in Sec.~\ref{sec:inefficient-detection}.

The dynamics proceed in discrete time steps $\Delta t$. For inefficient
detection, each step is further decomposed into two substeps of duration
$\eta \Delta t$ and $\left(1 - \eta\right)\Delta t$, respectively. The first
substep implements measurement-conditioned evolution; its numerical treatment is
presented in Sec.~\ref{sec:cond-evol} below. The second substep implements
unconditional evolution, corresponding to an average over measurement outcomes;
the associated algorithm is detailed in Sec.~\ref{sec:unconditional-evolution}.
Additional implementation details for the no-jump evolution during the first
substep are provided in Sec.~\ref{sec:no-jump-evolution}.

For generality, we formulate the algorithm in terms of arbitrary gain and loss
jump operators,
\begin{equation}
  \label{eq:general-gain-loss-jump-operators}
  \hat{c}_{+, l} = \sqrt{2} \sum_{l' = 1}^L B_{+, l, l'}
  \hat{\psi}_{l'}^{\dagger}, \quad
  \hat{c}_{-, l} = \sqrt{2} \sum_{l' = 1}^L B_{-, l, l'} \hat{\psi}_{l'}.  
\end{equation}
Moreover, for later use, we denote the $l$th row and column of a matrix $A$ by
$A_{l, *}$ and $A_{*, l}$, respectively, and introduce
\begin{equation}
  \label{eq:M-+---l}
  M_{+, l} = \left( B_+^{\transpose} \right)_{*, l} B_{+, l, *}^{*}, \qquad
  M_{-, l} = \left( B_-^{\dagger} \right)_{*, l} B_{-, l, *},
\end{equation}
along with the Hermitian bath matrices
\begin{equation}
  M_+ = \sum_{l = 1}^L M_{+, l} = B_+^{\transpose} B_+^{*}, \qquad M_- = \sum_{l
    = 1}^L M_{-, l} = B_-^{\dagger} B_-.
\end{equation}
The corresponding expressions for local gain and loss as in
Eq.~\eqref{eq:gain-loss-jump-operators} are obtained by setting
\begin{equation}
  \label{eq:B-gain-loss}
  B_{\pm} = \sqrt{\gamma_{\pm}/2}.
\end{equation}

\subsection{Conditional evolution}
\label{sec:cond-evol}

During the first time substep, the evolution is conditioned on the measurement
outcome, and the algorithm samples different outcomes stochastically. However,
we do not directly implement the trajectory evolution described in
Secs.~\ref{sec:quantum-jump-processes}
and~\ref{sec:inefficient-detection}. Instead, exploiting the locality of the
jump processes, we construct a more efficient scheme.

To begin, we outline why the first-order-in-time quantum-jump
method~\cite{Daley2014} of Secs.~\ref{sec:quantum-jump-processes}
and~\ref{sec:inefficient-detection} cannot be applied directly. The probability
that any jump occurs during a single time step, given the system state
$\hat{\rho}(t)$, is the sum of the individual jump probabilities in
Eq.~\eqref{eq:P-alpha-inefficient}. With $\alpha=\pm$ denoting gain and loss and
$l$ the site index,
\begin{equation}
  P_{\alpha, l} = \tr \! \left[ \mathcal{K}_{1, \alpha, l} \hat{\rho}(t) \right].
\end{equation}
Substituting here the jump superoperator~\eqref{eq:K-1-superoperator} together
with Eq.~\eqref{eq:K-1-0-a} and the general gain and loss jump
operators~\eqref{eq:general-gain-loss-jump-operators}, one obtains
\begin{equation}
  \begin{split}    
    P_{+, l} & = 2 \eta \Delta t \tr \! \left\{ M_{+, l} \left[ 1 - D(t) \right]
    \right\} = \eta \gamma_+ \Delta t \left[ 1 - D_{l, l}(t) \right], \\ P_{-,
      l} & = 2 \eta \Delta t \tr \! \left[ M_{-, l} D(t) \right]
    = \eta \gamma_- \Delta t D_{l, l}(t),
  \end{split}
\end{equation}
where $D(t)$ is the single-particle density
matrix~\eqref{eq:single-particle-density-matrix}, the matrices $M_{\pm,l}$ are
defined in Eq.~\eqref{eq:M-+---l}, and the final equalities apply in the case of
local gain and loss. The total jump probability during the time step form $t$ to
$t + \Delta t$ becomes
\begin{equation}
  P_{\mathrm{jump}}
  = \sum_{\alpha=\pm} \sum_{l=1}^L P_{\alpha, l}
  = \eta \Delta t \left\{ L \gamma_+ - \Delta \gamma \tr[D(t)] \right\},
\end{equation}
where $\tr[D(t)] = \langle \hat{N}(t) \rangle$ is the mean particle number.
Since $\langle\hat{N}(t)\rangle$ is typically of order $L$, enforcing
$P_{\mathrm{jump}} \ll 1$, as required for the validity of the first-order
method~\cite{Daley2014}, forces $\Delta t \sim L^{-1}$, which is computationally
prohibitive. To circumvent this limitation, we treat the local measurement
processes at different sites independently. We first describe the resulting
algorithm and then discuss its validity. Although our approach differs from the
standard first-order method outlined above in several respects, we argue below
that both schemes are equivalent to first order in the time step.

The algorithm employed in our numerical simulations proceeds as follows. During
each time step, all lattice sites $l \in \{ 1, \dotsc, L \}$ are visited in a
randomly chosen order. At each site $l$, we determine whether a jump occurs and,
if so, update the state of the system according to the type of the jump that
takes place. The probability for a jump to occur at site $l$ is
\begin{equation}
  \label{eq:P-jump-l}
  P_{\mathrm{jump}, l}
  = \sum_{\alpha = \pm} P_{\alpha, l}
  = \eta \Delta t \left[ \gamma_+ - \Delta \gamma D_{l, l}(t) \right].
\end{equation}
For each site we draw a uniformly distributed random number $r_1 \in [0,1]$; a
jump occurs if $r_1 < P_{\mathrm{jump}, l}$. A second random number
$r_2 \in [0,1]$ determines whether the jump is gain or loss, with loss selected
if $r_2 < P_{-,l}/P_{\mathrm{jump},l}$. In practice, we select $\Delta t$ such
that the jump probability at a single site, estimated from
Eq.~\eqref{eq:P-jump-l} with $\eta = 1$ and half filling as
$P_{\mathrm{jump}, l} = \gamma \Delta t$, is $1.25 \, \%$. We verified that the
results remain unchanged upon doubling or halving $\Delta t$.

When a jump occurs, the state is transformed as
\begin{equation}
  \hat{\rho}(t) \to
  \left. \mathcal{K}_{1,\pm,l}\hat{\rho}(t) \middle/ \tr \! \left[
      \mathcal{K}_{1,\pm,l}\hat{\rho}(t) \right] \right..
\end{equation}
This transformation preserves Gaussianity. Hence the induced change in the
single-particle density matrix can be computed using Wick’s theorem. For gain of
a fermion at site $l$,
\begin{equation}
  \begin{split}
    D(t) \to
    D(t) + \frac{\left[ 1 - D(t) \right] M_{+, l} \left[ 1 - D(t) \right]}
    {\tr\{ M_{+, l} \left[ 1 - D(t) \right] \}}.
  \end{split}
\end{equation}
and for a fermion loss at site $l$,
\begin{equation}
  D(t) \to D(t) - \frac{D(t) M_{-, l} D(t)}{\tr[M_{-, l} D(t)]}.
\end{equation}
When gain and loss occur locally at single lattice sites, multiplication by
$M_{\pm,l}$ merely selects specific rows and columns of $D(t)$, allowing the
updates to be implemented without explicit matrix multiplications. In
particular, for gain at site $l$, using the notation of Eq.~\eqref{eq:M-+---l},
\begin{equation}
  D(t) \to
  D(t) + \frac{[1 - D(t)]_{*,l} [1 - D(t)]_{l,*}}{1 - D_{l,l}(t)}.
\end{equation}
and for loss at site $l$,
\begin{equation}
  D(t) \to
  D(t) - \frac{D_{*,l}(t) D_{l,*}(t)}{D_{l,l}(t)}.
\end{equation}

As stated above, in each time step, this procedure is applied to all lattice
sites in random order. Thus, our algorithm permits multiple jumps to occur
during the same time step at different lattice sites. By contrast, the standard
first-order method allows only a single jump to occur at any given time.

After sweeping through the chain and applying all jumps, we carry out continuous
no-jump evolution generated by $\hat{K}_{1,0}$ in Eq.~\eqref{eq:K-1-0-a}. This
evolution acts on every lattice site, including those at which jumps occurred.
However, the resulting error is negligible because a given site undergoes only
finitely many jumps in any finite time~\cite{Wiseman2010}. The continuous
evolution is
\begin{equation}
  \label{eq:no-jump-evolution}
  \hat{\rho}(t+\eta\Delta t)
  = \frac{\mathcal{K}_{1,0}\hat{\rho}(t)}
      {\tr\!\left[ \mathcal{K}_{1,0}\hat{\rho}(t) \right]}
  = \frac{\hat{K}_{1,0}\hat{\rho}(t)\hat{K}_{1,0}^{\dagger}}
      {\tr\!\left[ \hat{K}_{1,0}\hat{\rho}(t)\hat{K}_{1,0}^{\dagger} \right]}.
\end{equation}
As shown in Sec.~\ref{sec:no-jump-evolution}, this can be recast in terms of the
single-particle density matrix as
\begin{multline}
  \label{eq:no-jump-evolution-D}
  D(t + \eta \Delta t)
  = U(\eta \Delta t) \\
  \times \left\{
      D(t) + \mathrm{e}^{-\eta\Delta\gamma\Delta t} \left[ 1 - D(t) \right]
    \right\}^{-1}
    D(t)
    U^{\dagger}(\eta \Delta t),
\end{multline}
where $U(t) = \mathrm{e}^{-\mathrm{i} H t}$. This completes the first time
substep.

Before turning to the unconditional evolution in the second substep, we briefly
discuss the validity of our algorithm. Our approach amounts to applying the
standard first-order method independently and in parallel at each lattice site.
This procedure is clearly justified for uncoupled lattice sites in the absence
of hopping, that is, for $J = 0$ in Eq.~\eqref{eq:Hamiltonian-matrix}. When
fermions are allowed to hop between sites, a jump at one lattice site affects
the statistics of subsequent jumps at other sites. In our approach, this effect
is not accounted for when multiple jumps occur simultaneously within the same
time step. However, a jump at site $l$ can influence other jumps within a given
time step only inside a light cone of width $\sim J \Delta t$. Moreover, even
within this light cone, the resulting correction is itself of order $\Delta t$.
We therefore expect our approach to reproduce the standard first-order method up
to higher-order corrections in the time step.

While we have not attempted a fully rigorous analysis of the error introduced by
treating local measurements independently, we performed numerical benchmarks.
We found observables that are linear in the state to agree quantitatively with
the unconditional evolution that can be computed numerically exactly as
described in Sec.~\ref{sec:unconditional-evolution}. Furthermore, for balanced
and efficient fermion counting with $\gamma_+ = \gamma_-$ and $\eta = 1$, our
results agree well with those of Ref.~\cite{Starchl2025}, obtained using a
numerically exact method~\cite{Daley2014}. The implementation in
Ref.~\cite{Starchl2025} relies on the rates being balanced,
$\gamma_+ = \gamma_-$, in which case fermion counting can be rephrased as a
random generalized measurement. In that case, the waiting time between jumps can
be sampled from an exponential distribution, and the evolution between jumps can
be implemented through exact diagonalization.

\subsection{Unconditional evolution}
\label{sec:unconditional-evolution}

Unconditional evolution during the second time substep is governed by
Eq.~\eqref{eq:K-2}. This equation corresponds to the first-order expansion in
$\Delta t$ of the evolution over the interval $\left( 1 - \eta \right) \Delta t$
generated by the Liouvillian~\eqref{eq:Liouvillian}. In our numerical
implementation, we treat this interval exactly by evolving the state under the
master equation
\begin{equation}
  \frac{\diff \hat{\rho}(t)}{\diff t} = \mathcal{L} \hat{\rho}(t),
\end{equation}
for a duration $\left( 1 - \eta \right) \Delta t$. This master equation can be
recast as a closed equation of motion for the single-particle density
matrix~\eqref{eq:single-particle-density-matrix}, which admits an exact
solution~\cite{Starchl2024}.

We consider the general quadratic Hamiltonian in Eq.~\eqref{eq:Hamiltonian} and
the gain-loss jump operators in
Eq.~\eqref{eq:general-gain-loss-jump-operators}. The single-particle density
matrix obeys
\begin{equation}
  \label{eq:d-D-d-t}
  \frac{\diff D(t)}{\diff t}
  = - \imag \left[ Z D(t) - D(t) Z^{\dagger} \right] + 2 M_+,
\end{equation}
with
\begin{equation}
  Z = H - \imag M, \qquad M = M_- + M_+.
\end{equation}
The solution to Eq.~\eqref{eq:d-D-d-t} with initial condition $D_0$ at $t = 0$
is
\begin{equation}
  \label{eq:unconditional-evolution}
  D(t) = Q(t) D_0 Q(t)^{\dagger} + R(t),
\end{equation}
where
\begin{equation}
  Q(t) = \e^{-\imag Z t}.
\end{equation}
To specify $R(t)$, write $Z = V \Lambda V^{-1}$, with
$\Lambda = \diag(\lambda_1, \dotsc, \lambda_L)$. Define the elementwise or
Hadamard product $A \circ B$ via
\begin{equation}
  \left( A \circ B \right)_{l,l'} = A_{l,l'} B_{l,l'}.
\end{equation}
Then
\begin{equation}
  R(t)
  = - 2 \imag \,
    V \left[ \left( V^{-1} M_+ V^{-\dagger} \right) \circ K(t) \right] V^{\dagger},
\end{equation}
where
\begin{equation}
  K_{l,l'}(t)
  = \frac{1 - \e^{-\imag \left( \lambda_l - \lambda_{l'}^{*} \right) t}}
         {\lambda_l - \lambda_{l'}^{*}}.
\end{equation}

As noted above, our numerical implementation uses
Eq.~\eqref{eq:unconditional-evolution} to evolve the single-particle density
matrix over the second substep, from the beginning time $t$ to
$t + \left( 1 - \eta \right) \Delta t$.

\subsection{No-jump evolution}
\label{sec:no-jump-evolution}

It remains to derive Eq.~\eqref{eq:no-jump-evolution-D}. For the jump operators
in Eq.~\eqref{eq:general-gain-loss-jump-operators}, the effective
Hamiltonian~\eqref{eq:H-eff} becomes
\begin{equation}  
  \hat{H}_{\mathrm{eff}} = \hat{H} - \imag \left[ \hat{H}' + \tr(M_+) \right].
\end{equation}
where
\begin{equation}
  \hat{H}' = - \sum_{l, l' = 1}^L \hat{\psi}_l^{\dagger} \Delta M_{l, l'}
  \hat{\psi}_{l'}, \qquad \Delta M = M_+ - M_-.
\end{equation}
For local gain and loss, this expression reduces to
\begin{equation}
  \label{eq:H-eff-gain-loss}
  \hat{H}_{\mathrm{eff}} = \hat{H} + \frac{\imag}{2} \left( \Delta \gamma
    \hat{N} - L \gamma_+ \right).
\end{equation}
Up to first order in $\Delta t$, the no-jump Kraus operator in
Eq.~\eqref{eq:K-1-0-a} can therefore be written as
\begin{equation}
  \hat{K}_{1, 0} = \e^{- \imag
    \eta \hat{H}_{\mathrm{eff}} \Delta t} = \e^{- \imag \eta \hat{H} \Delta t}
  \e^{- \eta \hat{H}' \Delta t} \e^{- \eta \tr(M_+) \Delta t}.
\end{equation}
The final exponential drops out due to the normalization in
Eq.~\eqref{eq:no-jump-evolution} and may hence be omitted. The remaining two
factors generate unitary and non-unitary evolution, respectively. We define
\begin{equation}
  \hat{U}(t) = \e^{- \imag \hat{H} t}, \qquad
  \hat{V}(t) = \e^{- \hat{H}' t} = \e^{\Delta \gamma \hat{N} t/2}
  = \hat{V}^{\dagger}(t),
\end{equation}
so that
\begin{equation}
  \hat{K}_{1, 0} = \hat{U}(\eta \Delta t) \hat{V}(\eta \Delta t).
\end{equation}
The evolution in Eq.~\eqref{eq:no-jump-evolution} then factorizes as
\begin{equation}
  \label{eq:factorized-evolution-first-substep}
  \hat{\rho}(t + \eta \Delta t) = \hat{U}(\eta \Delta t) \frac{\hat{V}(\eta
    \Delta t) \hat{\rho}(t) \hat{V}(\eta \Delta t)}{\tr \! \left[ \hat{V}(\eta
      \Delta t) \hat{\rho}(t) \hat{V}(\eta \Delta t) \right]}
  \hat{U}^{\dagger}(\eta \Delta t).
\end{equation}

We first analyze the evolution generated by $\hat{V}(t)$. Consider the general
expression
\begin{equation}
  \label{eq:nonunitary-evolution}
  \hat{\rho}(t) = \frac{\hat{V}(t) \hat{\rho}_0 \hat{V}(t)}{\tr \! \left[
      \hat{V}(t) \hat{\rho}_0 \hat{V}(t) \right]},
\end{equation}
which reduces to the form required in
Eq.~\eqref{eq:factorized-evolution-first-substep} when we set
$t = \eta \Delta t$ and $\hat{\rho}_0 = \hat{\rho}(t)$. Differentiating with
respect to $t$ yields
\begin{equation}
  \frac{\diff \hat{\rho}}{\diff t} = - \left\{ \hat{H}', \hat{\rho} \right\}
  + 2 \tr \! \left( \hat{H}' \hat{\rho} \right) \hat{\rho}.
\end{equation}
This equation can be rewritten as an evolution equation for the single-particle
density matrix~\eqref{eq:single-particle-density-matrix}.  Since $\hat{\rho}(t)$
in Eq.~\eqref{eq:nonunitary-evolution} is a normalized product of Gaussian
operators, it remains a normalized Gaussian state. Wick’s theorem therefore
applies, leading to
\begin{equation}
  \frac{\diff D}{\diff t} = \left\{ \Delta M, D \right\} - 2 D \Delta M D =
  \Delta \gamma \left( 1 - D \right) D.
\end{equation}

To solve this equation, we assume that $D$ is invertible and introduce the
matrix
\begin{equation}
  Y = D^{-1} - 1, \qquad D = \left( 1 + Y \right)^{-1}.
\end{equation}
Using
\begin{equation}
  \frac{\diff D^{-1}}{\diff t} = - D^{-1} \frac{\diff D}{\diff t} D^{-1},
\end{equation}
we obtain
\begin{equation}  
  \frac{\diff Y}{\diff t} = \frac{\diff D^{-1}}{\diff t} = - \left\{ \Delta M, Y
  \right\}.  
\end{equation}
The solution of this equation reads
\begin{equation}
  Y(t) = \e^{- \Delta M t} Y_0 \e^{- \Delta M t},
\end{equation}
which yields
\begin{equation}
  \begin{split}
    D(t) & = \left[ 1 + \e^{- \Delta M t} \left( D_0^{-1} - 1 \right) \e^{-
        \Delta M t} \right]^{-1} \\
         & = \left[ D_0 + \e^{- \Delta \gamma t}
      \left( 1 - D_0 \right) \right]^{-1} D_0.
  \end{split}
\end{equation}

In Eq.~\eqref{eq:factorized-evolution-first-substep}, we require
$t = \eta \Delta t$ and $D_0 = D(t)$. Finally, also the unitary evolution in
Eq.~\eqref{eq:factorized-evolution-first-substep} can be reformulated in terms
of the single-particle density matrix using the approach of
Sec.~\ref{sec:unconditional-evolution}. We thus obtain
Eq.~\eqref{eq:no-jump-evolution-D}.


%

\end{document}